\documentclass[aps,twocolumn,superscriptaddress,showpacs,floatfix,10pt]{revtex4-2}
\usepackage{amsfonts}
\usepackage{amsmath}
\usepackage{bbm}
\usepackage{amssymb}
\usepackage{graphicx}%
\usepackage{footmisc}
\usepackage{bm}
\usepackage{amsthm}
\usepackage{enumitem}

\usepackage{braket}
\usepackage{hyperref}
\usepackage{cleveref}
\usepackage{longtable,booktabs}
\usepackage{subfigure}

\def\Xint#1{\mathchoice
    {\XXint\displaystyle\textstyle{#1}}%
    {\XXint\textstyle\scriptstyle{#1}}%
    {\XXint\scriptstyle\scriptscriptstyle{#1}}%
    {\XXint\scriptscriptstyle\scriptscriptstyle{#1}}%
      \!\int}
\def\XXint#1#2#3{{\setbox0=\hbox{$#1{#2#3}{\int}$}
    \vcenter{\hbox{$#2#3$}}\kern-.5\wd0}}
\def\dashint{\Xint-}

\newcommand{\abs}[1]{\lvert#1\rvert}
\newcommand{\norm}[1]{\lVert#1\rVert}

\begin{document}

\title{Nonperturbative Resolvent Hierarchies for Dense-Spectrum Interacting Many-Body Systems}

\author{Zhiqiang Huang}
\email{zqhuang@hubu.edu.cn}
\affiliation{School of Physics, Hubei University, Wuhan 430062, China.}
\author{Qing-yu Cai}
\email{qycai@hainanu.edu.cn}
\affiliation{Center for Theoretical Physics, Hainan University, Haikou 570228, China}
\affiliation{School of Information and Communication Engineering, Hainan University, Haikou 570228, China}

\date{\today}

\begin{abstract}
	We develop a nonperturbative framework for generic nonintegrable many-body systems that reorganizes the conventional many-body expansion at the level of diagonal Green's functions. Starting from exact projection identities and the spectral representation of the resolvent, we derive a recursive self-energy hierarchy in which cross-correlated propagation processes are systematically rewritten in terms of diagonal resolvents. Under a diagonal closure approximation, the hierarchy truncates self-consistently while remaining systematically improvable.

The framework combines two complementary nonperturbative structures. First, a Lanczos continued-fraction representation provides a recursive single-resolvent backbone that naturally generates non-Lorentzian spectral features beyond standard self-consistent Born approximations. Second, an exact projected multi-resolvent hierarchy introduces nonlocal frequency couplings through products of resolvents and their Hilbert transforms. These multi-resolvent contributions mix parity sectors under energy reflection and constitute the leading microscopic mechanism for spectral skewness, which is absent in parity-preserving single-resolvent closures.

To solve the resulting self-consistent equations, we employ a hierarchy of Lorentzian, Gaussian, and hybrid Voigt-type ansätze together with an effective Faddeeva self-energy representation that preserves analyticity and causality. Within this scheme, spectral broadening, spectral tails, and higher-order fluctuation effects emerge from the interplay between continued-fraction recursion and multi-resolvent correlations.

The approach requires neither small expansion parameters nor selective diagrammatic resummations. Instead, it is closed through ETH-type statistical assumptions appropriate for dense chaotic spectra. The resulting framework provides a unified route from microscopic interactions to emergent spectral structures and fluctuation phenomena, revealing a progressive enrichment of resolvent dynamics from single-pole self-consistent approximations to continued-fraction renormalization and ultimately to multi-resolvent interference effects.

\end{abstract}

\maketitle
\section{Introduction}
The theoretical description of quantum many-body systems has long relied on perturbative expansions in a small parameter. While highly successful in weakly interacting regimes, such expansions suffer from fundamental limitations, including the breakdown of low-order truncations in systems with dense spectra and strong correlations \cite{D52,BW69}. In generic nonintegrable many-body systems, the energy spectrum becomes exponentially dense with system size, and local expansions around a reference point are insufficient for capturing global spectral properties.

A central limitation of conventional approaches lies not only in the presence of a small expansion parameter, but in \textbf{how} correlations are organized. Standard perturbative methods expand physical quantities locally—typically in powers of the interaction strength—and truncate the resulting series at low orders. Higher-order terms quickly proliferate, and even when resummations are attempted (e.g., through diagrammatic techniques), the underlying expansion remains tied to a perturbative ordering that is often uncontrolled in nonintegrable regimes where the density of states is exponentially large.

Rather than introducing a different expansion variable, this work reorganizes the structure of the problem at the level of the resolvent. The resolvent (Green's function) provides a natural language for encoding such global information, as its analytic structure is fully determined by the poles corresponding to the eigenvalues of the Hamiltonian \cite{E06}. In conventional many-body theory, resolvent-based approaches typically lead to the Dyson equation and diagrammatic expansions of the self-energy \cite{FW12}, often truncated at low orders such as in the self-consistent Born approximation. While successful in certain regimes, these approaches rely on perturbative truncations and generally fail to capture higher-order correlations in strongly interacting systems.

In parallel, the Eigenstate Thermalization Hypothesis (ETH) has emerged as a cornerstone for understanding thermalization in isolated quantum systems \cite{D91,S94,RDM08}. ETH provides a statistical description of matrix elements, suggesting that off-diagonal components behave as effectively random variables with well-defined variance structure. This insight enables the replacement of microscopic complexity with statistical regularity, forming the basis of many modern approaches to quantum chaos and thermalization \cite{AKP16}.

In this work, we combine these two perspectives into a unified methodological framework. The central result is an exact projected recursive reorganization of cross-correlated contributions into a projected multi-resolvent hierarchy. Under a diagonal closure approximation, this hierarchy is formally expressed in terms of diagonal Green's functions alone. This hierarchy is subsequently treated under ETH-type statistical assumptions and solved using a hierarchical ansatz strategy. The framework is built on the following principles:

\begin{itemize}
    \item \textbf{Resolvent-based formulation.} We start from the spectral representation of the resolvent, which is an exact identity rather than an approximation. This allows us to write closed recursive equations that directly encode the system's global features, such as level broadening and the distribution of eigenstate overlaps \cite{HHTG24}.
    \item \textbf{Statistical treatment of local fluctuations.} Following the spirit of ETH, we treat off-diagonal matrix elements statistically under ETH-type assumptions, effectively capturing their fluctuating nature. This allows us to average over rapidly fluctuating cross-correlated terms and obtain a closed set of mean-field equations, while the full hierarchy retains controlled fluctuations.
    \item \textbf{Systematic recursive hierarchy of cross-correlated terms.} Crucially, we do not discard the cross-correlated contributions entirely. Instead, we derive an exact recursive re-expansion into projected diagonal resolvents, which under a cavity-type diagonal closure reduces to an expansion in terms of ordinary diagonal resolvents. This expansion reveals that the mean-field result is the leading order, while higher orders generate contributions that control distribution tails, branch splitting, and higher-order moments.
    \item \textbf{Hierarchical ansatz strategy.} The resulting self-consistent equations are solved through a hierarchy of ans\"atzes—Lorentzian for the bulk, Gaussian for the tails, and a hybrid Lorentzian–Gaussian (Voigt) form for a unified description—providing a quantitatively accurate characterization of the full distribution.
\end{itemize}

The hierarchical structure is systematically improvable: higher-order multi-resolvent contributions can be incorporated without resorting to finite-order truncations or diagrammatic resummations. This distinguishes our approach from conventional methods such as the self-consistent Born approximation (SCBA) \cite{E06,FW12}, cumulant expansions \cite{K62}, or linked-cluster techniques. Rather than producing an increasingly complex set of diagrams, the projected hierarchy is exact; replacing projected resolvents by full diagonal resolvents yields an approximate closure in terms of diagonal resolvents alone (see \cref{sec:cross_expansion} and Appendices). In contrast to diagrammatic approaches, the present hierarchy is generated algebraically from resolvent identities and projection operators, and is organized by the multiplicity of diagonal resolvents rather than by perturbative order. This work provides a systematic reorganization of many-body expansions in resolvent space, establishing a direct connection between microscopic interactions and macroscopic statistical behavior. In this sense, the framework provides a direct link between microscopic correlations and emergent spectral structures in generic nonintegrable systems. The framework is particularly suited for the dense, chaotic spectra of nonintegrable systems and offers a unified, analytically controlled approach to nonintegrable many-body systems beyond the reach of traditional perturbation theory.

The remainder of this paper is organized as follows. \Cref{SME} develops the resolvent self-consistent equation, introduces the statistical averaging based on ETH, derives the recursive expansion of cross-correlated terms, and presents the hierarchical ansatz strategy for solving the resulting equations. \Cref{CTM} compares the framework with traditional methods such as perturbation theory, SCBA, and ETH. \Cref{CAG} discusses the conditions of applicability and possible generalizations. \Cref{CON} concludes with a summary and outlook. Several appendices provide detailed derivations, including the structural distinction from finite diagrammatic resummations (\cref{SDDR}), the Hilbert transform of the hybrid ansatz (\cref{HTEA}), the effective self-energy representation (\cref{app:analytic_foundations}), and the limitations of the constant self-energy approximation (\cref{app:const_selfenergy}).

\section{The Structure of the Methodology}\label{SME}
\subsection{Resolvent Self-Consistent Equation and Global Analyticity}\label{RSCE}

Consider a system \( S \) and a bath \( B \) that are initially independent, with the unperturbed Hamiltonian \( H_0 = H_S + H_B \).  
Assume their initial states are energy eigenstates \( \ket{\phi_{\mu i}} = \ket{\phi^{S}_{i}} \otimes \ket{\phi^{B}_{\mu}} \), satisfying \( H_0 \ket{\phi_{\mu i}} = a_{\mu i} \ket{\phi_{\mu i}} \), where \( a_{\mu i} = E_i + \epsilon_\mu \).  
When an interaction \( V \) is introduced, the total Hamiltonian becomes \( H = H_0 + V \), with eigenstates \( \ket{\psi_n} \) obeying \( H \ket{\psi_n} = \lambda_n \ket{\psi_n} \).

Our starting point is the diagonal resolvent
\begin{equation}
    \mathcal{R}_{\mu i}(z) = \bra{\phi_{\mu i}} \frac{1}{z - H} \ket{\phi_{\mu i}},
\end{equation}
which is analytic in the complex plane except for simple poles at the eigenvalues \(\lambda_n\) of \(H\). Its boundary value on the real axis defines the spectral measure:
\begin{equation}
    \frac{1}{\pi} \Im \mathcal{R}_{\mu i}(x - \mathrm{i}0^+) = \sum_n p^{\mu i}_n \delta(x - \lambda_n),
\end{equation}
where \(p^{\mu i}_n = |\braket{\psi_n|\phi_{\mu i}}|^2\).

We employ the resolvent identity
\begin{equation}
    \frac{1}{z-H} = \frac{1}{z-H_0} + \frac{1}{z-H_0} V \frac{1}{z-H},
\end{equation}
together with the projection operator \( \Phi_{\mu i} = I - \ket{\phi_{\mu i}}\bra{\phi_{\mu i}} \).
Following a standard projection procedure, one obtains the exact identity
\begin{equation}\label{eq:CSEQ}
    \mathcal{R}_{\mu i}(z) = \frac{1}{z - a_{\mu i} - V_{\mu i} - \mathcal{G}_{\mu i}(z)},
\end{equation}
where \( V_{\mu i} = \bra{\phi_{\mu i}} V \ket{\phi_{\mu i}} \), and the self-energy is given by
\begin{equation}\label{SE1}
    \mathcal{G}_{\mu i}(z) = \bra{\phi_{\mu i}} V \Phi_{\mu i} \frac{1}{z- \Phi_{\mu i}H \Phi_{\mu i}} \Phi_{\mu i} V \ket{\phi_{\mu i}}.
\end{equation}
This equation is formally analogous to a Dyson equation, but its significance here is different: it provides an \emph{exact projected identity, non-perturbative decomposition} of the resolvent in which all coupling to the rest of the Hilbert space is encoded in the projected resolvent \(\mathcal{G}_{\mu i}(z)\).  
Unlike perturbative expansions, no truncation has been made, and the full analytic structure of \(\mathcal{R}_{\mu i}(z)\) is preserved.

In particular, the eigenvalues \(\lambda_n\) are determined by the nonlinear condition
\begin{equation}
    z - a_{\mu i} - V_{\mu i} - \mathcal{G}_{\mu i}(z) = 0,
\end{equation}
which shows that the pole structure of the resolvent is encoded self-consistently through the energy dependence of \(\mathcal{G}_{\mu i}(z)\). This representation therefore captures the global analytic structure of the spectrum rather than a local expansion around a reference point.

The above self-consistent equation admits an equivalent non-perturbative representation in a Krylov (Lanczos) basis. Starting from the initial state \( |\phi_0\rangle \equiv |\phi_{\mu i}\rangle \), the Lanczos algorithm generates an orthonormal basis \( \{|\phi_n\rangle\} \) in which the Hamiltonian becomes tridiagonal:
\begin{equation}
    H|\phi_n\rangle = b_n|\phi_{n-1}\rangle + a_n|\phi_n\rangle + b_{n+1}|\phi_{n+1}\rangle,
\end{equation}
with \( b_0 = 0 \), \( a_n = \langle\phi_n|H |\phi_n\rangle \), and \( b_{n+1} = \langle\phi_{n+1}|H|\phi_n\rangle \).
Defining the semi-infinite tail subspace $H^{[\ge n]}$, i.e., the restriction of $H$ to $\mathrm{span}\{|\phi_n\rangle, |\phi_{n+1}\rangle, \ldots\}$ with \( R_n(z) = \langle\phi_n|(z- H^{[\ge n]})^{-1}|\phi_n\rangle \), one obtains the exact recurrence
\begin{equation}
    R_n(z) = \frac{1}{z - a_n - b_{n+1}^2 R_{n+1}(z)}.
\end{equation}
In particular, \( \mathcal{R}_{\mu i}(z) = R_0(z) \), leading to the continued-fraction representation
\begin{equation}
    \mathcal{R}_{\mu i}(z)
    =
    \cfrac{1}{z - a_0 - \cfrac{b_1^2}{z - a_1 - \cfrac{b_2^2}{z - a_2 - \ddots}}}.
\end{equation}
This representation is \emph{exactly equivalent} to the projected resolvent equation above and provides a complementary reorganization of the same physics. While the projection formalism expresses the self-energy in terms of couplings to the rest of the Hilbert space, the Lanczos representation absorbs all orders of interaction into the coefficients \(a_n\) and \(b_n\), yielding a one-dimensional recursive structure.

It is important to emphasize that these two formulations serve different purposes.  
The projected resolvent equation makes explicit the internal structure of the self-energy and forms the basis for the hierarchical expansion developed below. In contrast, the Lanczos continued fraction (LCF) provides a compact non-perturbative encoding of the same information within a single resolvent. In particular, the Lanczos representation, while exact, compresses all correlation information into the scalar coefficients \(a_n, b_n\) and thereby obscures the explicit multi-resolvent structure that is essential for analyzing fluctuations beyond mean field.

Thus, the projection-based formulation exposes the multi-resolvent correlation structure, while the Lanczos representation reorganizes it into an effective one-dimensional recursion. The two are fully equivalent at the exact level but lead to different approximation schemes and physical insights.

\subsection{Random Phase Hypothesis and Statistical Averaging}

The self-energy decomposition derived in  \cref{SE1} reads exactly
\begin{equation}\label{SEP1}
    \mathcal{G}_{\mu i}(z) =\mathcal{G}^{\mathrm{OD}}_{\mu i}(z) + \Delta\mathcal{G}_{\mu i}(z) +\mathcal{G}^{\mathrm{CC}}_{\mu i}(z) ,
\end{equation}
where the cross-correlated part is
\[
\mathcal{G}^{\mathrm{CC}}_{\mu i}(z) = \sum_{\substack{\nu j \neq \xi k \neq \mu i}} V_{\mu i,\nu j} V_{\xi k,\mu i} \bra{\phi_{\nu j}} \frac{1}{z-\Phi_{\mu i}H\Phi_{\mu i}} \ket{\phi_{\xi k}},
\]
and the difference between the projected diagonal resolvent $   \mathcal{R}^{(\mu i)}_{\nu j}(z) = \bra{\phi_{\nu j}} \frac{1}{z -\Phi_{\mu i} H\Phi_{\mu i}} \ket{\phi_{\nu j}}$ and the full diagonal resolvent gives
\[
 \Delta\mathcal{G}_{\mu i}(z) = \sum_{\nu j \neq \mu i} \abs{V_{\mu i,\nu j}}^2 [\mathcal{R}^{(\mu i)}_{\nu j}(z)-\mathcal{R}_{\nu j}(z)].
\]
represents the correction of backward return processes. Its properties are discussed in more detail in Appendix~\ref{app:DCA}; here we regard this contribution as negligible.
The off-diagonal (mean-field) part is
\[
\mathcal{G}^{\mathrm{OD}}_{\mu i}(z) = \sum_{\nu j \neq \mu i} \abs{V_{\mu i,\nu j}}^2 \mathcal{R}_{\nu j}(z).
\]
This separation \cref{SEP1} is exact and does not rely on any truncation.  The first term describes diagonal propagation channels, while the second captures interference between distinct intermediate states.

To make progress, we invoke the random phase hypothesis, a cornerstone of the ETH.  For generic nonintegrable systems, the matrix elements of a local operator $O$ in the energy eigenbasis of $H_0$ take the form \cite{AKP16}:
\[
\abs{\bra{\phi_\mu} O \ket{\phi_\nu}}^2 = e^{-S(\epsilon^+_{\mu\nu})} f^2(\epsilon^+_{\mu\nu},\delta) \abs{R_{\mu\nu}}^2,\quad \mu\neq\nu,
\]
where $e^{S(\epsilon)}$ is the density of states, $f$ is a smooth function, and $R_{\mu\nu}$ are random variables with zero mean and unit variance.  For distinct indices they are statistically independent \cite{M04}:
\[
\mathbb{E}(R_{\mu\nu}R_{\xi\mu}) = 0,\quad \mu\neq\nu\neq\xi.
\]

For the system-bath interaction $V$ (a sum of local operator products), this property allows an ensemble average over random phases.  Under the random-phase hypothesis, the cross-correlated contributions average out:
\[
\mathbb{E}\bigl[\mathcal{G}^{\mathrm{CC}}_{\mu i}(z)\bigr] \approx 0,
\]
leaving only the mean-field term:
\begin{equation}\label{eq:GRE}
	\mathcal{G}_{\mu i}(z) \approx \mathcal{G}^{\mathrm{OD}}_{\mu i}(z) =\sum_{\nu j \neq \mu i} \abs{V_{\mu i,\nu j}}^2 \mathcal{R}_{\nu j}(z)\
\end{equation}
and the consistent equation gives
\begin{equation}\label{Dyson}
    \mathcal{R}_{\mu i}(z) = \frac{1}{z - a_{\mu i} - V_{\mu i} - \sum_{\nu j \neq \mu i} \abs{V_{\mu i,\nu j}}^2 \mathcal{R}_{\nu j}(z)}.
\end{equation}
This is the \textit{mean-field approximation} of our framework.  It yields a closed equation for the coarse-grained spectral structure and should be understood as a projection of the full resolvent hierarchy (Sec.~\ref{sec:cross_expansion}) onto its single-resolvent sector.  The full hierarchy, which retains all cross-correlated terms, will be analyzed in the next subsection.  Importantly, ETH-type assumptions are used only at the level of statistical averaging; the exact projected recursive expansion of $\mathcal{G}^{\mathrm{CC}}_{\mu i}(z)$ derived below does not rely on them.

\subsection{The self-consistent equation under mean-field approximation}
Substituting \eqref{eq:GRE} into \eqref{eq:CSEQ} establishes a self-consistent equation for \( \mathcal{R}_{\mu i}(z) \). To connect this to the probability distribution \( p^{\mu i}_n \), we exploit the completeness relation of perturbed eigenstates:
\begin{equation}\label{PEEX}
	\mathcal{R}_{\mu i}(z) = \sum_n \frac{p^{\mu i}_n}{z - \lambda_n}.
\end{equation}
Using the Sokhotski-Plemelj identity:
\begin{equation}
	\frac{1}{x - \mathrm{i}0^+} = \mathcal{P}\left(\frac{1}{x}\right) + \mathrm{i}\pi \delta(x),
\end{equation}
the imaginary component of \( \mathcal{R}_{\mu i} \) maps to the spectral measure:
\begin{equation}\label{IMRtoP}
	\frac{1}{\pi} \Im\, \mathcal{R}_{\mu i}(x - \mathrm{i}0^+) = \sum_n p^{\mu i}_n \delta(x - \lambda_n).
\end{equation}
This defines \( p^{\mu i}_n \) via energy binning:
\begin{align}\label{pmnd}
	p^{\mu i}_n = \int_{(\lambda_{n-1} + \lambda_n)/2}^{(\lambda_n + \lambda_{n+1})/2} \! dx \, \frac{1}{\pi} \Im\, \mathcal{R}_{\mu i}(x - \mathrm{i}0^+)\notag \\
	=\frac{1}{e^{S(\lambda_n)}} \frac{1}{\pi} \Im\, \mathcal{R}_{\mu i}(\lambda_n - \mathrm{i}0^+).
\end{align}
Defining the weighted spectral function 
\(f^{\mu i}(x) \equiv e^{S(x)} p^{\mu i}(x) = \frac{1}{\pi}\Im \mathcal{R}_{\mu i}(x - \mathrm{i}0^{+})\) and using  the continuum approximation \( \sum_m \to \int d\lambda \, e^{S(\lambda)} \),
the boundary value of the resolvent satisfies the Kramers-Kronig (dispersion) relation in the compact form
\begin{equation}\label{RkkR}
 \frac{1}{\pi} \mathcal{R}_{\mu i}(\omega - \mathrm{i}0^{+}) = H[f^{\mu i}](\omega) + \mathrm{i} f^{\mu i}(\omega) ,
\end{equation}
where the Hilbert transform 
\begin{equation}\label{Hilbert}
    H(f)(\lambda) := \frac{1}{\pi} \dashint d\lambda' f(\lambda')/(\lambda - \lambda').
\end{equation}
 The relation \eqref{RkkR} is exact and will be used repeatedly throughout the hierarchical ansatz strategy.

The analytic continuation \( H \to H + \mathrm{i}0^+ \) shifts \( \mathcal{R}_{\mu i}(x) \to \mathcal{R}_{\mu i}(x - \mathrm{i}0^+) \). From \eqref{eq:CSEQ},
\begin{equation}
	\mathcal{R}_{\mu i}(x - \mathrm{i}0^+) = \frac{1}{x - a_{\mu i} - V_{\mu i} - \mathcal{G}_{\mu i}(x - \mathrm{i}0^+)}.
\end{equation}
Combining with \eqref{pmnd}, the probability distribution becomes:
\begin{equation}\label{PAG}
	p^{\mu i}_n =  \frac{1}{\pi e^{S(\lambda_n)}} \frac{\Im\, \mathcal{G}_{\mu i}(\lambda_n - \mathrm{i}0^+)}{[\Delta^{\mu i}_n - \Re\, \mathcal{G}_{\mu i}(\lambda_n)]^2 + [\Im\, \mathcal{G}_{\mu i}(\lambda_n - \mathrm{i}0^+)]^2},
\end{equation}
where \( \Delta^{\mu i}_n = \Delta^{\mu i}(\lambda_n):= \lambda_n - a_{\mu i} - V_{\mu i} \). Equation \eqref{PAG} is an exact identity derived from the spectral representation and the Sokhotski-Plemelj formula, provided $\mathcal{G}_{\mu i}(z)$ is the exact projected correlation kernel. In practice, we have access only to approximate forms of this kernel, such as the mean-field expression \eqref{eq:GRE} or the truncations introduced below. Substituting an approximate self-energy into \cref{PAG} generally does not guarantee the correct normalization $\sum_n p_n^{\mu i}=1$ nor the proper analytic properties of the resolvent. Therefore, we treat \cref{PAG} not as a direct formula for the probabilities, but as a self-consistency condition that must be satisfied together with the approximate expression for $\mathcal{G}_{\mu i}$. Concretely, we seek a spectral distribution $p_{\mu i}(\lambda)$ (or equivalently a resolvent $\mathcal{R}_{\mu i}$) such that when it is used to construct $\mathcal{G}_{\mu i}^{\text{approx}}$ via the chosen approximation (e.g., the mean-field expression \eqref{eq:GRE} or the hierarchical ansatz described below), the resulting right-hand side of \cref{PAG} reproduces the same distribution. This self-consistency is at the heart of the hierarchical ansatz strategy presented in \cref{HAS}.

By \eqref{eq:GRE} and \eqref{IMRtoP}, we have
\begin{equation}\label{IMGAP}
	\frac{1}{\pi} \Im\, \mathcal{G}_{\mu i}(\lambda_n - \mathrm{i}0^+) = \sum_{\nu j \neq \mu i} \abs{V_{\mu i,\nu j}}^2 \sum_m p^{\nu j}_m \delta(\lambda_n - \lambda_m) .
\end{equation}
The real part of $\mathcal{G}_{\mu i}$ is obtained via the Kramers-Kronig (dispersion) relation, which follows from the analyticity of $\mathcal{G}_{\mu i}(z)$ in the lower half-plane. Substituting the expression (\ref{IMGAP}) for $\Im \mathcal{G}_{\mu i}$ into this relation yields
\begin{align}\label{REGAP}
\Re \mathcal{G}_{\mu i}(\lambda_n) 
&=  \dashint dx \frac{1}{\lambda_n - x} 
   \sum_{\nu j \neq \mu i} |V_{\mu i,\nu j}|^2 \sum_m p^{\nu j}_m \delta(x - \lambda_m) \nonumber \\
&=   \sum_{\nu j \neq \mu i} |V_{\mu i,\nu j}|^2 \sum_m p^{\nu j}_m 
   \dashint dx \frac{\delta(x - \lambda_m)}{\lambda_n - x} \nonumber \\
&=  \sum_{\nu j \neq \mu i} |V_{\mu i,\nu j}|^2 \sum_{m \neq n} \frac{p^{\nu j}_m}{\lambda_n - \lambda_m}.
\end{align}
In the last step we used the identity $\dashint dx \, \delta(x - \lambda_m) / (\lambda_n - x) = 1/(\lambda_n - \lambda_m)$ for $m \neq n$, and the term $m = n$ is excluded by the principal-value prescription.  

Under the continuum approximation, combining \cref{RkkR,eq:GRE}, we obtain 
\begin{equation}\label{TTGAP}
	\frac{1}{\pi}  \mathcal{G}_{\mu i}(\lambda_n - \mathrm{i}0^+)  =\sum_{\nu j \neq \mu i} |V_{\mu i,\nu j}|^2[H(f^{\nu j})+\mathrm{i}f^{\nu j}],
\end{equation}
where $f^{\nu j}(\lambda):= p^{\nu j}(\lambda) e^{S(\lambda)}$.

If the cross terms are not neglected, then by referring to \cref{IMGAP} and adopting the continuum approximation, one readily obtains
\begin{equation}\label{IMGAPTT}
\frac{1}{\pi}\Im \mathcal{G}_{\mu i}(\lambda_n - \mathrm{i}0^+)
= \left| \sum_{\nu j \neq \mu i} V_{\mu i,\nu j} \braket{\phi_{\nu j}|\psi_n} \right|^2 e^{S(\lambda_n)} .
\end{equation}
This result shows that, once cross-correlated contributions are retained, the imaginary part of the self-energy is governed by the coherent superposition of transition amplitudes rather than an incoherent sum of probabilities. The expression \eqref{IMGAPTT} captures the exact effect of cross-correlated terms but involves a coherent superposition of transition amplitudes, making it difficult to analyze directly. To systematically handle these contributions, we turn to a recursive expansion that organizes the cross terms order by order.

\subsection{Systematic Recursive Hierarchy of Cross-Correlated Terms}
\label{sec:cross_expansion}
While the cross-correlated terms vanish on average, they are not identically zero. They give rise to fluctuations around the mean-field solution and are essential for describing the distribution tails, branch splitting, and higher-order moments. To make the structure of the cross-correlated contribution explicit beyond the mean-field projection, we now derive an exact recursive re-expression of  $\mathcal{G}_{\mu i}^{\mathrm{CC}}(z)$  into the projected resolvent hierarchy.

Starting from the projection identity derived in \cref{RSCE},
\begin{equation}\label{PRID}
    \Phi_{\mu i} \frac{1}{z - H} \ket{\phi_{\mu i}} = \Phi_{\mu i} \frac{1}{z - \Phi_{\mu i}H\Phi_{\mu i}} \Phi_{\mu i} V \ket{\phi_{\mu i}} \mathcal{R}_{\mu i}(z),
\end{equation}
Inserting the identity  $I = \sum_{\nu j} (\Phi_{\nu j}+\Pi_{\nu j})\dots \Pi_{\nu j}$ yields
\begin{align}\label{offRexp}
	\Phi_{\mu i} \frac{1}{z - H} \ket{\phi_{\mu i}}=\sum_{\nu j\neq \mu i}\ket{\phi_{\nu j}}\mathcal{R}^{(\mu i)}_{\nu j}(z)\mathcal{R}_{\mu i}(z)V_{\nu j,\mu i}\notag\\
	+\sum_{\nu j\neq \mu i}\Phi_{\mu i} \Phi_{\nu j} \frac{1}{z - \Phi_{\mu i} H \Phi_{\mu i}} \ket{\phi_{\nu j}}\mathcal{R}_{\mu i}(z)V_{\nu j,\mu i}.
\end{align}
With it, the cross term can be rewritten as
\begin{align}\label{CCRes}
    \mathcal{G}^{\mathrm{CC}}_{\mu i}(z) &= \sum_{\nu j} \bra{\phi_{\mu i}} V \Phi_{\mu i} \Phi_{\nu j} \frac{1}{z - \Phi_{\mu i} H \Phi_{\mu i}} \Pi_{\nu j} \Phi_{\mu i} V \ket{\phi_{\mu i}} \notag\\
    &=\mathcal{G}^{(3)}_{\mu i}(z) + \mathcal{G}^{\mathrm{(res)}}_{\mu i}(z)
\end{align}
where the leading third-order term is
\begin{equation}\label{eq:G3}
    \mathcal{G}^{(3)}_{\mu i}(z) = \sum_{\xi k \neq \nu j \neq \mu i} V^{(3)}_{\mu i,\xi k,\nu j} \mathcal{R}^{(\mu i)}_{\nu j}(z) \mathcal{R}^{(\mu i,\nu j)}_{\xi k}(z),
\end{equation}
where $V^{(3)}_{\mu i,\xi k,\nu j}:=V_{\mu i,\xi k} V_{\xi k,\nu j} V_{\nu j,\mu i}$. \Cref{eq:G3} represents the leading element of a recursive hierarchy of cross-correlated processes. This structure reveals that higher-order correlations are not arbitrary corrections, but follow a well-defined algebraic pattern that can be extended systematically to all orders.

The remainder $\mathcal{G}^{\mathrm{(res)}}_{\mu i}(z)$ has the same structure as the original expression and can be expanded recursively.
The recursive structure generates a hierarchy of higher-order terms:
\begin{equation}\label{HHOT}
    \mathcal{G}^{\mathrm{CC}}_{\mu i}(z) = \sum_{\ell \ge 3} \mathcal{G}^{(\ell)}_{\mu i}(z),
\end{equation}
where $\ell$ denotes the order in the interaction. Equation~\eqref{HHOT} establishes a hierarchical organization of
cross-correlated contributions, which constitutes one of the central results of this work. The explicit form of the fourth-order term
$\mathcal{G}^{(4)}_{\mu i}$ and the general recursive pattern are derived in Appendix~\ref{app:fourth_order}. Rather than being treated perturbatively or discarded, these terms form a structured expansion that remains closed at the level of resolvents. This hierarchy provides a nonperturbative framework for incorporating fluctuations beyond the mean-field description. From the ETH perspective, $\mathcal{G}^{(3)}_{\mu i}$ corresponds to third-order correlations of matrix elements, while higher-order terms encode higher moments of the ETH distribution \cite{FK19}.  While the projected hierarchy is exact, its closure to ordinary diagonal 
resolvents is approximate; its practical solution requires additional 
approximations, such as the hierarchical ansatz strategy discussed in 
\cref{HAS}.

Under the diagonal closure approximation (DCA), i.e., neglecting cavity 
corrections from backward Krylov-sector return processes, whose contribution 
is suppressed by off-diagonal propagation amplitudes scaling as $e^{-S}$ in 
the ETH regime (see \cref{app:DCA} for a detailed justification). And the 
projected diagonal resolvent may be approximated by the full diagonal resolvent:
\begin{align}\label{APPR2DR}
    \sum_{\xi k \neq \nu j \neq \mu i} V^{(3)}_{\mu i,\xi k,\nu j} \mathcal{R}^{(\mu i)}_{\nu j}(z) \mathcal{R}^{(\mu i,\nu j)}_{\xi k}(z)\approx \notag\\
     \sum_{\xi k \neq \nu j \neq \mu i} V^{(3)}_{\mu i,\xi k,\nu j} \mathcal{R}_{\nu j}(z) \mathcal{R}_{\xi k}(z)=: \mathcal{G}^{(3),\text{D}}_{\mu i}(z) .
\end{align}
The third-order term \eqref{eq:G3} can be evaluated directly using \eqref{RkkR}. 
Since $\mathcal{R}_{\xi k}\mathcal{R}_{\nu j}$ is analytic in the lower half-plane, its boundary value is the product of the boundary values of each factor. Therefore
\begin{align}\label{TTG3}
	 \frac{1}{\pi^2}  \mathcal{G}^{(3),\text{D}}_{\mu i}(\lambda - \mathrm{i}0^+)=\sum_{\substack{\xi k \neq \nu j \neq \mu i}}V^{(3)}_{\mu i,\xi k,\nu j}\notag\\
	 \times[H(f^{\nu j})+\mathrm{i} f^{\nu j}][H(f^{\xi k })+\mathrm{i} f^{\xi k }].
\end{align}
Crucially, the imaginary part of $\mathcal{G}^{(3)}$ contains terms of the form $f_{\nu j} H[f_{\xi k}]$, which represent a nonlocal convolution of spectral densities from different resolvent channels.
Unlike the diagonal (mean-field) contribution, these products of a function and its Hilbert transform acquire an odd-parity component under frequency reflection.
Consequently, the leading multi-resolvent correction intrinsically mixes parity sectors, generating an odd component in the self-energy.
As proved in Appendix~\ref{app:no_skewness_mf}, the SCBA functional preserves parity; hence this parity mixing is the minimal mechanism capable of generating spectral skewness   beyond parity-preserving single-resolvent closures.
The nonlocal convolution structure implies that the self-energy at a given frequency is influenced by a broad range of intermediate states, making it a distinct fingerprint of multi-resolvent interference. A detailed analysis of how this multi-resolvent structure modifies the
spectral function and its tails, in contrast to single-resolvent (SCBA-type)
approaches, is given in Appendix~\ref{SCMS}.

We note in passing that expanding the cavity propagator's Z-factor
denominator yields a formal $|V|^{4}$ multi-resolvent correction
$\mathcal{R}_{\nu j}^{2}\mathcal{R}_{\mu i}$ that also mixes parity
sectors.  Numerical analysis indicates, however, that its contribution
is quantitatively subdominant as a source of spectral skewness; a
brief discussion is deferred to Appendix~\ref{app:cavity_V4}.

\subsection{The solution of self-consistent equation: Hierarchical Ansatz Strategy}\label{HAS}

Substituting \eqref{REGAP} and \eqref{IMGAP} into \eqref{PAG} yields a closed self-consistent equation for \( p^{\mu i}_n \). This formalism enables iterative solutions:
(1) Initialization: Assume a trial distribution \( p^{\mu i}_n \).
(2) Update: Compute \( \mathcal{G}_{\mu i} \) via \eqref{IMGAP} and \eqref{REGAP}.
(3) Convergence: Reconstruct \( p^{\mu i}_n \) through \eqref{PAG}.  Since the self-consistent equation captures the global properties, this iterative approach is theoretically capable of efficiently yielding a solution that closely aligns with the actual distribution in terms of global behavior.

An alternative pathway to determine the statistical behavior of \( p^{\mu i}_n \) involves solving self-consistent equations. However, even with the mean-field approximation, solving the self-consistent equations directly is challenging. We therefore employ a hierarchical strategy, solving for the probability distribution $p^{\mu i}(\lambda)$ using physically motivated ans\"atzes. Given the inherent stochasticity of \( p^{\mu i}_n \), we focus on statistical averaging within energy shells. Partitioning eigenstates into energy intervals:
\begin{equation}
	\mathcal{M}_{E,\Delta} = (E - \Delta/2, E + \Delta/2),
\end{equation}
the smoothed probability distribution becomes:
\begin{equation}\label{SMPD}
	p^{\mu i}(\lambda) = \mathbb{E}(p^{\mu i}_n) := \frac{1}{d_{\mathcal{M}}} \sum_{\lambda_m \in \mathcal{M}_{\lambda,\Delta}} p^{\mu i}_m,
\end{equation}
where \( d_{\mathcal{M}} = e^{S(\lambda)} \Delta \) denotes the Hilbert space dimension within the shell. 

Under the assumption that $p_n^{\mu i}$ varies smoothly within the energy window $\mathcal{M}_{\lambda,\Delta}$ and that the density of states is exponentially large, the discrete spectral measure can be replaced by a continuous representation:
\begin{equation}
\sum_n p^{\mu i }_n\,\delta(x - \lambda_n)
\;\longrightarrow\;
e^{S(x)}\, p^{\mu i}(x).
\end{equation}
Consequently, \cref{IMRtoP} reduces to the coarse-grained form
\begin{equation}
\frac{1}{\pi}\,\Im\,\mathcal{R}_{\mu i}(x - \mathrm{i}0^+)
\approx
e^{S(x)}\, p_{\mu i}(x),
\end{equation}
which yields
\begin{equation}\label{SPIR}
p^{\mu i}(x)=\frac{1}{\pi e^{S(x)}}\,\Im\,\mathcal{R}_{\mu i}(x -  \mathrm{i}0^+).
\end{equation}
This relation provides a smooth characterization of the spectral weight distribution. The validity of this replacement relies on the existence of a separation of scales: the coarse-graining window $\Delta$ is taken to be much larger than the mean level spacing, but still small compared to macroscopic energy scales. This procedure is consistent with standard coarse-graining approaches used in the context of ETH and random matrix theory.

The coarse-graining procedure described above admits an equivalent formulation that makes the continuum approximation mathematically transparent.
Instead of taking the boundary value $\mathrm{i}0^+$ from the outset, one may retain a
{\em finite} imaginary shift $\eta>0$ and consider
\begin{equation}\label{eq:finite_eta}
\frac{1}{\pi}\,\Im\,\mathcal{R}_{\mu i}(x-\mathrm{i}\eta)
= \sum_n p^{\mu i}_n\,
\frac{\eta/\pi}{(x-\lambda_n)^2+\eta^2}.
\end{equation}
Each $\delta$-peak of the spectral measure~\eqref{IMRtoP} is thereby regularized into a
Lorentzian of half-width $\eta$.  
When $\eta$ is {\em larger than the mean level spacing} $\delta_{\rm LS}\sim e^{-S(x)}$,
neighboring Lorentzians overlap and the discrete sum becomes indistinguishable from a
continuous integral:
\begin{align}\label{eq38}
\sum_n p^{\mu i}_n\,\frac{\eta/\pi}{(x-\lambda_n)^2+\eta^2}
\notag\\
\xrightarrow{\;\eta\,\gg\,\delta_{\rm LS}\;}\int d\lambda\; e^{S(\lambda)}\,p^{\mu i}(\lambda)\,
\frac{\eta/\pi}{(x-\lambda)^2+\eta^2}.
\end{align}
In the limit $\eta\to 0^+$ {\em after} the continuum limit
$S\to\infty$ (where $\delta_{\rm LS}\to 0$), one recovers
\begin{equation}
\frac{1}{\pi}\,\Im\,\mathcal{R}_{\mu i}(x-\mathrm{i}0^+)
= e^{S(x)}\,p^{\mu i}(x),
\end{equation}
which is precisely the coarse-grained relation~\eqref{SPIR}.
Thus, the continuum approximation underlying the entire framework is
equivalent to the statement that there exists a
window of $\eta$ satisfying
\begin{equation}\label{eq:eta_window}
\delta_{\rm LS}\;\ll\;\eta\;\ll\;\Gamma_{\mu i},
\end{equation}
where $\Gamma_{\mu i}$ is the characteristic spectral width encoded in
$\mathcal{G}_{\mu i}$.
The lower bound ensures sufficient overlap of neighboring levels,
while the upper bound guarantees that the artificial broadening
$\eta$ does not distort the {\em physical} line shape determined by
the self-energy.

Three points deserve emphasis.
First, the finite-$\eta$ picture is a {\em regularization device},
not a modification of the physical theory; all final results are
understood in the ordered limit $\lim_{\eta\to0^+}\lim_{S\to\infty}$,
so the continuum mainline of the paper is unaffected.
Second, for {\em any} finite $\eta>0$ the resolvent
$\mathcal{R}_{\mu i}(z)$ with $z=x-\mathrm{i}\eta$ remains analytic in
the lower half-plane, so the Kramers--Kronig (Hilbert-transform)
structure~\eqref{RkkR} and the self-consistent formalism of
Sec.~\ref{RSCE} hold without modification.
Third, the scale-separation condition~\eqref{eq:eta_window} provides a
quantitative criterion for the validity of the continuum approximation
in finite systems: as long as the physical broadening exceeds the mean
level spacing, the coarse-grained spectral function $p^{\mu i}(x)$ is
a faithful representation of the underlying discrete distribution.

This perspective supplies the continuum approximation with a rigorous
finite-size foundation while keeping the analytic
(Hilbert/Kramers--Kronig) structure fully intact, and without reducing
the theory to a discrete Green-function treatment.

\subsubsection{Lorentzian Ansatz (Bulk)}\label{LABK}
Guided by equation \eqref{PAG} and supported by numerical evidence, we adopt the minimal Lorentzian ansatz formulation for the bulk region (where the values are relatively large):
\begin{equation}\label{eq:Lorentz}
	p^{\mu i}(\lambda) =\frac{1}{\pi e^{S(\lambda)}} \Im\left(\frac{1}{\delta \lambda_{\mu i}-\mathrm{i} \chi_{\mu i}}\right)
	=\frac{1}{ e^{S(\lambda)}} L^{\mu i}(\lambda),
\end{equation}
where $\delta \lambda_{\mu i}:= \lambda -a_{\mu i} - \Delta_{\mu i}$, $L^{\mu i}(\lambda)=L(\delta\lambda_{\mu i};\chi_{\mu i})$ and $L(x;\chi)=\frac{\chi}{\pi(\chi^2+x^2)}$ is the Lorentzian distribution.
The form \eqref{eq:Lorentz} inherently satisfies normalization due to Lorentzian properties:
\begin{equation}  \label{Norm}
	\sum_n p^{\mu i}_n = \int d\lambda \, \frac{1}{\pi} \frac{\chi_{\mu i}}{\delta \lambda_{\mu i}^2 + \chi_{\mu i}^2} = 1.
\end{equation}
The ansatz reduces the self-consistent equations for \( p^{\mu i}_n \) to determining parameters \( \chi_{\mu i} \) (width) and \( \Delta_{\mu i} \) (shift). It should be noted that the distribution (\ref{eq:Lorentz}) is only a smooth distribution about $\lambda$ in a statistical sense (global behavior). It only represents the statistical behavior of $ p^{\mu i}_n$. Therefore, the self-consistent equations corresponding to the parameters $\chi_{\mu i}$ and $\Delta_{\mu i}$ are also only valid in a statistical sense.

Before proceeding to the self-consistent equations, it is instructive to 
connect the Lorentzian ansatz to the Kramers-Kronig structure of 
Eq.~\eqref{RkkR}. From the ansatz~\eqref{eq:Lorentz}, one has
\begin{equation}
    f^{\nu j}(\lambda) = e^{S(\lambda)}p^{\nu j}(\lambda) 
    = \frac{1}{\pi}\,\Im\frac{1}{\delta\lambda_{\nu j}-\mathrm{i}\chi_{\nu j}}.
\end{equation}
Its Hilbert transform follows immediately from the analytic structure:
\begin{equation}
    H[f^{\nu j}](\lambda) 
    = \frac{1}{\pi}\,\Re\frac{1}{\delta\lambda_{\nu j}-\mathrm{i}\chi_{\nu j}}
    = \frac{1}{\pi}\frac{\delta\lambda_{\nu j}}{\delta\lambda_{\nu j}^2+\chi_{\nu j}^2}.
\end{equation}
Hence the combination that enters the mean-field self-energy~\eqref{TTGAP} 
reduces to the diagonal resolvent itself:
\begin{equation}\label{eq:R_Lorentz_bridge}
    \pi\bigl[H(f^{\nu j}) + \mathrm{i}f^{\nu j}\bigr]
    = \frac{1}{\delta\lambda_{\nu j} - \mathrm{i}\chi_{\nu j}}
    = \mathcal{R}_{\nu j}(\lambda-\mathrm{i}0^{+}).
\end{equation}
Equation~\eqref{eq:R_Lorentz_bridge} makes explicit that, for the Lorentzian 
ansatz, the self-energy kernel~\eqref{TTGAP} is a weighted sum of single-pole 
resolvents---the characteristic analytic structure. In this sense, SCBA amounts 
to projecting the full many-body resolvent hierarchy onto the single-pole 
sector. This single-pole form will be contrasted with the richer Voigt-type 
resolvent decomposition in Sec.~\ref{SLGA}.

The self-consistent equations for the parameters $\chi_{\mu i}$ and $\Delta_{\mu i}$ remain highly complex. Here we implement the most straightforward decoupling. By comparing Eqs. \eqref{PAG} and \eqref{eq:Lorentz}, and since for any nonzero function $ r(\lambda)$,
\begin{equation}\label{exL}
	\pi L^{\mu i}(\lambda)=\frac{\chi_{\mu i}\times r(\lambda)}{(\delta \lambda_{\mu i}^r)^2 +r^2(\lambda)\chi^2_{\mu i}},
\end{equation}
where
\begin{equation}
	(\delta \lambda_{\mu i}^r)^2 :=  (\lambda-a_{\mu i}-\Delta_{\mu i})^2r(\lambda)+r(\lambda)\chi^2_{\mu i}-r^2(\lambda)\chi^2_{\mu i},
\end{equation}
it follows that
\begin{align}\label{ParaConsiso}
	r(\lambda)\chi_{\mu i}                      & = \mathbb{E}\big(\Im\, \mathcal{G}_{\mu i}(\lambda - \mathrm{i}0^+)\big), \notag \\
	\Delta^{\mu i}(\lambda)- \delta \lambda_{\mu i}^r & =  \mathbb{E}\big(\Re\, \mathcal{G}_{\mu i}(\lambda)\big).
\end{align}
Assuming that for different indices $\mu i$, after choosing an appropriate $\lambda$, we may set $r(\lambda)=1$, the self-consistent system simplifies to
\begin{align}\label{ParaConsis}
	\chi_{\mu i}   & = \mathbb{E}\big(\Im\, \mathcal{G}_{\mu i}(\lambda - \mathrm{i}0^+)\big), \notag \\
	\Delta_{\mu i} & = V_{\mu i} + \mathbb{E}\big(\Re\, \mathcal{G}_{\mu i}(\lambda)\big).
\end{align}
Substituting Eqs.\eqref{IMGAP}, \eqref{REGAP}, \eqref{eq:Lorentz} and using the known Hilbert-transform result for the Lorentz distribution yields explicit equations:
\begin{align}
	\chi_{\mu i}               & = \sum_{\nu j \neq \mu i} \abs{V_{\mu i,\nu j}}^2 \frac{\chi_{\nu j}}{\delta \lambda_{\nu j}^2 + \chi_{\nu j}^2}, \label{consist}            \\
	\Delta_{\mu i} - V_{\mu i} & = \sum_{\nu j \neq \mu i} \abs{V_{\mu i,\nu j}}^2 \frac{\delta \lambda_{\nu j}}{\delta \lambda_{\nu j}^2 + \chi_{\nu j}^2}. \label{eq:consist2}
\end{align}
These coupled equations enable iterative determination of \( \chi_{\mu i} \) and \( \Delta_{\mu i} \) for specific interaction profiles \( \abs{V_{\mu i,\nu j}}^2 \). This solution is the self-consistent mean-field (SCBA) solution, valid for the bulk of the distribution. 

\subsubsection{Gaussian Ansatz (Tail)}

The Lorentzian ansatz provides a good description of the spectral function near 
the central region, but becomes less accurate in the tails when $\lambda$ deviates 
significantly from $a_{\mu i}$. This deviation originates from the energy dependence 
of the interaction kernel. In particular, the effective interaction strength
$F^2_{i,j}(\epsilon_\mu, \mathcal{M}_{\epsilon_\nu,\Delta})
:= \sum_{\epsilon_\kappa \in \mathcal{M}_{\epsilon_\nu,\Delta}} 
|V_{\mu i,\kappa j}|^2$
typically decays rapidly as $|\epsilon_\mu-\epsilon_\nu|$ increases. As a result, 
the spectral tails decay faster than the Lorentzian form.

This behavior can be naturally captured by a Gaussian ansatz. We emphasize that, 
unlike the Lorentzian profile arising in the strict SCBA limit with 
energy-independent couplings, the Gaussian form does not originate from the 
structure of the SCBA functional itself. Instead, it reflects the presence of an 
intrinsic energy scale in the interaction kernel. When the coupling matrix 
elements and/or the effective density of states exhibit a smooth energy dependence, which often well approximated by a Gaussian envelope in many-body systems and the mean-field 
self-energy inherits this scale.

Motivated by this, we adopt the Gaussian ansatz
\begin{equation}\label{eq:Gaussian}
	p^{\mu i}(\lambda) 
	=\frac{1}{e^{S(\lambda)}\sigma_{\mu i}}
	\varphi\!\left(\frac{\delta \lambda'_{\mu i}}{\sigma_{\mu i}}\right),
\end{equation}
where $\varphi(x)=\frac{1}{\sqrt{2\pi}} e^{-x^2/2}$ and 
$\delta \lambda'_{\mu i}:= \lambda -a_{\mu i} - \Delta'_{\mu i}$. 
The parameters $\Delta'_{\mu i}$ and $\sigma_{\mu i}$ characterize the shift and 
width of the tail, respectively.
Importantly, this Gaussian ansatz remains strictly reflection-symmetric about 
its center. As established in Appendix~\ref{app:no_skewness_mf}, the SCBA 
self-consistent functional preserves parity under energy reflection and therefore 
cannot generate odd (skewness) components in the spectral function. The Gaussian 
form thus represents symmetric broadening induced by the energy dependence of 
the interaction, rather than a signature of multi-resolvent correlations.

We now derive the corresponding self-consistency condition in the tail regime. 
For large $|\lambda-\epsilon_\mu|$, both $\Im\, \mathcal{G}$ and 
$\Re\, \mathcal{G}$ decay rapidly and become much smaller than 
$|\Delta^{\mu i}(\lambda)|$. In this regime, the self-consistent equation 
(\cref{PAG}) simplifies to
\begin{equation}
	e^{S(\lambda)}p^{\mu i}(\lambda)
	\approx \frac{1}{\pi} 
	\frac{\Im\, \mathcal{G}_{\mu i}(\lambda - \mathrm{i}0^+)}
	{[\Delta^{\mu i}(\lambda)]^2}.
\end{equation}
Substituting the Gaussian ansatz \eqref{eq:Gaussian} into 
\eqref{IMGAP} yields
\begin{equation} \label{gauscons}
	\frac{1}{\pi} \Im\, \mathcal{G}_{\mu i}(\lambda - \mathrm{i}0^+)
	= \sum_{\nu j \neq \mu i} 
	|V_{\mu i,\nu j}|^2 
	\varphi\!\left(\frac{\delta \lambda'_{\nu j}}{\sigma_{\nu j}}\right)
	\frac{1}{\sigma_{\nu j}}.
\end{equation}
Combining this with \eqref{eq:Gaussian}, we obtain the consistency condition
\begin{equation}\label{tailconsis}
	\frac{[\Delta^{\mu i}(\lambda)]^2}{\sigma_{\mu i}}
	\varphi\!\left(\frac{\delta\lambda'_{\mu i}}{\sigma_{\mu i}}\right)
	\approx \sum_{j}\int d\epsilon_\nu 
	f^2_{i,j}(\epsilon_\mu,\delta)\,
	\varphi\!\left(\frac{\delta\lambda'_{\nu j}}{\sigma_{\nu j}}\right)
	\frac{1}{\sigma_{\nu j}},
\end{equation}
where $\delta=\epsilon_\mu-\epsilon_\nu$. 

Here we have assumed that the interaction strength can be statistically averaged, 
so that its microscopic correlations with the tail distribution are negligible. 
This allows us to introduce a smooth effective interaction profile
\begin{equation}\label{intesmo}
	\mathbb{E}(|V_{\mu i,\nu j}|^2)
	= e^{-S(\epsilon_\nu)}f^2_{i,j}(\epsilon_\mu,\delta).
\end{equation}

Similar to the central-region consistency equations (\cref{consist,eq:consist2}), 
Eq.~\eqref{tailconsis} provides a closed scheme for determining the parameters 
$\sigma_{\mu i}$ and $\Delta'_{\mu i}$ iteratively for a given interaction profile.

We therefore interpret the Gaussian ansatz as a natural extension of the SCBA 
description to systems with structured (energy-dependent) interactions. It captures 
the emergence of a finite energy scale and faster-than-Lorentzian decay in the 
tails, while remaining confined to the single-resolvent sector prior to the 
inclusion of multi-resolvent correlations.

\subsubsection{Hybrid Lorentzian-Gaussian (LG) Ansatz (symmetry)}\label{SLGA}

The Lorentzian ansatz captures the bulk of the distribution at the strict mean-field level, while the Gaussian ansatz accounts for the energy dependence of the interaction matrix elements in the tails.
To obtain a unified description across all energy scales, we combine the two into a hybrid Voigt-type profile.
Crucially, the absence of spontaneous skewness in the SCBA self-consistency (proved in Appendix~\ref{app:no_skewness_mf}) forces the Lorentzian and Gaussian components to share the same center,
which eliminates the asymmetry parameter and reduces the profile to the symmetric form
\begin{equation}\label{eq:LGdistr_sym}
  p^{\mu i}(\lambda)
  = \frac{G^{\mu i}(\lambda)\,L^{\mu i}(\lambda)}{e^{S(\lambda)}\,V^{\mu i}},
\end{equation}
where now both factors are centred at \(a_{\mu i}+\Delta_{\mu i}\):
  $G^{\mu i}(\lambda)=G(\delta\lambda_{\mu i};\sigma_{\mu i})$,
and $G(x;\sigma)=e^{-x^2/(2\sigma^2)}/(\sqrt{2\pi}\sigma)$ is the Gaussian distribution.
The normalization factor is the standard Voigt profile
\begin{equation}
    V^{\mu i}=V(0;\sigma_{\mu i},\chi_{\mu i})
    =\int dx'\,G(x';\sigma_{\mu i})L(x';\chi_{\mu i}),
\end{equation}
which ensures
\begin{equation}
    \sum_n p^{\mu i}_n = \int d\lambda \, e^{S(\lambda)} p^{\mu i}(\lambda) = 1.
\end{equation}

The symmetric LG distribution is characterized by three parameters: the peak position \(a_{\mu i}+\Delta_{\mu i}\), the Lorentzian width \(\chi_{\mu i}\), and the Gaussian width \(\sigma_{\mu i}\). It reduces to the pure Lorentzian \eqref{eq:Lorentz} in the limit \(\sigma_{\mu i}\to\infty\), and to the Gaussian \eqref{eq:Gaussian} in the limit \(\chi_{\mu i}\to0\). The profile exhibits a Lorentzian sharp peak with a Gaussian thin tail, while the mirror symmetry about the center is strictly preserved, consistent with the SCBA parity theorem.

The composite profile $G\times L/V$  is characterized by three parameters $(a_{\mu i}+\Delta_{\mu i},\chi_{\mu i},\sigma_{\mu i})$, which correspond roughly to the peak position, full width at half maximum (FWHM), and variance of the distribution, respectively. 
Substituting \cref{eq:LGdistr_sym} into the self-consistency condition \eqref{PAG}, and following the same decoupling procedure as for the Lorentzian case, we obtain the parameter equations.  When $\lambda$ is not too far from the center (so that neither $\Im\, \mathcal{G}$ nor  $\Re\, \mathcal{G}$ is negligible), we obtain
\begin{align}
	r(\lambda)\chi_{\mu i}=\Im\, \mathcal{G}_{\mu i}(\lambda - \mathrm{i}0^+)\times G^{\mu i}(\lambda)/V^{\mu i}\notag \\
	\delta \lambda_{\mu i}^r=[\Delta^{\mu i}(\lambda) - \Re\, \mathcal{G}_{\mu i}(\lambda)]\times G^{\mu i}(\lambda)/V^{\mu i}.
\end{align}
If a suitable $\lambda$ is chosen for each $\mu i$, we should be able to set $ r(\lambda)=1$, thereby simplifying the self-consistent equations to
\begin{align}
    \chi_{\mu i} &= \Im\,\mathcal{G}_{\mu i}(\lambda - \mathrm{i}0^+) \times \frac{G^{\mu i}(\lambda)}{V^{\mu i}}, \label{EAFC1_sym}\\
    \lambda-a_{\mu i}-\Delta_{\mu i} &= \bigl[\lambda-a_{\mu i}-V_{\mu i} - \Re\,\mathcal{G}_{\mu i}(\lambda)\bigr] \times \frac{G^{\mu i}(\lambda)}{V^{\mu i}}. \label{EAFC2_sym}
\end{align}
When \(\lambda\) is far from the center, both \(\Im\,\mathcal{G}\) and \(\Re\,\mathcal{G}\) decay rapidly, becoming much smaller than \(|\Delta^{\mu i}(\lambda)|\). In this regime,
\begin{equation}\label{EAFCfar_sym}
    [\Delta^{\mu i}(\lambda)]^2 \frac{G^{\mu i}(\lambda)L^{\mu i}(\lambda)}{V^{\mu i}} \approx \frac{1}{\pi}\,\Im\,\mathcal{G}_{\mu i}(\lambda - \mathrm{i}0^+).
\end{equation}

The self-energy components are evaluated using the mean-field expressions \eqref{IMGAP} and \eqref{REGAP} with the symmetric LG ansatz:
\begin{align}
    \frac{1}{\pi}\Im\,\mathcal{G}^{\mathrm{OD}}_{\mu i}(\lambda - \mathrm{i}0^+) = \sum_{\nu j \neq \mu i} \abs{V_{\mu i,\nu j}}^2 \, \frac{G^{\nu j}(\lambda)L^{\nu j}(\lambda)}{V^{\nu j}}, \label{IMGui_sym}\\
    \Re\,\mathcal{G}^{\mathrm{OD}}_{\mu i}(\lambda) = \sum_{\nu j \neq \mu i} \abs{V_{\mu i,\nu j}}^2 \dashint d\lambda_m \, \frac{G^{\nu j}(\lambda_m)L^{\nu j}(\lambda_m)}{V^{\nu j}\times(\lambda-\lambda_m)}. \label{REGui_sym}
\end{align}
According to \cref{HTEA}, the real part reduces to the compact form
\begin{equation}\label{REGui_sym_simple}
    \Re\,\mathcal{G}_{\mu i}(\lambda) = 	\sum_{\nu j \neq \mu i} \abs{V_{\mu i,\nu j}}^2\frac{\delta\lambda_{\nu j}+\chi_{\nu j} \delta D}{\delta\lambda_{\nu j}^2+\chi_{\nu j}^2},
\end{equation}
where $\delta D= D(\delta\lambda_{\nu j};\sigma_{\nu j},0)/V^{\nu j}$ and $D$ is the dispersion profile, defined as the Hilbert transform of the Voigt profile. 
The expression \eqref{REGui_sym_simple} is manifestly odd under reflection \(\delta\lambda_{\nu j}\to-\delta\lambda_{\nu j}\), as required by the SCBA parity structure.
Equations \eqref{EAFC1_sym}--\eqref{REGui_sym_simple} constitute a closed set of self-consistent equations for the three parameters \((\Delta_{\mu i},\chi_{\mu i},\sigma_{\mu i})\) within the SCBA framework. They smoothly interpolate between the Lorentzian bulk and the Gaussian tail, while the strict reflection symmetry  is built in as a consequence of the parity-preserving nature of the SCBA functional.

\subsection{Beyond Mean-Field: Non-Perturbative Structure and Fluctuations}
\label{sec:beyond_mf}

\subsubsection{Lanczos continued fraction as a non-perturbative backbone}\label{sec:lanczos}

As noted in Sec.~\ref{RSCE}, the diagonal resolvent $\mathcal{R}_{\mu i}(z)$ admits an exact continued-fraction representation via the Lanczos (Krylov) algorithm. Starting from $|\phi_0\rangle \equiv |\phi_{\mu i}\rangle$, the Lanczos basis $\{|\phi_n\rangle\}$ tridiagonalizes the Hamiltonian:
\begin{equation}
 H|\phi_n\rangle = b_n|\phi_{n-1}\rangle + a_n|\phi_n\rangle + b_{n+1}|\phi_{n+1}\rangle,
\end{equation}
with $b_0=0$. Defining $R_n(z)=\langle\phi_n|(z-  H^{[\ge n]})^{-1}|\phi_n\rangle$, one obtains the exact recurrence
\begin{equation}
R_n(z) = \frac{1}{z-a_n-b_{n+1}^2R_{n+1}(z)},
\end{equation}
so that $\mathcal{R}_{\mu i}(z)=R_0(z)$ and
\begin{equation}
\mathcal{R}_{\mu i}(z)
= \cfrac{1}{z - a_0 - \cfrac{b_1^2}{z - a_1 - \cfrac{b_2^2}{z - a_2 - \ddots}}}.
\label{eq:cf_main}
\end{equation}

This representation is fully equivalent to the projected resolvent equation \eqref{eq:CSEQ}, but reorganizes the same information in a fundamentally different way: all many-body correlations are encoded into the scalar sequences $\{a_n,b_n\}$, yielding an effective one-dimensional recursion. The projection-based formulation exposes the explicit multi-resolvent structure of the self-energy, while the Lanczos continued fraction compresses this structure into a single recursive chain. The two are complementary at the exact level and lead to different approximation schemes.

Crucially, the continued fraction does not rely on any expansion in the interaction. The Lanczos coefficients $a_n$ and $b_n$ are determined by the full Hamiltonian and contain all orders of $V$ exactly. Truncating the continued fraction at a finite depth is equivalent to approximating the Hamiltonian by a finite band matrix, providing a non-perturbative closure scheme that preserves analyticity and causality. Truncation and self-consistent completion are discussed in detail in Appendix~\ref{app:cf_closure}. 

The continued-fraction recursion is exact for the projected tail resolvents $R_n(z) $, while the replacement of the infinite continued fraction by a finite-depth truncation constitutes a diagonal closure approximation that neglects backward Krylov-sector return processes.

The existence of this exact, non-perturbative backbone motivates the analysis that follows: it supplies the recursive structure that, even before multi-resolvent correlations are introduced, already transcends conventional mean-field treatments.

\subsubsection{Two complementary nonperturbative structures beyond SCBA}
\label{sec:TLB}

The mean-field approximation (Eq.~\eqref{eq:GRE}) yields a closed self-consistent equation that, under the standard assumptions of a flat density of states and energy-independent couplings, reduces precisely to the SCBA, with a frequency-independent self-energy and a pure Lorentzian spectral function. Note that a mean-field treatment can produce symmetric non-Lorentzian tails if the input couplings $|V_{\mu i,\nu j}|^2$ themselves possess an energy dependence; in such cases the spectral function passively inherits these features, while the self-energy remains a linear functional of the resolvent.

The present framework transcends SCBA in a fundamentally more structural way, restoring two complementary ingredients that SCBA lacks regardless of the coupling profile: a recursive, non-perturbative propagation structure within the single-resolvent sector, and genuine multi-resolvent correlations that couple different frequency scales. We organize these as successive mechanisms beyond mean field.

\textbf{Continued-fraction closure.} A first, decisive departure from SCBA is achieved by exploiting the exact continued-fraction representation of the resolvent. Truncating this structure at a finite depth and imposing self-consistency yields a non-perturbative closure that remains confined to the diagonal (single-resolvent) sector yet qualitatively alters the propagation physics. The simplest such closure, still neglecting all cross-correlated terms ($\mathcal{G}^{\mathrm{CC}}$), is given by Eq.~\eqref{eq:2oclosure} in Appendix~\ref{app:hierarchy_closure}:
\begin{equation}\label{CFC2}
\mathcal{R}_{\mu i}(z) \approx \frac{1}{z - a_{\mu i,0} - \frac{b_{\mu i,1}^{2}}{z - a_{\mu i,1} - b_{\mu i,2}^{2}\,\mathcal{G}_{\mu i,2}^{\mathrm{OD}}(z)}} .
\end{equation}
The structural distinction from SCBA is fundamental: the resolvent appears \emph{recursively in the denominator}, making the effective self-energy a nonlinear functional of $\mathcal{R}_{\mu i}(z)$. Even for constant microscopic couplings, this enforces an intrinsic frequency dependence that SCBA cannot produce, actively generating a finite non-Lorentzian scale rather than passively inheriting one.
This closure is therefore a non-perturbative extension of mean field that reorganizes an infinite subset of higher-order virtual processes into a structurally richer single-resolvent backbone, while preserving analyticity and causality.

In the language of analytic structure, the limitation of SCBA transcends its status as a low-order approximation: SCBA compresses the entire many-body propagation into a single retarded pole, yielding an effectively single-pole relaxation structure associated with exponential decay in the time domain. The continued-fraction closure qualitatively breaks this single-pole constraint by making the effective self-energy a nonlinear functional of the resolvent, thereby generating intrinsic non-Lorentzian spectral scales even without invoking multi-resolvent correlations.

\textbf{Multi-resolvent correlations.} While the continued-fraction closure introduces intrinsic frequency dependence, it remains confined to the single-resolvent sector. The full fluctuation physics emerges when multi-resolvent correlations are incorporated. As derived in Sec.~\ref{sec:cross_expansion}, the exact projected correlation kernel admits a hierarchical expansion
\begin{equation}
\mathcal{G}_{\mu i}(z) = \mathcal{G}^{\mathrm{OD}}_{\mu i}(z)+  \Delta\mathcal{G}_{\mu i}(z)+ \sum_{\ell\ge 3} \mathcal{G}^{(\ell)}_{\mu i}(z),
\end{equation}
where the leading term
\begin{equation}
\mathcal{G}_{\mu i}^{(3)}(z) = \sum_{\xi k\neq\nu j\neq\mu i} V_{\mu i,\xi k}V_{\xi k,\nu j}V_{\nu j,\mu i}  \mathcal{R}^{(\mu i)}_{\nu j}(z) \mathcal{R}^{(\mu i,\nu j)}_{\xi k}(z)
\end{equation}
introduces an explicit product of two resolvents and, consequently, nonlocal frequency couplings via Hilbert transforms (Eq.~\eqref{TTG3}). This multi-resolvent structure is the origin of features absent in any single-resolvent closure:
\begin{itemize}
    \item \textbf{Spectral asymmetry (skewness):} The imaginary part of $\mathcal{G}^{(3)}$ possesses odd frequency parity when the coupled spectral functions differ, directly inducing an asymmetric line shape.
    \item \textbf{Refinement of spectral tails:} The real part of $\mathcal{G}^{(3)}$ provides frequency-dependent corrections to the tail decay, enriching the non-Lorentzian behavior established by the continued-fraction backbone.
\end{itemize}
These multi-resolvent effects are a systematic layer of fluctuation corrections built on top of the non-perturbative backbone: the continued fraction sets the overall width and non-Lorentzian scale, while the multi-resolvent hierarchy supplies the odd-parity components and tail refinements.

SCBA with energy-dependent couplings can produce symmetric non-Lorentzian tails (e.g.\ the symmetric Voigt profile of Sec.~\ref{HAS}), but as proved in Appendix~\ref{app:no_skewness_mf}, SCBA can never generate spectral asymmetry: its self-energy functional is parity-preserving, enforcing $\Delta'_{\mu i} = \Delta_{\mu i}$. The continued-fraction closure \eqref{eq:2oclosure} provides an intrinsic non-Lorentzian scale, yet remains single-resolvent and therefore also preserves symmetry. Genuine skewness ($\Delta'_{\mu i} \neq \Delta_{\mu i}$) arises only from the multi-resolvent hierarchy, whose odd-parity interference terms are the unique microscopic source of spectral asymmetry. These distinct capabilities are summarized in Table~\ref{tab:layer_capabilities}. Numerical confirmation that multi-resolvent contributions can improve
self-consistency beyond the single-resolvent level is provided in \cref{FIG4}.

\begin{table*}[htbp]
\centering
\caption{Capabilities of the three theoretical levels with respect to non-Lorentzian spectral features. 
Continued fraction generates symmetric non-Lorentzian scales; Multi-resolvent hierarchy provides the leading mechanism for skewness.}
\label{tab:layer_capabilities}
\begin{tabular}{|p{3.5cm}|c|c|c|}
\hline
\textbf{Feature} & \textbf{SCBA (constant $V$)} & \textbf{SCBA (energy-dep.\ $V$)} & \textbf{Present framework} \\
\hline
Lorentzian bulk & \checkmark & \checkmark & \checkmark \\
\hline
Symmetric non-Lorentzian tails & -- & \checkmark & \checkmark \\
\hline
Intrinsic non-Lorentzian scale (const.\ $V$) & -- & -- & \checkmark \\
\hline
Spectral asymmetry (skewness) & -- & -- & \checkmark \\
\hline
\textbf{Physical origin} & \textbf{Local, parity-preserving} & \textbf{Local, energy-dependent} & \textbf{Recursion + Multi-R parity mixing} \\
\hline
\end{tabular}
\end{table*}

The hierarchical ansatz strategy of Sec.~\ref{HAS} encodes the combined effects of both mechanisms: the non-perturbative closure determines the overall width and tail scale, while the multi-resolvent hierarchy supplies the odd-parity components responsible for asymmetry. In this unified framework, the continued-fraction closure provides an accurate non-perturbative skeleton for solving the self-consistent equation, whereas the multi-resolvent hierarchy supplies the crucial source terms that introduce nonlocal physics such as skewness. Standard SCBA fails precisely because it lacks both of these essential structural ingredients—nonlinear recursion and nonlocal coupling—which the present framework restores in a controlled and complementary manner.

\subsection{Hierarchical Ans\"atze and Effective Self-Energy Closure}
\label{sec:ansatz_closure}

The analysis of Sec.~\ref{sec:beyond_mf} motivates us to synthesize the non-perturbative backbone and multi-resolvent fluctuations into a single, practical scheme. To capture these effects, we are naturally led to consider spectral ans\"atze of increasing complexity, culminating in an effective self-energy closure.

\subsubsection{Asymmetric Lorentzian-Gaussian (LG) Ansatz and Its Direct Solution}

The symmetric LG ansatz of Sec.~\ref{SLGA} inherits the parity-preserving nature of the SCBA functional and therefore cannot describe the skewness induced by multi-resolvent correlations ($\mathcal{G}^{(3)}$, etc.). To incorporate asymmetry, we must relax the constraint $\Delta'_{\mu i} = \Delta_{\mu i}$ and allow the Gaussian and Lorentzian components to centre at different energies. This leads to the asymmetric hybrid Voigt-type profile:
\begin{equation}\label{eq:LGdistr_final}
    p^{\mu i}(\lambda) = \frac{G^{\mu i}_A(\lambda) L^{\mu i}(\lambda)}{e^{S(\lambda)} V^{\mu i}_A},
\end{equation}
where $G^{\mu i}_A(\lambda)=G(\delta\lambda'_{\mu i};\sigma_{\mu i})$, $L^{\mu i}(\lambda)=L(\delta\lambda_{\mu i};\chi_{\mu i})$, and $V^{\mu i}_A=V(\Delta_{\mu i}-\Delta'_{\mu i};\sigma_{\mu i},\chi_{\mu i})$ is the asymmetric Voigt normalization. The profile is characterized by four parameters: the Lorentzian centre $a_{\mu i}+\Delta_{\mu i}$, the Gaussian centre $a_{\mu i}+\Delta'_{\mu i}$, the Lorentzian width $\chi_{\mu i}$, and the Gaussian width $\sigma_{\mu i}$. The inequality $\Delta'_{\mu i} \neq \Delta_{\mu i}$ breaks mirror symmetry, consistent with the numerical results of Ref.~\cite{HC25}.

The self-consistent equations for these parameters follow from inserting \cref{eq:LGdistr_final} into \cref{PAG} and using the resolvent structure. Crucially, the required real and imaginary parts of $\mathcal{R}_{\nu j}$ can be obtained by directly generalizing the symmetric Voigt results of \cref{SLGA} and \cref{HTEA}. This generalization is straightforward and only requires the substitutions $G\to G_A$, $V\to V_A$, and the replacement of the symmetric dispersion function $\delta D$ by its asymmetric counterpart:
\begin{align}
    \delta D_A= [D(\delta\lambda'_{\nu j};\sigma_{\nu j},0)-D(\Delta_{\nu j}-\Delta'_{\nu j};\sigma_{\nu j},\chi_{\nu j})]/V^{\nu j}_A .
\end{align}
This compactly encodes how the real-part dispersion relation is modified when the Gaussian centre is shifted relative to the Lorentzian centre.

Following this prescription, we define the quantity $\delta w$ that naturally appears in the resolvent decomposition,
\begin{equation}
    \delta w =\left[w\!\Big(\frac{\delta\lambda'_{\nu j}}{\sqrt{2}\sigma_{\nu j}}\Big)- w\!\left(z_0^{\nu j}\right)\right]/\Re w\!\left(z_0^{\nu j}\right),
\end{equation}
with $z_0^{\nu j}:=\frac{\Delta_{\nu j}-\Delta'_{\nu j}+\mathrm{i}\chi_{\nu j}}{\sqrt{2}\sigma_{\nu j}}$. One can then show that $\delta w = G^{\nu j}(\lambda)/V^{\nu j}-1+\mathrm{i}\,\delta D_A$, which directly links the complex resolvent structure to the shift in the dispersion function. After some algebra, the boundary value of the resolvent decomposes as

    \begin{equation}\label{KKR_final}
        \mathcal{R}_{\nu j}(\lambda - \mathrm{i}0^{+})
        = \frac{1}{\delta\lambda_{\nu j}- \mathrm{i}\chi_{\nu j} }
        + \mathrm{i}\frac{\chi_{\nu j}\delta w^*}{\abs{\delta\lambda_{\nu j}-\mathrm{i}\chi_{\nu j} }^2}.
    \end{equation}
    
Equation~\eqref{KKR_final} is the LG counterpart of the Lorentzian 
bridge relation~\eqref{eq:R_Lorentz_bridge}. 
The structural difference between the two closures becomes transparent when 
the LG correction is rewritten in terms of retarded and advanced resolvents:
\begin{equation}\label{eq:LG_RA_decomp}
    \mathcal{R}^{\mathrm{LG}}_{\nu j}(\lambda-\mathrm{i}0^{+})
    = \Bigl(1+\frac{\delta w^{*}}{2}\Bigr)
      \frac{1}{\delta\lambda_{\nu j}-\mathrm{i}\chi_{\nu j}}
 -
      \frac{\delta w^{*}}{2}\,
      \frac{1}{\delta\lambda_{\nu j}+\mathrm{i}\chi_{\nu j}}
\end{equation}
where the second term is anadvanced-type analytic component absent in the 
Lorentzian closure.
Equation~\eqref{eq:LG_RA_decomp} reveals the essential physics of the LG 
ansatz in a unified manner:
\begin{enumerate}
    \item \textbf{Causal-structure enrichment.} The Lorentzian closure 
    contains only a retarded pole $(\delta\lambda-\mathrm{i}\chi)^{-1}$. 
    The LG correction introduces an advanced pole $(\delta\lambda+\mathrm{i}\chi)^{-1}$ 
    with complex weight $-\delta w^{*}/2$. The resolvent is no longer purely 
    retarded: it develops a retarded/advanced interference structure.

    \item \textbf{Parity mixing and skewness.} The retarded and advanced 
    sectors transform oppositely under energy reflection 
    $\delta\lambda\to-\delta\lambda$. Their mixing through $\delta w$ (whose 
    imaginary part is sourced by the centre offset $\Delta'-\Delta$) 
    generates odd-parity contributions in $\Im\mathcal{R}$, i.e.\ spectral 
    skewness. In the symmetric limit $\Delta'=\Delta$, $\delta w$ is purely 
    real and the decomposition~\eqref{eq:LG_RA_decomp} does not generate odd-parity contributions.

    \item \textbf{Gaussian tail encoding.} The weight $\delta w$ involves 
    the Faddeeva function $w(z)$, which encodes the Gaussian suppression of 
    spectral tails through the scale $\sigma_{\nu j}$. The advanced admixture 
    is thus the analytic mechanism by which the Gaussian envelope modifies 
    the resolvent structure.

    \item \textbf{Smooth Lorentzian limit.} When $\sigma_{\nu j}\to\infty$, 
    $w(z)\to 1$ and $\delta w\to 0$, so 
    $\mathcal{R}^{\mathrm{LG}}\to (\delta\lambda-\mathrm{i}\chi)^{-1}$. 
    The advanced sector decouples and the single-pole retarded structure 
    of the SCBA is recovered.
\end{enumerate}
The logical chain connecting microscopic correlations to emergent spectral 
asymmetry is now explicit:
\begin{align}
    \text{odd-parity hierarchy}\;\longrightarrow\;
    \Delta'-\Delta\;\longrightarrow\;
    \Im\delta w\;\notag\\
    \longrightarrow\;
    \text{parity-section mixing}\;\longrightarrow\;
    \text{skewness}\notag.
\end{align}

When $\lambda$ is not too far from the centre, the self-consistency conditions \cref{EAFC1_sym,EAFC2_sym} generalize to
    \begin{align}\label{DVC_final}
        \Delta_{\mu i}-V_{\mu i}+ \mathrm{i} \chi_{\mu i}
        \sim (\lambda-a_{\mu i}-V_{\mu i})\Bigl(1-\frac{G^{\mu i}(\lambda)}{V^{\mu i}}\Bigr)
       \notag\\
	   +\Bigl[\, \sum_{\nu j \neq \mu i} \abs{V_{\mu i,\nu j}}^2 \mathcal{R}_{\nu j}
        +\sum_{\xi k \neq \nu j \neq \mu i} V^{(3)}_{\mu i,\xi k,\nu j} \mathcal{R}_{\nu j}\mathcal{R}_{\xi k}\Bigr]\frac{G^{\mu i}(\lambda)}{V^{\mu i}} .
    \end{align}
Together with the far-tail condition (the asymmetric version of \cref{EAFCfar_sym}), \cref{DVC_final} constitutes a closed set of equations for $(a_{\mu i}+\Delta_{\mu i}, a_{\mu i}+\Delta'_{\mu i}, \chi_{\mu i}, \sigma_{\mu i})$. 

\textbf{However, solving these equations directly is formidably difficult.} The difficulty is not merely the proliferation of parameters, but the structural complexity of the coupled system: \cref{DVC_final} nonlinearly couples the discrete indices $(\mu i, \nu j, \xi k)$ and, through the Hilbert transforms implicit in $\mathcal{R}_{\nu j}$ and $\mathcal{R}_{\nu j}\mathcal{R}_{\xi k}$, mixes integral and algebraic operations. This renders direct iterative solution computationally prohibitive for all but the smallest systems.

\subsubsection{ Effective Self-Energy Representation}\label{ESER}
The impasse encountered above motivates a strategic shift in perspective. Instead of formulating the ansatz at the level of the probability density $p^{\mu i}$ and deriving its parameter equations from \cref{PAG}, we go back one step and construct an ansatz directly for the \emph{self-energy} $\mathcal{G}_{\mu i}$. This approach proves more fundamental and leads to a mathematically simpler closure scheme.

The Faddeeva structure that we shall adopt admits a deeper theoretical motivation beyond its role as a convenient causal ansatz. Within the multi-resolvent hierarchy, products $\mathcal{R}_\nu\mathcal{R}_\xi$ couple distinct pole contributions whose positions are determined by the eigenvalues of the interacting Hamiltonian. Under the ETH-type statistical treatment, the matrix elements that control these pole positions fluctuate, and when the pole positions $\lambda_\nu$ themselves are statistically distributed. As is natural for the dense spectra of nonintegrable systems, the ensemble average over Gaussian-distributed pole positions naturally motivates the Faddeeva function: $\int d\lambda_\nu\, e^{-\lambda_\nu^2}(\lambda-\lambda_\nu+i\chi)^{-1} \propto w(z)$. The Faddeeva self-energy is therefore not an \emph{ad hoc} choice but the natural consequence of multi-resolvent propagation under statistical regularity. In the language of complex analysis, the Faddeeva function corresponds to a Gaussian-weighted effective distribution of pole contributions---a distributed singularity structure rather than an isolated single pole, and it is precisely this effective distribution of pole contributions that captures the emergent non-Lorentzian broadening.

We propose an effective self-energy of the form
\begin{equation}\label{afg_final}
    \mathcal{G}_{\mu i}(\lambda) = \Delta^{\text{eff}}_{\mu i}-V_{\mu i} +\Sigma_{\mu i}(\lambda),
\end{equation}
where the frequency dependence is carried by an analytically continued error function,
\begin{equation}
    \Sigma_{\mu i}(\lambda)= \mathrm{i}\chi^{\text{eff}}_{\mu i} w\!\Big(\frac{-\delta \lambda'_{\mu i}}{\sqrt{2}\sigma_{\mu i}}\Big).
\end{equation}
Here $w(z)$ is the Faddeeva function. The corresponding spectral function follows directly from the Dyson equation \cref{eq:CSEQ}:
\begin{equation}\label{ofag_final}
    p^{\mu i}_n = \frac{1}{e^{S(\lambda_n)}} \frac{1}{\pi} \Im \frac{1}{\lambda_n - a_{\mu i}-\Delta^{\text{eff}}_{\mu i} -\Sigma_{\mu i}(\lambda_n)}.
\end{equation}

This representation possesses several crucial advantages:
\begin{itemize}
    \item \textbf{Causality and normalization:} Since $w(z)$ is analytic in the complex plane and satisfies $w(z)=\mathrm{i}/(\sqrt{\pi}z)+O(z^{-3})$ as $|z|\to\infty$, the function $F(z)=[z-a_{\mu i}-V_{\mu i}-\mathcal{G}_{\mu i}(z)]^{-1}$ is guaranteed to be a causal (retarded) Green's function. Its spectral function automatically satisfies the sum rule $\int_{-\infty}^{\infty}dx \frac{1}{\pi}\Im F(x-\mathrm{i}0^+)=1$ for any positive parameters $\sigma_{\mu i}$ and $\chi_{\mu i}$, ensuring correct normalization without additional constraints. A systematic analysis of the admissible functional forms for the self-energy, based on the analyticity and Kramers-Kronig structure, is presented in \cref{app:analytic_foundations}.
    \item \textbf{Analytic tractability:} The self-energy is expressed in closed analytic form, eliminating the need for explicit Hilbert transforms in the defining equations.
    \item \textbf{Physical transparency:} The real and imaginary parts of the Faddeeva function describe, respectively, the Gaussian-tailed dispersion and absorption, providing a clear separation of broadening mechanisms.
\end{itemize}

As shown in detail in \cref{app:analytic_foundations}, the spectral function \cref{ofag_final} is approximately equivalent to the asymmetric Voigt profile \cref{eq:LGdistr_final}, with the effective parameters $(\Delta^{\text{eff}}_{\mu i}, \chi^{\text{eff}}_{\mu i})$ fixed by matching the peak position and peak height. For the parameter ranges relevant to the numerical model of Ref.~\cite{HC25}, the two representations are nearly indistinguishable. Thus, the effective self-energy representation provides an equivalent but computationally more economical parametrization of the same physics.

\subsubsection{Closed Self-Consistent Equations with Multi-Resolvent Corrections}

The true power of the effective self-energy representation emerges when it is combined with the exact resolvent structure. By inserting \cref{afg_final} into the defining equation \cref{eq:CSEQ} and using the hierarchical expansion of the self-energy from Sec.~\ref{sec:cross_expansion}, we obtain a compact self-consistency condition:
\begin{align}\label{gacon_final}
    \Delta^{\text{eff}}_{\mu i}-V_{\mu i} + \mathrm{i}\chi^{\text{eff}}_{\mu i} w\!\Big(\frac{-\delta \lambda'_{\mu i}}{\sqrt{2}\sigma_{\mu i}}\Big) \notag\\
    = \sum_{\nu j \neq \mu i} \abs{V_{\mu i,\nu j}}^2 \mathcal{R}_{\nu j}
    + \sum_{\xi k \neq \nu j \neq \mu i} V^{(3)}_{\mu i,\xi k,\nu j} \mathcal{R}_{\nu j}\mathcal{R}_{\xi k}.
\end{align}
In contrast to \cref{DVC_final}, \textbf{this equation does not require a case distinction} between the central and far-tail regimes, and its structure is symmetric with respect to the real and imaginary parts. The right-hand side is evaluated using the explicit resolvent decomposition \cref{KKR_final} or, in statistically homogeneous regimes, can be further simplified using ETH-type self-averaging approximations as described in Appendix~\ref{app:hierarchy_closure}.

Equation \cref{gacon_final} represents the central practical result of the hierarchical ansatz strategy. The left-hand side encodes an asymmetric line shape with the minimal number of physically transparent parameters: the effective level $\Delta^{\text{eff}}_{\mu i}$, the effective width $\chi^{\text{eff}}_{\mu i}$, and the Gaussian scale $\sigma_{\mu i}$ (which also controls the skewness through the offset $\delta\lambda'_{\mu i}$). The right-hand side incorporates both the mean-field (SCBA) contribution and the leading multi-resolvent correction, whose odd frequency parity provides the microscopic source of asymmetry in the spectral line shape. 

\subsubsection{Parity decomposition, Hilbert-transform structure, and skewness self-consistency}\label{sec:parity_decomp}

The origin of spectral asymmetry becomes particularly transparent when the
hierarchical expansion is analyzed together with the analytic structure of the
resolvent. We begin by decomposing the effective self-energy according to its
parity under reflection about the effective center.

\medskip
\noindent\textbf{Parity decomposition and the skewness order parameter.}
Let
\begin{equation}
x := \lambda - a_{\mu i} - \Delta^{\mathrm{eff}}_{\mu i},
\end{equation}
and parameterize the displacement between the Gaussian and Lorentzian centers by
\begin{equation}
\delta s_{\mu i} := \Delta'_{\mu i} - \Delta^{\mathrm{eff}}_{\mu i}.
\end{equation}
The quantity $\delta s_{\mu i}$ therefore measures the skewness of the spectral
line shape. In the symmetric SCBA sector one has $\delta s_{\mu i}=0$, whereas
$\delta s_{\mu i}\neq0$ signals the emergence of odd-parity multi-resolvent
correlations.

Using the effective self-energy ansatz introduced in Eq.~\eqref{afg_final},
\begin{equation}
\Sigma_{\mu i}(\lambda)
=
\mathrm{i}\chi^{\mathrm{eff}}_{\mu i}
\,
w\!\left(
-\frac{x-\delta s_{\mu i}}
{\sqrt{2}\,\sigma_{\mu i}}
\right),
\end{equation}
we expand around the symmetric point $\delta s_{\mu i}=0$. Defining
\begin{equation}
z_{\mu i}:=-\frac{x}{\sqrt{2}\,\sigma_{\mu i}},
\qquad
\eta_{\mu i}:=
\frac{\delta s_{\mu i}}
{\sqrt{2}\,\sigma_{\mu i}},
\end{equation}
the Faddeeva function admits the expansion
\begin{equation}
w(z_{\mu i}-\eta_{\mu i})
=
w(z_{\mu i})
-\eta_{\mu i}\,w'(z_{\mu i})
+
\mathcal{O}(\eta_{\mu i}^2).
\end{equation}

Accordingly, the self-energy naturally separates into even- and odd-parity sectors,
\begin{equation}
\Sigma_{\mu i}
=
\Sigma^{(e)}_{\mu i}
+
\Sigma^{(o)}_{\mu i}
+
\mathcal{O}(\eta_{\mu i}^2),
\end{equation}
with
\begin{align}
\Sigma^{(e)}_{\mu i}
&=
\mathrm{i}\,\chi^{\mathrm{eff}}_{\mu i}\,w(z_{\mu i}),
\\
\Sigma^{(o)}_{\mu i}
&=
-\mathrm{i}\,\chi^{\mathrm{eff}}_{\mu i}
\,\eta_{\mu i}\,
w'(z_{\mu i}).
\end{align}
The even sector reproduces the parity-preserving SCBA-type structure discussed in
Sec.~\ref{SLGA} and Appendix~\ref{app:no_skewness_mf}. The odd sector
$\Sigma^{(o)}_{\mu i}$ represents a parity-mixing correction whose microscopic
origin is exposed by the Hilbert-transform structure of the resolvent hierarchy.

\medskip
\noindent\textbf{Hilbert-transform structure and parity mixing.}
The boundary value of the diagonal resolvent satisfies the exact relation
Eq.~\eqref{RkkR},
\begin{equation}
\frac{1}{\pi}
\mathcal{R}_{\mu i}(\lambda-\mathrm{i}0^+)
=
H[f_{\mu i}](\lambda)
+
\mathrm{i}\,f_{\mu i}(\lambda),
\end{equation}
where $f_{\mu i}(\lambda)=e^{S(\lambda)}p_{\mu i}(\lambda)$. A crucial property of
the Hilbert transform is that it exchanges parity sectors:
\begin{equation}
H[f^{(e)}] = f^{(o)},
\qquad
H[f^{(o)}] = -f^{(e)},
\label{Hilbert_parity}
\end{equation}
where $f^{(e)}(-x)=f^{(e)}(x)$ and $f^{(o)}(-x)=-f^{(o)}(x)$. Consequently, a
purely symmetric spectral function generates a resolvent whose real and imaginary
parts possess opposite parity---precisely the structure underlying the symmetric
SCBA solution. Within the single-resolvent sector, the Hilbert-transform
structure remains closed under the parity-preserving mapping
$f^{(e)} \to H[f^{(e)}]^{(o)} \to f^{(e)}$, and therefore preserves reflection
symmetry.

This closure is manifest in the single-pole resolvent 
$\mathcal{R}=1/(\delta\lambda-\mathrm{i}\chi)$, whose real and imaginary 
parts are, respectively, odd and even---the Hilbert-transform pair 
generated by a symmetric $f$. The LG resolvent~\eqref{KKR_final}, 
by contrast, acquires an extra term $\propto\delta w^{*}$ that explicitly 
mixes these parity sectors whenever $\Im\delta w\neq 0$. The Lorentzian 
closure preserves a strict separation between even and odd parity sectors 
through a purely retarded single-pole structure, whereas the LG correction 
introduces retarded/advanced mixing that couples these sectors through the 
complex fluctuation amplitude $\delta w$. This transition from a purely retarded single-pole
structure to a retarded/advanced-interfering analytic
decomposition provides a useful perspective on how the
LG/Faddeeva closure extends the single-pole SCBA picture.

By contrast, the leading multi-resolvent term
\begin{equation}
\mathcal{G}^{(3)}_{\mu i}\approx\mathcal{G}^{(3),\text{D}}_{\mu i}
=
\sum_{\xi k\neq\nu j\neq\mu i}
V^{(3)}_{\mu i,\xi k,\nu j}
\;\mathcal{R}_{\nu j}\,\mathcal{R}_{\xi k}
\end{equation}
mixes distinct Hilbert-transform sectors. Using Eq.~\eqref{RkkR},
\begin{equation}
\mathcal{R}_{\nu j}\,\mathcal{R}_{\xi k}
=
\pi^2
\Bigl(
H[f_{\nu j}] + \mathrm{i}f_{\nu j}
\Bigr)
\Bigl(
H[f_{\xi k}] + \mathrm{i}f_{\xi k}
\Bigr),
\end{equation}
whose imaginary part becomes
\begin{equation}
\Im(\mathcal{R}_{\nu j}\,\mathcal{R}_{\xi k})
=
\pi^2
\Bigl(
f_{\nu j}\,H[f_{\xi k}]
+
H[f_{\nu j}]\,f_{\xi k}
\Bigr).
\label{RR_imag_H}
\end{equation}
Equation~\eqref{RR_imag_H} shows explicitly that the leading fluctuation
correction is generated by interference between a spectral component and the
Hilbert transform of another component. Since the Hilbert transform exchanges
parity sectors, Eq.~\eqref{RR_imag_H} necessarily generates odd-parity
contributions even when the underlying spectral densities are individually
symmetric.

Physically, this interference admits a transparent causal interpretation.
The Hilbert transform $H[f]$ encodes the dispersive (virtual) part of the
propagation, while $f$ itself represents the absorptive (on-shell) part,
as dictated by the Kramers--Kronig relation \eqref{RkkR}. The product
$f_\nu H[f_\xi] + H[f_\nu] f_\xi$ in \eqref{RR_imag_H} therefore describes
interference between real and virtual propagation processes traversing
\emph{distinct} intermediate channels ($\nu$ and $\xi$). Spectral skewness
thus originates from nonlocal interference between different causal
propagation paths---the system simultaneously propagates through multiple
intermediate states, multiple intermediate frequency channels, and multiple virtual
channels, and it is the quantum interference among these channels that
breaks mirror symmetry in the resulting line shape. This establishes
skewness as a direct spectroscopic signature of multi-resolvent coherence
in the frequency domain.

\medskip
\noindent\textbf{Skewness self-consistency condition.}
Matching the odd-parity sector of Eq.~\eqref{gacon_final} therefore yields the
skewness self-consistency condition
\begin{equation}
-\mathrm{i}\,\chi^{\mathrm{eff}}_{\mu i}
\,\eta_{\mu i}\,
w'(z_{\mu i})
\sim
\sum_{\xi k\neq\nu j\neq\mu i}
V^{(3)}_{\mu i,\xi k,\nu j}
\left(
\mathcal{R}^{(e)}_{\nu j}\,\mathcal{R}^{(o)}_{\xi k}
+
\mathcal{R}^{(o)}_{\nu j}\,\mathcal{R}^{(e)}_{\xi k}
\right),
\label{skew_self}
\end{equation}
or, equivalently in terms of the spectral density,
\begin{equation}
\delta s_{\mu i}
\propto
\sum_{\xi k\neq\nu j\neq\mu i}
V^{(3)}_{\mu i,\xi k,\nu j}
\Bigl(
f_{\nu j}\,H[f_{\xi k}]
+
H[f_{\nu j}]\,f_{\xi k}
\Bigr)_{\mathrm{odd}},
\label{skew_hilbert}
\end{equation}
where the subscript ``odd'' denotes projection onto the odd-parity sector.
Equation~\eqref{skew_hilbert} identifies spectral skewness as a \emph{direct
manifestation of parity mixing induced by the analytic Hilbert-transform
structure of the multi-resolvent hierarchy}. In this sense, skewness is not
merely a consequence of higher-order interactions, but specifically of the
nonlocal frequency coupling generated by analyticity itself.

Consequently, $\delta s_{\mu i}=0$ is recovered automatically whenever the
hierarchy is projected onto the single-resolvent (SCBA) sector. The Lanczos
continued-fraction representation provides the geometric counterpart of this
analytic structure: a stationary Krylov chain with constant Lanczos coefficients
$a_n$ preserves reflection symmetry, whereas spectral skewness reflects a
breakdown of this stationarity, schematically associated with gradients
$a_{n+1}-a_n\neq0$. Within the present hierarchy, the multi-resolvent
correlations provide the microscopic origin of this effective ``drift'' along
the Krylov chain, while the skewness parameter $\delta s_{\mu i}$ measures its
spectral manifestation. This establishes that parity-preserving
continued-fraction closures, even when non-perturbative, cannot generate a
non-zero $\delta s_{\mu i}$; skewness is a specific signature of nonlocal
multi-resolvent correlations.

For completeness, we note that a much simpler skewed Cauchy profile can be
obtained directly from the odd component of the self‑energy in the
wide-band limit (Appendix~\ref{app:skewed_cauchy}).
In that limit the mean-field spectral function becomes a pure Lorentzian,
and the leading multi-resolvent correction reduces to a rational function;
the resulting line shape takes the minimal form
$p(\lambda)\propto(1+\alpha x)/[(x+\delta)^2+\Gamma^2]$.
This model is, however, overly restrictive: it requires the constant-self-energy
approximation to hold for the SCBA part, and therefore cannot describe the
Gaussian tails or the intrinsic frequency scales that are central to the
effective Faddeeva representation developed here.
Hence it should be regarded as a pedagogical limiting case rather than a
generic result of the present framework.

\subsubsection{Limitations and Outlook}

Despite its advantages, \cref{gacon_final} is not fully self-contained in its present form: the right-hand side explicitly depends on a finite energy window through the resolvent products $\mathcal{R}_{\nu j}\mathcal{R}_{\xi k}$, whereas the left-hand side is a purely local function of $\lambda$. This structural asymmetry signals that the simple Faddeeva self-energy \cref{afg_final}, while capturing the leading multi-resolvent effects, cannot exhaust all the nonlocal correlations generated by the infinite hierarchy.

As demonstrated in \cref{app:analytic_foundations}, a systematic way to construct more refined ans\"atze is to start from the analytic structure of $\mathcal{G}_{\mu i}$ itself. Any admissible self-energy must be analytic in the lower half-plane, and its real and imaginary parts must satisfy the Kramers--Kronig relations in the form $-\Im g^{\mu i}(\lambda) = H(\Re g^{\mu i})(\lambda)$. The Faddeeva function $g^{\mu i}(\lambda)=w(-\delta\lambda'/\sqrt{2}\sigma)$ is the simplest extension of the constant ($g(\lambda) =1$) case that introduces a non-Lorentzian scale while preserving causality. A natural next step is to consider a linear combination of two such functions,
\begin{equation}
    g^{\mu i}(\lambda)= w\!\Big(\frac{-\delta \lambda^{(1)}_{\mu i}}{\sqrt{2}\sigma^1_{\mu i}}\Big)+ w\!\Big(\frac{-\delta \lambda^{(2)}_{\mu i}}{\sqrt{2}\sigma^2_{\mu i}}\Big),
\end{equation}
which provides additional flexibility to model finite-band effects and more complex tail structures. The analysis of the resulting six-parameter self-consistent equations lies beyond the scope of the present work and is left for future investigation.

In summary, the hierarchical ansatz strategy developed in this section bridges the exact but computationally intractable multi-resolvent hierarchy of Sec.~\ref{sec:cross_expansion} and the practical need for quantitative spectral predictions. The effective self-energy representation \cref{gacon_final}, together with the resolvent decomposition \cref{KKR_final}, constitutes a closed, systematically improvable framework for computing spectral line shapes, distribution tails, and fluctuation effects in generic nonintegrable many-body systems.

\section{Comparison with Traditional Methods}\label{CTM}

To clarify the role and advantages of the proposed methodology, it is useful to contrast it with established approaches. 

\begin{table*}[htbp]
\centering
\caption{Comparison of theoretical frameworks for many-body systems.}
\label{tab:comparison}
\begin{tabular}{|p{4.5cm}|p{3.5cm}|p{3cm}|p{4.5cm}|}
\hline
\textbf{Method} & \textbf{Expansion Basis} & \textbf{Applicability} & \textbf{Key Feature of Our Method} \\
\hline
Perturbation Theory & Local Taylor & Weak coupling, large gaps & Global analytic structure via pole expansion \\
\hline
Non-equilibrium Green's Functions / Diagrammatic & Feynman diagrams & Needs infinite resummation & Self-consistent closure via random-phase averaging \\
\hline
Random Matrix Theory (RMT) & Fully random & Featureless chaos & Structured randomness via ETH \\
\hline
Eigenstate Thermalization Hypothesis (ETH) & Mechanism explanation & Nonintegrable systems & Quantitative, predictive framework with fluctuation expansion \\
\hline
\end{tabular}
\end{table*}

These contrasts are summarized in Table~\ref{tab:comparison}. 
Our methodology bridges the gap between the universal but qualitative ETH and the quantitative but perturbative diagrammatic methods.  It inherits ETH's physical insight that local observables thermalize and translates it into closed self-consistent equations yielding quantitative predictions for global quantities like entropy.  Unlike perturbation theory, it does not rely on a small parameter; unlike RMT, it incorporates the structured randomness dictated by the system's Hamiltonian.  The systematic recursive hierarchy of cross-correlated terms provides a structured and systematically improvable way to go beyond mean field, capturing fluctuations and tails.

To clarify the conceptual distinction from conventional self-consistent approaches, we compare our formulation with the SCBA, which belongs to the class of finite diagrammatic resummations.  In SCBA, the self-energy is truncated at second order: $\Sigma_{\mu i}^{\rm SCBA}(z)=\sum_{\nu j}|V_{\mu i,\nu j}|^2 G_{\nu j}(z)$.  By contrast, the present framework retains the full cross-correlated terms $\mathcal{G}_{\mu i}^{\rm CC}(z)$ containing products of multiple resolvents.  This distinction is already apparent at the level of the self-energy decomposition introduced in Sec.~\ref{RSCE}: the mean-field term $\mathcal{G}_{\mu i}^{\mathrm{OD}}$ yields a self-consistent structure closely related to SCBA, while the full hierarchy includes $\mathcal{G}_{\mu i}^{\mathrm{CC}}$ and generates a recursive hierarchy not captured by any finite diagrammatic class.  A detailed structural analysis is given in \cref{SDDR}.  From this perspective, SCBA is the leading-order statistical approximation obtained by neglecting $\mathcal{G}^{\mathrm{CC}}_{\mu i}(z)$.

The present framework should not be viewed as a refinement of self-consistent Born-type approximations, but as a reorganization of many-body theory in which both mean-field and fluctuation contributions are treated within a unified, resolvent-based structure.

\section{Conditions of Applicability and Generalization}\label{CAG}
\subsection{Conditions for Validity}
The success of this methodology rests on several conditions, which are typical of generic, nonintegrable many-body systems:
\begin{itemize}
    \item \textbf{Nonintegrability}: The system must be chaotic enough to satisfy the random-phase condition.
    \item \textbf{Random-Phase Condition}: The phases of off-diagonal matrix elements must be sufficiently random to allow cross-term cancellation at the mean-field level. The recursive expansion shows that this condition ensures that higher-order terms are parametrically suppressed but not zero—they control the fluctuations.
    \item \textbf{Exponential Density of States}: The framework relies on the canonical scaling of the Hilbert space dimension, which is exponential in system size, to convert sums into integrals and to justify the smoothing procedure.
\end{itemize}
Under these conditions, the method provides a robust and predictive tool.

\subsection{Finite-system considerations and spectral resolution}

The hierarchy developed in this work is naturally formulated in the continuum (thermodynamic) limit, where discrete spectra are coarse-grained through the $\eta\to 0^+$ prescription.
For finite systems, one instead considers the broadened resolvent
\begin{equation}
\mathcal{R}_{\mu i}(\omega-\mathrm{i}\eta) = \sum_n \frac{p_n^{\mu i}}{\omega-\lambda_n-\mathrm{i}\eta}.
\end{equation}
Although the analytic structure of both the continued-fraction recursion and the hierarchy-constrained self-energy formally survives under $z\to z-\mathrm{i}\eta$, the practical effectiveness of hierarchy corrections depends sensitively on spectral resolution.

Two complementary implementations are available. In the \emph{direct finite-$\eta$ approach}, the broadened resolvent is inserted directly into the self-consistent hierarchy. Alternatively, one may first reconstruct a smooth coarse-grained spectral function and then analyze the corresponding continuum resolvent (\emph{indirect approach}). The latter is often advantageous for finite systems because the hierarchy corrections involve correlated multi-resolvent structures whose physical interpretation becomes increasingly sensitive to spectral coarse graining.

A particularly important limitation arises in the overbroadened regime. For sufficiently large $\eta$, the resolvent loses sensitivity to higher spectral moments, while nonlinear hierarchy corrections and deeper continued fractions still attempt to resolve increasingly fine spectral structures. Consequently, the effective advantage of hierarchy corrections is systematically reduced, and the coarse-grained SCBA structure can become comparatively more robust.

The hierarchy corrections are therefore most naturally interpreted within the scale-separation window
\begin{equation}
\delta_{\rm LS}\ll\eta\ll W,
\end{equation}
where $\delta_{\rm LS}$ denotes the local level spacing and $W$ the characteristic spectral bandwidth. A detailed analysis, including the asymptotic large-$\eta$ expansion and concrete implementation guidance, is presented in \cref{sec:finite_eta_limitations}.

\subsection{Generalization to Other Systems}
The methodology is general and can be adapted to:
\begin{itemize}
    \item \textbf{Fermionic/Bosonic Systems}: By appropriately defining the basis states $\ket{\phi_{\mu i}}$ as Fock states and using the corresponding ETH form for fermionic or bosonic operators.
    \item \textbf{Out-of-Equilibrium Situations}: The resolvent-based equations can be generalized to non-equilibrium Green's functions (Keldysh formalism), potentially capturing entropy dynamics during thermalization.
    \item \textbf{Open Quantum Systems}: The framework can be extended to study the steady state of a system coupled to a Markovian bath, by modifying the self-energy term to include bath-induced dissipation.
    \item \textbf{Systems with Additional Symmetries}: The recursive expansion can be adapted to incorporate the effects of conserved quantities, leading to branch structures similar to those observed here.
\end{itemize}
\section{Numerical Results}\label{NumR}

To validate our theoretical framework, we conduct numerical simulations on an Ising spin chain with both transverse and longitudinal magnetic fields. The system Hamiltonian is given by:
\begin{equation}
    H = \sum_{k=1}^{N}\left(-\sigma_z^k \otimes \sigma_z^{k+1} + g \sigma_x^k + h \sigma_z^k\right),
\end{equation}
where \( g = 1.05 \) and \( h = 0.1 \) in our implementation. This system exhibits nonintegrability except when either \( g \) or \( h \) vanishes. While periodic boundary conditions are adopted for simplicity, our conclusions remain valid for arbitrary boundary conditions.
We construct the composite system with \( N = 15 \) spins:
\begin{itemize}
    \item  System \( S \): Single spin governed by \( H_S = g \sigma_x^S + h \sigma_z^S \)
    \item Bath \( B \): Remaining 14 spins following \( H_E = -\sum_{k=1}^{13}\sigma_z^k \otimes \sigma_z^{k+1} + \sum_{k=1}^{14}\left(g \sigma_x^k + h \sigma_z^k\right) \)
    \item System-bath interaction: \( V = - \sigma_z^S \otimes ( \sigma_z^1+\sigma_z^{14} )\)
\end{itemize}
The system energies approximately satisfy \( E_i \approx (-1)^i 1.055 \), distinguishing two energy branches.

We perform numerical simulations using this nonintegrable model. By directly diagonalizing the Hamiltonians \( H_0 \) and \( H \), we obtain their respective energy eigenstates. The transition probability distribution \( p^{\mu i}_n \) is calculated through eigenstate inner products \( \abs{\braket{\psi_n|\phi_{\mu i}}}^2 \). This distribution is subsequently smoothed by energy shell averaging, where probabilities within each energy interval \( \mathcal{M}_{\lambda,\Delta}= [\lambda - \Delta/2, \lambda + \Delta/2] \) are summed to form the binned probability \( P(\mathcal{M}_{\lambda,\Delta})= \sum_{\lambda_m \in \mathcal{M}_{\lambda,\Delta}} p^{\mu i}_m \). The fitting procedure applied to this binned distribution determines the characteristic parameters. 

\subsection{Self-consistency tests: SCBA versus continued-fraction closure}

A central theoretical prediction of our framework is that the continued-fraction (Lanczos) closure \eqref{CFC2} provides a more accurate self-consistent description than the standard SCBA \eqref{Dyson} at the single-resolvent level. As analyzed in Sec.~\ref{sec:TLB} and summarized in Table~\ref{tab:layer_capabilities}, the SCBA self-energy is a linear functional of the resolvent and preserves parity under energy reflection. The continued-fraction closure, by contrast, embeds the resolvent \emph{recursively in the denominator}, making the effective self-energy a nonlinear functional that intrinsically generates non-Lorentzian spectral scales---even for energy-independent microscopic couplings---without requiring multi-resolvent correlations.

To test this prediction, we extract the fitted Lorentzian parameters \((\Delta_{\mu i}, \chi_{\mu i})\) from the exact-diagonalization data and construct the corresponding smooth spectral function \(f^{\mu i}(\lambda)= \frac{1}{\pi}\frac{1}{\delta \lambda_{\mu i}-\mathrm{i} \chi_{\mu i}}\). Substituting this into the right-hand sides of \eqref{Dyson} (SCBA) and \eqref{CFC2} (LCF) yields two independent predictions for \(\Im\mathcal{R}_{\mu i}\), which can be compared against the directly computed value \(\Im\mathcal{R}_{\mu i}=1/\chi_{\mu i}\) obtained from the fitted width. The results are shown in \cref{FIGs2}(a).

We find that the continued-fraction closure \eqref{CFC2} exhibits markedly better self-consistency than the SCBA equation \eqref{Dyson}: the LCF prediction (red) lies systematically closer to the direct value (blue) than the SCBA prediction (orange). This directly corroborates the theoretical analysis of Sec.~\ref{sec:TLB}: the recursive denominator structure captures propagation processes that SCBA, being a linear functional, necessarily misses. Moreover, the systematic deviation of the SCBA curve explains why the self-consistency check reported in Ref.~\cite{HC25} showed noticeable discrepancies---that work employed the SCBA equation, whose inherent limitations at the single-resolvent level produce precisely the observed offset.

\subsection{Finite-\(\eta\) coarse graining and scale separation}\label{fiet}

A conceptually distinct test is provided by working at finite \(\eta\), i.e., evaluating the resolvent at a finite distance from the real axis. Using the exact probability distribution \(p^{\mu i}_n\) obtained from diagonalization, we construct \(\mathcal{R}_{\mu i}\) at finite \(\eta\) via the pole expansion \eqref{PEEX}. Substituting this directly computed \(\mathcal{R}_{\mu i}\) into the right-hand sides of \eqref{Dyson} and \eqref{CFC2} tests the self-consistency without any fitting ansatz, thereby probing the closure relations at a coarse-grained level where the effective energy resolution is set by \(\eta\). The results are shown in \cref{FIGs2}(b).

Strikingly, the relative performance of the two closures reverses compared to the \(\eta\to0\) limit: at finite \(\eta\) (here taken as \(100\times\) the mean level spacing, \(\approx 0.1228\)), the SCBA equation \eqref{Dyson} displays \emph{better} self-consistency than the continued-fraction closure \eqref{CFC2}. This crossover is a direct manifestation of scale separation. At large \(\eta\), the coarse-graining averages over many individual eigenstates, suppressing the fine-grained spectral structures (non-Lorentzian scales, multi-resolvent correlations) that distinguish the two closures, and the problem reduces to an effective mean-field description where the simpler SCBA functional is adequate. Conversely, as \(\eta\to 0\), the resolution becomes sufficient to resolve the intrinsic non-Lorentzian features, and the continued-fraction backbone---which natively encodes these scales---is required for accurate closure.

To quantify this crossover systematically, we compute, for both closures and as a function of \(\eta\), the 2-norm deviation \(\delta\mathcal{R}_{\mu i}\) between the left-hand and right-hand sides of the self-consistent equations, averaged over all bath states \(\mu\) and both system branches \(i=1,2\) within the energy window of \cref{FIGs2}. The result is displayed in \cref{FIG3}. Three regimes emerge: (i) at \(\eta\to 0\), the LCF closure is more accurate; (ii) at \(\eta\approx 0.01\) (comparable to the mean level spacing), the two closures perform comparably; (iii) at large \(\eta\), the SCBA closure regains the advantage. This \(\eta\)-dependent crossover confirms the theoretical picture developed in Sec.~\ref{RSCE}--\ref{sec:cross_expansion}: the continued-fraction hierarchy encodes fine-grained spectral correlations that become visible only below the Thouless energy scale, while above that scale the system is effectively described by a coarse-grained mean-field (SCBA) dynamics.

\subsection{Multi-resolvent signatures: skewness and non-Lorentzian tails}

Beyond the single-resolvent tests above, the full theoretical framework predicts specific signatures of multi-resolvent correlations that lie outside the descriptive reach of both SCBA and the continued-fraction closure. As established in Sec.~\ref{sec:TLB} and proved in Appendix~\ref{app:no_skewness_mf}, neither SCBA nor any single-resolvent closure can generate spectral skewness, because their self-energy functionals are strictly parity-preserving under energy reflection. Skewness arises uniquely from the multi-resolvent hierarchy, whose leading contribution \(\mathcal{G}^{(3)}_{\mu i}\) (Eq.~\eqref{eq:G3}) contains products of resolvents that mix parity sectors via Hilbert-transform convolutions.

The exact-diagonalization data of Ref.~\cite{HC25}---obtained on the same Ising model with identical parameters---provide direct evidence for both of these beyond-mean-field phenomena. The smoothed spectral functions \(f^{\mu i}\) extracted from the exact eigenstate overlaps display (i)~clear skewness, i.e., an asymmetry between the low- and high-energy wings of the distribution, and (ii)~Gaussian rather than Lorentzian decay in the far tails. Both features are qualitatively consistent with the theoretical predictions of our framework. Specifically, the spectral asymmetry is the direct observable consequence of the odd-parity component generated by \(\mathcal{G}^{(3)}_{\mu i}\) (see Sec.~\ref{sec:parity_decomp} for the explicit parity decomposition), while the Gaussian tail reflects the cumulative effect of the recursive continued-fraction backbone combined with the energy dependence of the effective interaction kernel (see Sec.~\ref{SLGA}--\ref{HAS}).

We note that, for the specific interaction structure \(V = -\sigma_z^S \otimes (\sigma_z^1+\sigma_z^{14})\) employed here, the leading multi-resolvent term \(\mathcal{G}^{(3)}_{\mu i}\) is numerically small due to the restricted connectivity of the coupling matrix; a quantitatively visible skewness signal may therefore require contributions up to \(\mathcal{G}^{(4)}_{\mu i}\) (products of three resolvents) or higher. 
A partial test of the hierarchy is nevertheless possible without solving the fully self-consistent equations~\eqref{DVC_final}:
one takes the asymmetric LG spectral functions
\(f^{\mu i}\) fitted to the exact eigenstate overlaps and substitutes them
directly into the right-hand sides of the SCBA and multi-resolvent closures.
This tests whether the \emph{functional form} of the multi-resolvent self-energy
improves the closure relation even when the input parameters are not themselves self-consistently determined. The result is shown in
\cref{FIG4}.

\begin{figure*}
    \centering
    \subfigure[]{\includegraphics[width=0.48\textwidth]{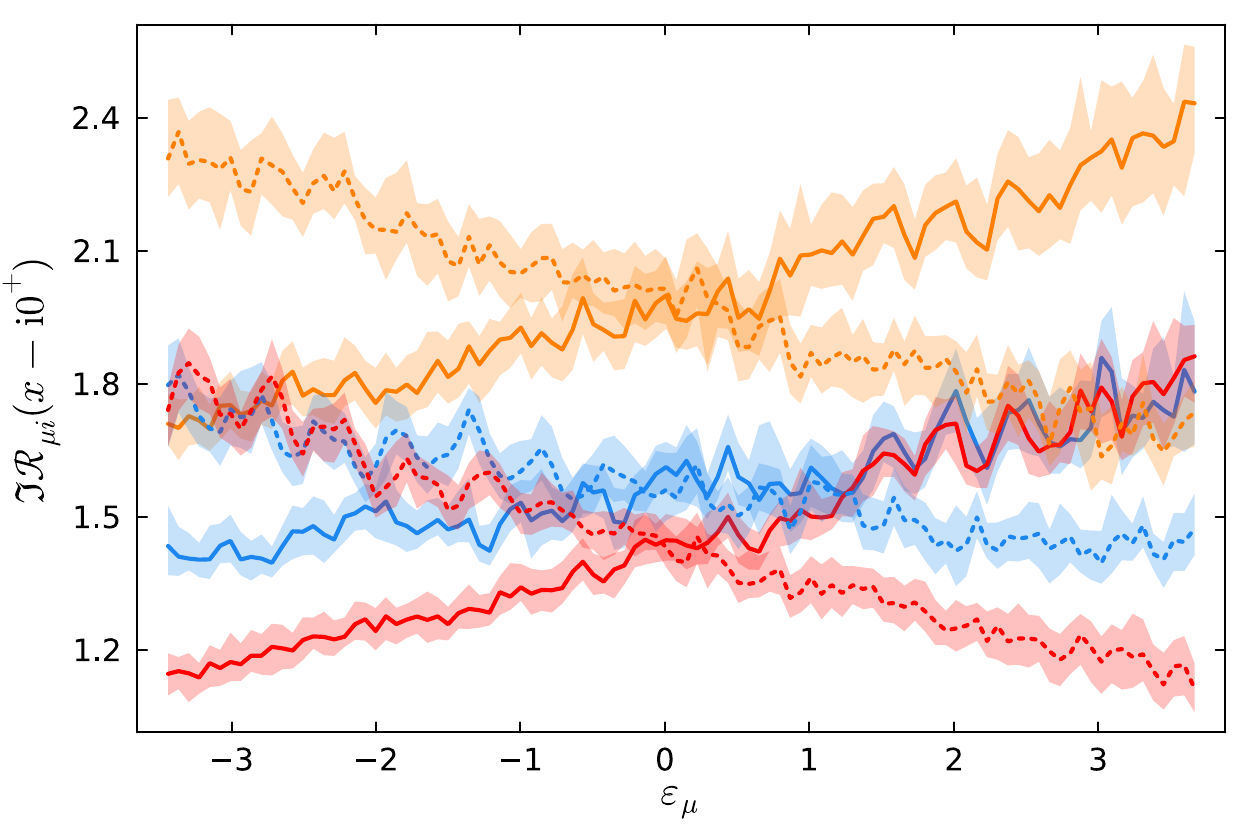}}
    \subfigure[]{\includegraphics[width=0.48\textwidth]{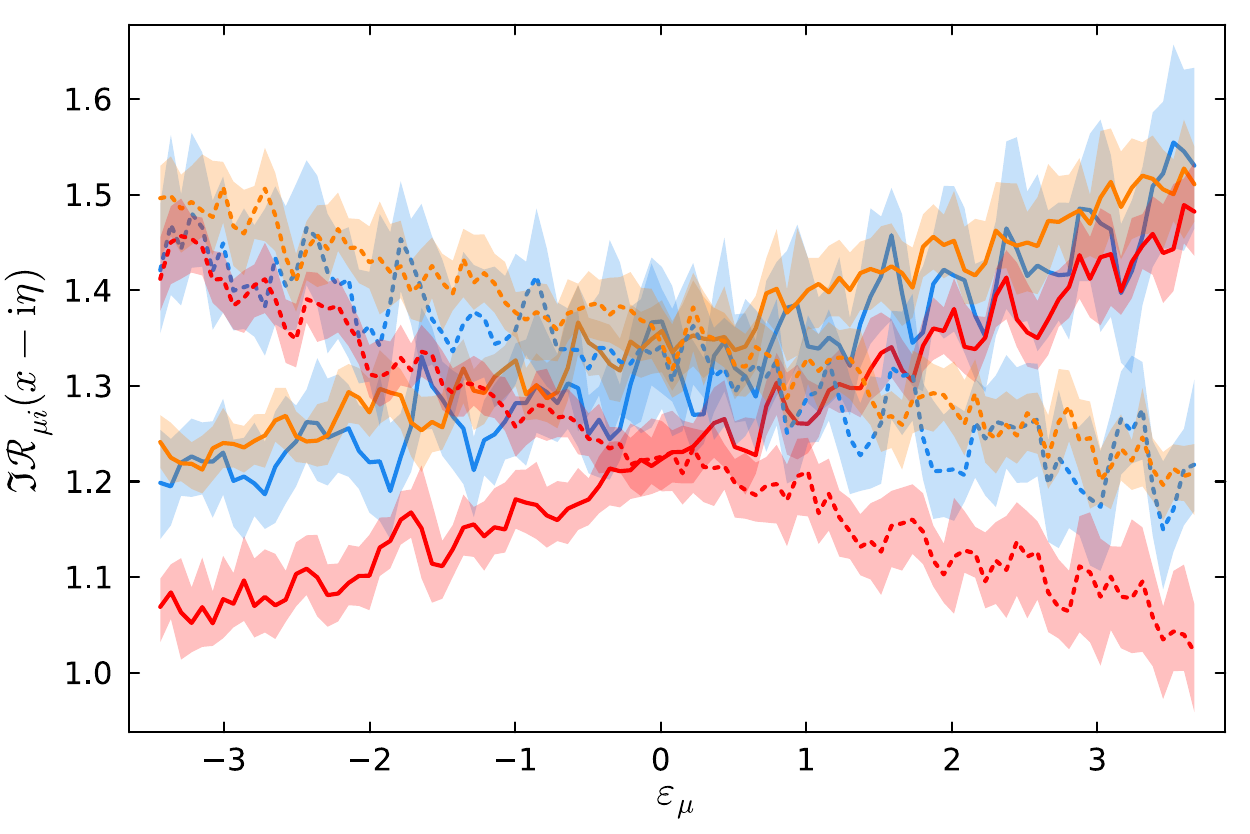}}
    \caption{
        Self-consistency tests of the SCBA and continued-fraction (LCF) closures.
        For each bath state index \(\mu\) and system branch \(i\), the imaginary part of the resolvent is evaluated at the fitted peak position \(x = a_{\mu i} + \Delta_{\mu i}\).
        Data are energy-binned over different \(\epsilon_\mu\); the  lines indicate the median value in each bin, while the shaded bands span the interquartile range (\(25\%\)--\(75\%\)).
        Dashed (solid) lines correspond to the system branch \(i=1\) (\(i=2\)).
        (a)~\(\eta\to0\) limit. Blue: direct evaluation \(\Im\mathcal{R}_{\mu i} = 1/\chi_{\mu i}\) from the Lorentzian fit parameters.
        Orange: right-hand side of the SCBA equation \eqref{Dyson} evaluated with the same fitted parameters.
        Red: right-hand side of the continued-fraction closure \eqref{CFC2} evaluated with the fitted parameters.
        The LCF closure is systematically closer to the direct value, confirming its superior self-consistency at high resolution.
        (b)~Finite-\(\eta\) test with \(\eta \approx 0.1228\) (\(100\times\) the mean level spacing).
        Blue: \(\Im\mathcal{R}_{\mu i}\) obtained directly from the exact probability distribution via \eqref{PEEX}.
        Orange/Red: right-hand sides of \eqref{Dyson} and \eqref{CFC2} evaluated with the directly computed \(\mathcal{R}_{\mu i}\).
        At this coarse resolution the SCBA closure regains the advantage, illustrating the scale-dependent crossover predicted by the theory.
    }
    \label{FIGs2}
\end{figure*}

\begin{figure}
    \centering
    \subfigure[]{\includegraphics[width=0.48\textwidth]{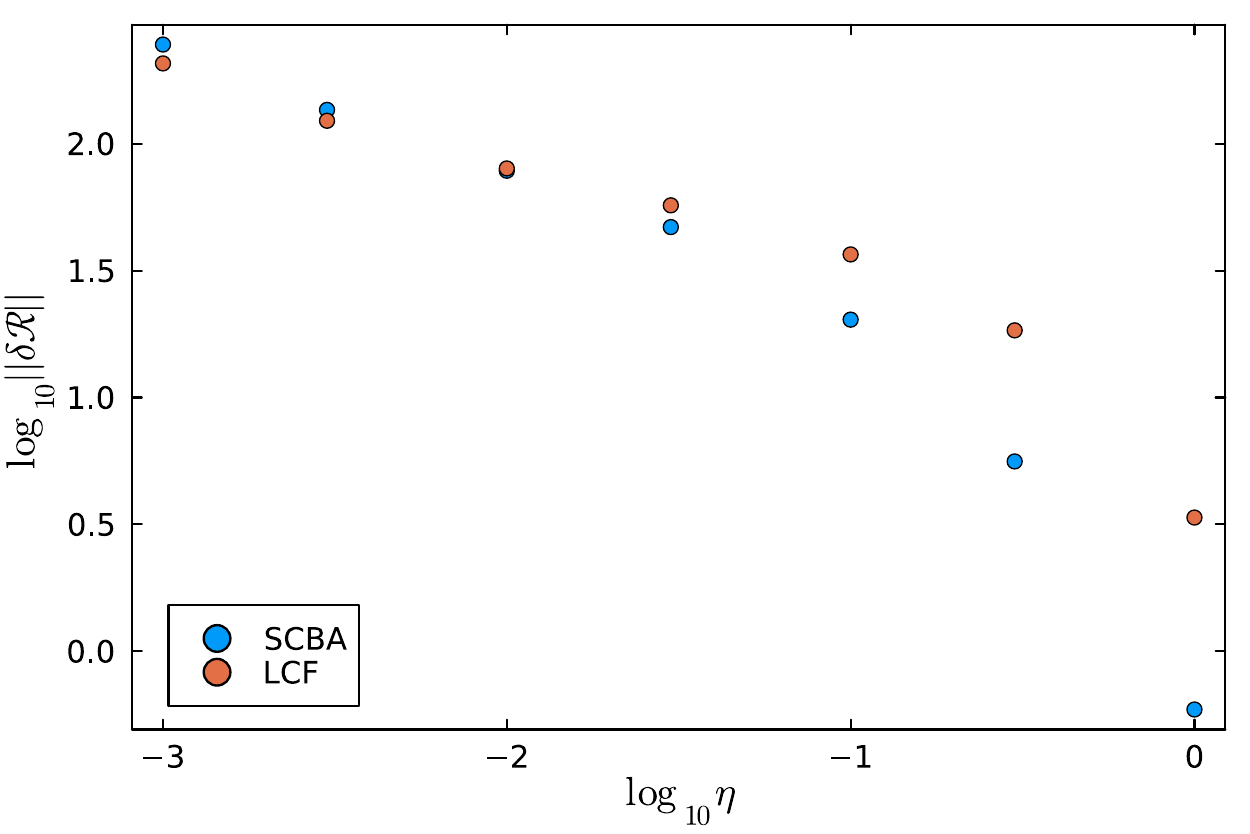}}
    \caption{
        Scale-dependent self-consistency error as a function of the coarse-graining parameter \(\eta\).
        For each \(\eta\), the deviation \(\delta\mathcal{R}_{\mu i}\) between the left- and right-hand sides of \eqref{Dyson} (SCBA, orange) and \eqref{CFC2} (LCF, red) is computed by direct substitution of the exact \(\mathcal{R}_{\mu i}\) from \eqref{PEEX}.
        The plotted quantity is the 2-norm of \(\delta\mathcal{R}_{\mu i}\) averaged over all bath states \(\mu\) and both system branches \(i=1,2\) within the energy window of \cref{FIGs2}.
        Three regimes are visible: (i)~\(\eta\to0\): LCF more accurate; (ii)~\(\eta\approx 0.01\): comparable performance; (iii)~large \(\eta\): SCBA more accurate.
        This crossover confirms that the continued-fraction hierarchy encodes fine-grained correlations resolvable only below the Thouless scale, while coarse-grained dynamics reverts to an effective mean-field (SCBA) description.
    }
    \label{FIG3}
\end{figure}

\begin{figure*}
    \centering
    \subfigure[]{\includegraphics[width=0.48\textwidth]{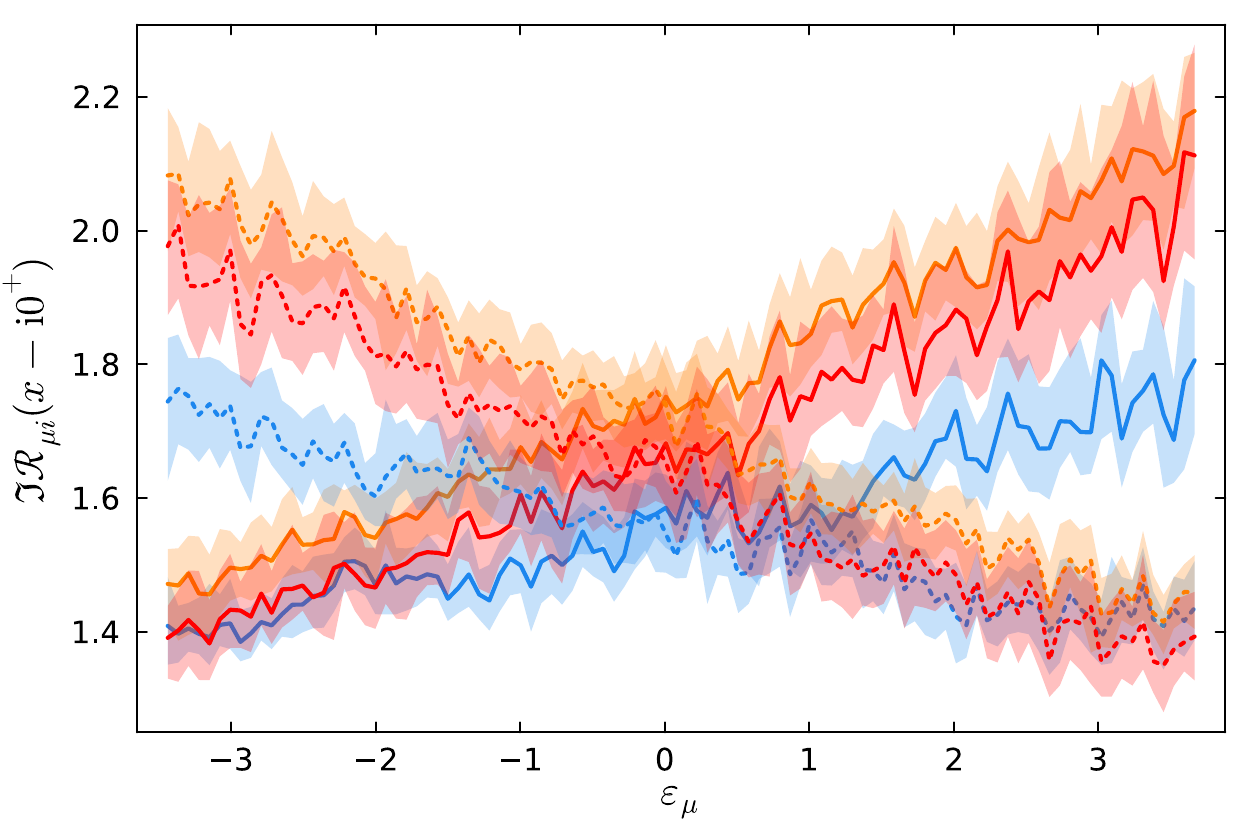}}
     \subfigure[]{\includegraphics[width=0.48\textwidth]{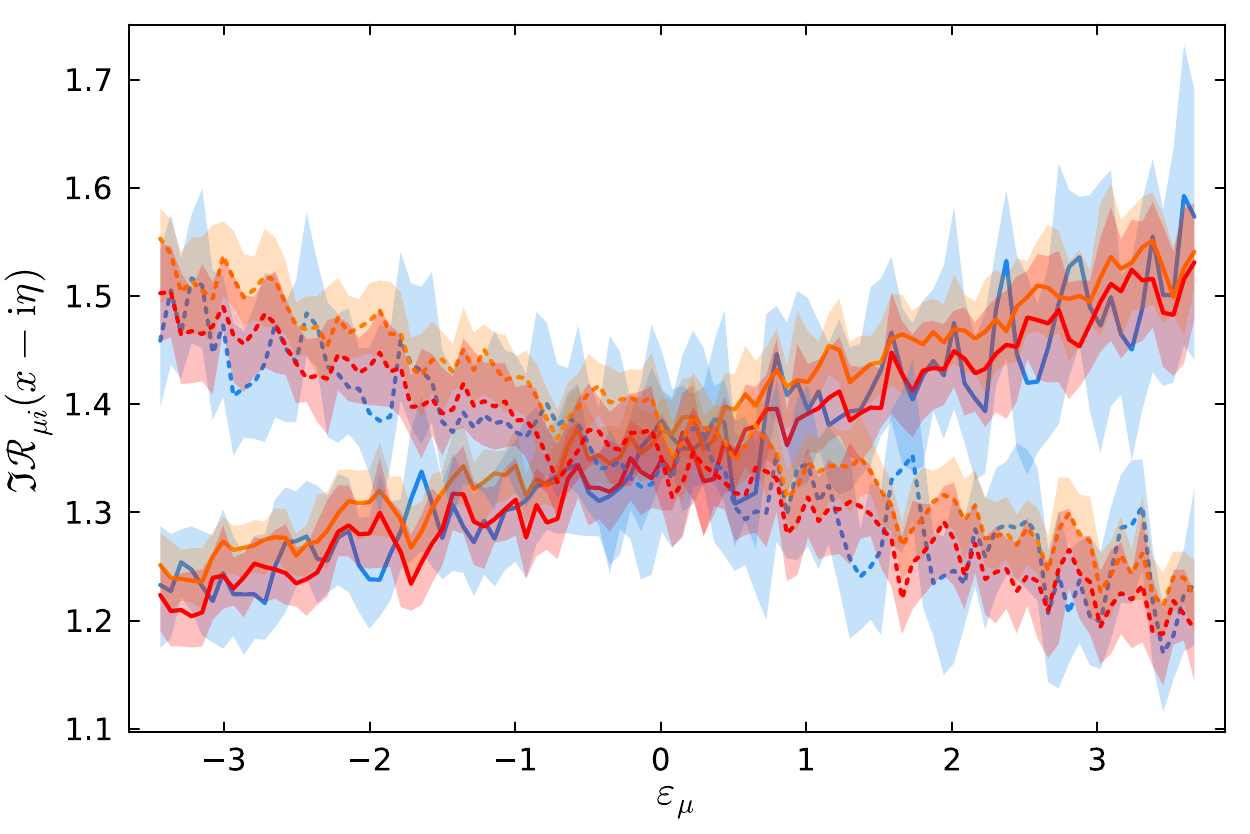}}
    \caption{
        Self-consistency test of the multi-resolvent hierarchy expansion.
        The plotting conventions (binning, median, interquartile bands, branch labeling) follow those of \cref{FIGs2}.
        For each $(\mu,i)$, the imaginary part of the resolvent is evaluated at the peak position
        $x = \lambda_p$ (cf.\ \cref{eq:voigt_peak_app}).
        (a)~$\eta\to0$ limit.  The calculation employs the asymmetric LG ansatz of \cref{eq:LGdistr_final}, with fitted parameters
        $(\Delta_{\mu i}, \Delta'_{\mu i}, \chi_{\mu i}, \sigma_{\mu i})$ obtained from the exact-diagonalization.  Blue: direct evaluation $\Im\mathcal{R}_{\mu i}(x)$ from the fitted asymmetric LG profile.
        Orange: right-hand side of the SCBA equation \eqref{Dyson} evaluated with the same fitted parameters.
        Red: right-hand side of the exact resolvent equation \eqref{eq:CSEQ} with the self-energy expanded
        up to the fourth-order diagonal-closure term $\mathcal{G}^{(4),\text{D}}_{\mu i}(z)$
        (see \cref{app:fourth_order}) and evaluated with the fitted parameters.
        The multi-resolvent expansion (red) lies systematically closer to the direct value (blue) than the
        SCBA prediction (orange), confirming that products of diagonal resolvents---the leading
        beyond-mean-field structure of the hierarchy---improve self-consistency already at
        the fourth-order level.
        (b)~Finite-$\eta$ test with $\eta \approx 0.1228$ ($100\times$ the mean level spacing).
        All quantities are evaluated using the directly computed $\mathcal{R}_{\mu i}$ from the exact
        probability distribution via \eqref{PEEX}. Blue: \(\Im\mathcal{R}_{\mu i}\) obtained directly from the exact probability distribution.
        Orange/Red: right-hand sides of \eqref{Dyson} and \eqref{eq:CSEQ} with the expanded
        up to $\mathcal{G}^{(4),\text{D}}$ 
        At this coarse resolution the multi-resolvent and SCBA closures become nearly indistinguishable
        from the direct value.
    }
    \label{FIG4}
\end{figure*}

\begin{figure}
    \centering
    \subfigure[]{\includegraphics[width=0.48\textwidth]{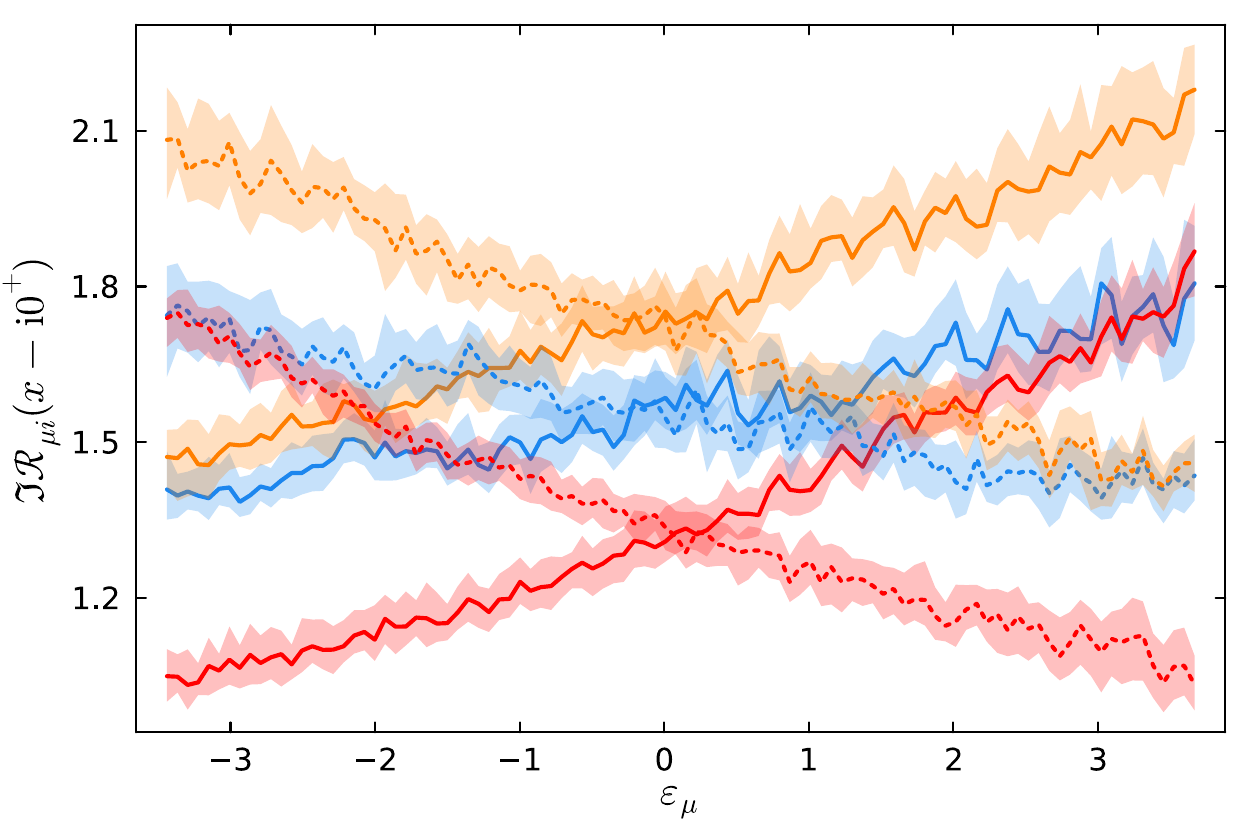}}
    \caption{
        Self-consistency tests of the SCBA and continued-fraction (LCF) closures using the asymmetric LG ansatz.  The plotting conventions (binning, median, interquartile bands, branch labeling) follow those of \cref{FIGs2}. For each $(\mu, i)$, the imaginary part of the resolvent is evaluated at the peak position $x = \lambda_p$ (cf. \cref{eq:voigt_peak_app}) with $\eta \to 0$.  The calculation employs the asymmetric LG ansatz of eq. (73), with fitted parameters $(\Delta_{\mu i}, \Delta'_{\mu i}, \chi_{\mu i}, \sigma_{\mu i})$ obtained from the exact-diagonalization data, identical to \cref{FIG4}(a). Blue: direct evaluation $\Im \mathcal{R}_{\mu i}(x)$ from the fitted asymmetric LG profile. Orange: right-hand side of the SCBA equation \eqref{Dyson} evaluated with the same fitted parameters. Red: right-hand side of the continued-fraction closure \eqref{CFC2} evaluated with the fitted parameters. 
    }
    \label{FIG5}
\end{figure}

Taken together, \cref{FIGs2,FIG3,FIG4} trace the progressive enrichment of the
self-consistency structure predicted by the theory: from SCBA (single resolvent,
linear functional), through the continued-fraction backbone (single resolvent,
nonlinear functional), to the multi-resolvent hierarchy (products of resolvents,
parity-mixing). At each level the closure relation moves systematically closer
to the exact diagonalization data. \Cref{FIGs2} establishes the superiority of the
continued-fraction closure over SCBA at high resolution and the scale-dependent
crossover; \cref{FIG3} quantifies this crossover as a function of \(\eta\); and
\cref{FIG4} demonstrates that multi-resolvent products---the signature of the
hierarchy beyond any single-resolvent description---provide an additional,
independent improvement to self-consistency, corroborating the hierarchical
organization of correlations developed in \cref{SME}.

\subsection{Decoupling Ansatz and Closure Limitations}

To systematically isolate the effects of the chosen spectral ansatz 
from the intrinsic capabilities of the closure methods, 
Fig.~\ref{FIG5} evaluates both the SCBA and continued-fraction (LCF) 
closures using the same asymmetric Lorentzian--Gaussian (LG) 
parameters employed in Fig.~\ref{FIG4}(a).
This direct comparison reveals a nuanced physical picture regarding 
the origins of the self-consistency error observed in Figs.~\ref{FIGs2}--\ref{FIG4}.

The most immediate observation from Fig.~\ref{FIG5} is the substantial 
improvement of the SCBA prediction compared to its performance under 
the pure Lorentzian ansatz in Fig.~\ref{FIGs2}(a).
The Lorentzian profile, whose cumulative distribution involves the 
standard $\arctan$ integral, inherently lacks the exponential decay 
required in the far spectral wings.
The markedly improved agreement of the SCBA closure under the 
asymmetric LG ansatz therefore indicates that a significant fraction 
of the discrepancy observed in Fig.~\ref{FIGs2}(a) originates from 
the limitations of the Lorentzian ansatz itself, rather than solely 
from the mean-field structure of the SCBA closure equation.
This constitutes a clean decoupling of the \emph{ansatz error} 
(choice of spectral profile) from the \emph{closure error} 
(approximations inherent in the self-consistency functional).

Interestingly, once the skewed spectral profile is explicitly 
incorporated through the asymmetric LG ansatz, the comparative 
advantage of the LCF recursion---which was prominent in 
Fig.~\ref{FIGs2}(a)---becomes significantly less pronounced.
In Fig.~\ref{FIG5}, the LCF closure exhibits only a modest deviation 
from the SCBA result.
This phenomenological observation strongly supports the theoretical 
division of labor proposed in Sec.~\ref{sec:TLB}:
the continued-fraction backbone and the multi-resolvent hierarchy 
represent orthogonal physical corrections.
The LCF primarily reorganizes local broadening effects and intrinsic 
spectral scales through one-dimensional recursion; its self-energy 
functional, like that of SCBA, remains parity-preserving under 
energy reflection (Appendix~\ref{app:no_skewness_mf}).
Consequently, even when the asymmetric parameters $(\Delta', \sigma)$ 
are supplied by the LG ansatz, the LCF closure cannot 
\emph{actively generate or sustain} spectral skewness through its 
self-consistency cycle---it can only passively accommodate the 
input asymmetry.
Spectral skewness, as formulated in Sec.~\ref{sec:cross_expansion}, is 
fundamentally driven by non-local multi-resolvent path correlations 
via products of the form $R(z_1) R(z_2)$, whose Hilbert-transform 
structure intrinsically mixes parity sectors.
This mechanism lies strictly outside the scope of any single-resolvent 
closure, whether SCBA or continued-fraction.

A further observation emerges when the Fig.~\ref{FIG5} data are 
evaluated against the finite-resolution benchmark established in 
Sec.~\ref{fiet}.
The continued-fraction prediction, although formally evaluated in 
the $\eta \to 0$ limit, systematically aligns closer with the 
finite-$\eta$ direct evaluation than with the microscopic 
$\eta \to 0$ target.
This suggests that the non-perturbative recursive structure of the 
LCF effectively acts as an \emph{intrinsic coarse-graining mechanism}:
the depth truncation of the continued fraction introduces an 
effective resolution scale, and the LCF natively generates a smoothed 
profile characteristic of the experimentally relevant, 
finite-resolution regime.
Rather than targeting the exact microscopic limit, the LCF backbone 
produces a spectral function that is already coarse-grained by its 
own recursive structure.
This observation provides a natural physical explanation for the 
scale-dependent crossover quantified in Fig.~\ref{FIG3}, and 
establishes a conceptual complementarity:
the LCF backbone supplies a \emph{built-in macroscopic cutoff}, 
while the multi-resolvent hierarchy supplies the 
\emph{microscopic interference effects} (skewness, tail refinement) 
that distinguish the true $\eta \to 0$ limit.

\section{Conclusion}\label{CON}
We have presented a self-contained methodological framework for analyzing the global properties of nonintegrable many-body systems. The framework rests on three pillars: (1) the resolvent's global analytic structure via pole expansion; (2) the statistical treatment of local fluctuations through the eigenstate thermalization hypothesis (ETH); and (3) a systematic recursive hierarchy of cross-correlated terms that generates higher-order corrections controlling tails, branch splitting, and fluctuations. The mean-field level provides a closed description as a projection of the full resolvent hierarchy onto the single-resolvent sector, yielding self-consistent equations for Lorentzian parameters. The recursive hierarchy, in turn, encodes correlated multi-resolvent fluctuations that systematically correct the mean-field picture, controlling tails, branch splitting, and higher-order correlations.

The analytic content of the theory can be organized as a progressive enrichment of singularity structure: SCBA (single pole) $\to$ continued-fraction recursion (pole renormalization, intrinsic non-Lorentzian scales) $\to$ multi-resolvent hierarchy (pole-pair interference, parity mixing, skewness) $\to$ Faddeeva closure (distributed pole cloud). Each level introduces a qualitatively new capability absent from all preceding levels, and the full framework unifies them within a single, systematically improvable hierarchy closed in terms of diagonal resolvents.

A hierarchical ansatz strategy: Lorentzian for the bulk, Gaussian for the tails, and a hybrid LG (Voigt) form for a unified description and translates these equations into tractable forms. Numerical validation on a nonintegrable Ising model shows reasonable agreement with the theoretical predictions, including the emergence of asymmetric line shapes and Gaussian tail decay, even at the lowest mean-field level \cite{HC25}.

Unlike conventional SCBA and other finite-diagrammatic resummations, our framework does not rely on a small expansion parameter. The explicit inclusion of multi-resolvent products (e.g., $\mathcal{G}^{(3)}\sim\mathcal{R}\mathcal{R}$) generates nonlocal frequency structures that are absent in single-resolvent approximations. These structures are responsible for non-Lorentzian spectral tails and skewness, as shown by the analytic properties of the Hilbert transform. Moreover, the Lanczos (Krylov) representation provides an equivalent non-perturbative reformulation: the continued-fraction form $R_n(z)=1/(z-a_n-b_{n+1}^2R_{n+1}(z))$ compresses all many-body correlations into scalar sequences $\{a_n,b_n\}$ and offers a complementary route to closure via truncation and self-similarity. The two formulations—resolvent hierarchy and continued fraction—are exact and complementary: the former exposes multi-resolvent correlations, the latter provides a non-perturbative backbone.

The validity of the methodology requires three conditions typical of generic nonintegrable systems: nonintegrability (chaotic dynamics), random-phase condition (sufficiently random off-diagonal matrix elements), and an exponential density of states (justifying coarse-graining). Under these conditions, the framework directly connects microscopic interactions to global statistical properties---including distribution tails and branch structures not captured by standard truncation methods---and provides quantitative predictions for key thermodynamic quantities, such as entropy, in the challenging regime of strong interactions and dense spectra.

Several directions emerge naturally from the analytic structures developed in this work.

\noindent\textit{(i) Singularity diffusion.}
The Faddeeva self-energy reveals that the resolvent's effective singularity structure is not an isolated pole but a Gaussian-weighted distribution of pole contributions. This raises the question of how the multi-resolvent hierarchy drives the transition from discrete poles toward distributed, branch-cut-like collective structures in the limit of dense spectra. A quantitative theory of this diffusion process that defining a pole density function and studying its flow under successive hierarchy levels, would connect the algebraic recursion of Sec.~\ref{sec:cross_expansion} to the emergent irreversibility encoded in continuous spectra.

\noindent\textit{(ii) Time-domain signatures.}
Although formulated in the frequency domain, the non-Lorentzian spectral structures produced by the hierarchy carry implications for real-time dynamics. Pure Lorentzian (SCBA) spectra are associated with exponentially decaying relaxation, whereas the Gaussian and Faddeeva profiles indicate departures from purely exponential decay, suggesting the emergence of memory effects. Whether the spectral skewness parameter $\delta s_{\mu i}$ can be rigorously connected to established measures of non-Markovianity remains an open question that would benefit from a systematic time-domain reformulation of the resolvent hierarchy.

\noindent\textit{(iii) Renormalization-group structure.}
The scale-dependent crossover demonstrated in \cref{FIG3}, where the continued-fraction closure is required at high resolution while SCBA suffices at coarse resolution, which  suggests that the resolvent hierarchy may admit an RG-like interpretation. The scale-separation condition $\delta_{\rm LS}\ll\eta\ll\Gamma_{\mu i}$ (\cref{eq38}) structurally resembles an RG shell decomposition. Clarifying this connection could situate the hierarchy within the broader framework of effective field theory for many-body spectra.

\noindent\textit{(iv) Statistical organization of poles.}
The present framework applies ETH-type statistical reasoning to off-diagonal matrix elements. A natural extension is whether analogous reasoning can be extended to the statistical organization of the resolvent poles themselves: under what conditions do pole positions and residues obey universal statistical laws, and how does the multi-resolvent hierarchy encode deviations from factorization? The Faddeeva structure hints at a factorized form; higher-order hierarchy corrections would encode correlations between pole positions and widths.

It unifies the treatment of typical behavior and statistical fluctuations, offering a valuable tool for theoretical investigations of quantum thermalization and emergent statistical mechanics. The practical utility of this approach is realized through the hierarchical ansatz strategy, which provides explicit spectral shapes from the full hierarchy.

\begin{acknowledgments}
    This work is supported by the National Natural Science Foundation of China under Grant No. 12305035 and Innovational Fund of Hainan Province (Grant No. KJRC2023B11).
\end{acknowledgments}

\appendix
\section{Structural distinction from finite diagrammatic resummations}\label{SDDR}

In this appendix we clarify why the self-consistent resolvent hierarchy developed in the main text cannot be generated by any diagrammatic resummation based on a finite diagrammatic class. Instead of a purely combinatorial counting argument, we compare the explicit algebraic structures and the rules by which they are generated.

\subsection{Representative finite diagrammatic resummations}

We take as the prototypical example the SCBA, which belongs to the class of finite diagrammatic resummations. In SCBA, the full Green's function (resolvent) satisfying the Dyson equation \cref{Dyson}.
The diagrammatic content of SCBA is well known: it sums all non-crossing diagrams (the “rainbow” series). More generally, a finite diagrammatic class is defined by a finite set of skeleton graphs (e.g., ladder, parquet) into which propagator and vertex insertions can be made. All such resummations yield self-energies that are \emph{additive} in the sense that the right-hand side of (\ref{Dyson}) contains only \emph{single} products of a vertex factor and a Green's function. Higher-order correlations appear only through the self-consistency of $G$, but the fundamental building blocks remain single insertions into a fixed set of topologies.

\subsection{Resolvent hierarchy of the present work}

In contrast, the self-consistent equations of the main text take the form
\begin{equation}
\mathcal{R}_{\mu i}(z) = \frac{1}{z - a_{\mu i} - V_{\mu i} - \mathcal{G}_{\mu i}(z)},
\label{eq:res_main}
\end{equation}
with $\mathcal{G}_{\mu i}(z) = \mathcal{G}_{\mu i}^{\rm OD}(z)+\Delta\mathcal{G}_{\mu i}(z) + \mathcal{G}_{\mu i}^{\rm CC}(z)$.
The off-diagonal (mean-field) term
\begin{equation}
\mathcal{G}_{\mu i}^{\rm OD}(z) = \sum_{\nu j \neq \mu i} |V_{\mu i,\nu j}|^2 \, \mathcal{R}_{\nu j}(z)
\label{eq:od}
\end{equation}
is structurally analogous to the SCBA self-energy. The crucial new ingredient is the cross-correlated term, whose leading contribution is
\begin{equation}
\mathcal{G}_{\mu i}^{(3)}(z) = \sum_{\xi k \neq \nu j \neq \mu i} V^{(3)}_{\mu i,\xi k,\nu j} \;  \mathcal{R}^{(\mu i)}_{\nu j}(z) \mathcal{R}^{(\mu i,\nu j)}_{\xi k}(z).
\label{eq:cc}
\end{equation}
This term contains a \emph{product of two full resolvents}, not just a single resolvent multiplied by a vertex factor. Moreover, the full hierarchy includes higher-order terms such as
\begin{equation}
\mathcal{G}_{\alpha}^{(4)}(z) \sim \sum_{\dots} V^4 \;  \mathcal{R}^{(\alpha)}_{\beta}(z) \mathcal{R}^{(\alpha, \beta)}_{\gamma}(z)\mathcal{R}^{(\alpha, \beta,\gamma)}_{\xi}(z),
\end{equation}
and so on. Each $\mathcal{G}^{(n)}_{\mu i}(z)$ is a finite sum of products of $m_n$ resolvents, with $m_n \ge 2$ for at least one term.

\subsection{Generation rules and structural difference}

The difference between the two frameworks is not merely the presence of higher-order terms in $V$; it lies in the \emph{generation rules}:

\begin{itemize}
\item In finite diagrammatic resummations (SCBA, ladder, parquet), all diagrams are built by inserting vertices and propagators into a fixed finite set of skeleton graphs. Consequently, the self-energy at any order can be expressed as a sum over diagrams each of which contains \emph{at most one} full propagator line connecting two vertex insertions in a topologically prescribed way. The algebraic structure is additive: $\Sigma = \sum_{\beta} |V|^2 \, G_\beta$ (or similar forms with more vertices but still a single chain of propagators).
\item In the present resolvent hierarchy, the cross-correlated terms $\mathcal{G}^{(n)}_{\mu i}(z)$ contain \emph{products of two or more  resolvents} already at the level of the bare algebraic expression, before any iteration. When these terms are inserted back into the definition of $\mathcal{R}_{\mu i}(z)$ via (\ref{eq:res_main}), the recursive substitution generates nested products of resolvents. For instance, substituting $\mathcal{R}_{\xi k}(z)$ from (\ref{eq:res_main}) into the product $\mathcal{R}_{\xi k}(z) \mathcal{R}_{\mu i}(z)$ in (\ref{eq:cc}) produces terms like
\begin{equation}
\mathcal{R}_{\xi k}  \mathcal{R}_{\nu j}\;\longrightarrow\; \frac{1}{z - a_{\xi k} - V_{\xi k} - \mathcal{G}_{\xi k}(z)} \;  \mathcal{R}_{\nu j},
\end{equation}
and $\mathcal{G}_{\xi k}(z)$ itself contains further products of resolvents. This leads to a \emph{recursive branching} structure: each resolvent factor can expand into a product of two or more resolvents, creating a hierarchy of nested products.
\end{itemize}

\subsection{Why finite diagrammatic classes cannot reproduce this structure}

A finite diagrammatic class is defined by a finite set of skeleton topologies. In such a class, all algebraic expressions are generated by inserting propagators and vertices into these skeletons. Crucially, the number of resolvent factors appearing in any term is bounded by the number of vertices in the skeleton plus the number of insertions, but the \emph{structure of nested products} is constrained by the topology of the skeleton.

In the resolvent hierarchy, however, the nested products arise from the recursive substitution of $\mathcal{R}$ into itself, without any pre-specified skeleton. At each recursion level, the product of two resolvents can generate a product of three or more resolvents, and this process continues indefinitely. This generates a family of algebraic structures that cannot be captured by any finite set of skeleton graphs, because the depth of nesting is not bounded by any fixed skeleton topology.

One might attempt to mimic the hierarchy by including an infinite set of skeleton graphs (e.g., all possible trees). However, such a construction would no longer constitute a \emph{finite} diagrammatic class. Moreover, it would lack the systematic closure property that is central to the present framework: the resolvent hierarchy is \emph{closed} in terms of the diagonal resolvents $\mathcal{R}_{\mu i}(z)$ themselves, whereas an infinite-skeleton expansion would involve an ever-growing set of auxiliary quantities.

\subsection{Physical implications}

This structural distinction has direct physical consequences:
\begin{itemize}
\item The product structure $\mathcal{R}_{\xi k}  \mathcal{R}_{\mu i}$ in $\mathcal{G}_{\mu i}^{(3),\text{D}}(z)$ provides, via the spectral representation, a natural mechanism for generating distribution tails and branch splitting. In finite diagrammatic resummations, such tails would require summing an infinite series of diagrams (e.g., high-order vertex corrections) to achieve comparable accuracy.
\item The recursive hierarchy provides a natural way to incorporate fluctuations beyond mean field while maintaining closure. In contrast, finite diagrammatic classes typically require a truncation at some diagrammatic order or the introduction of additional approximations (e.g., vertex approximations) to close the equations.
\end{itemize}

Therefore, the resolvent hierarchy is not merely an alternative way of organizing the same perturbative series; it constitutes a distinct theoretical structure that cannot be generated by any diagrammatic resummation based on a finite diagrammatic class.

\subsection{Summary}

We have shown that the self-consistent resolvent hierarchy of the main text differs fundamentally from finite-diagrammatic-class resummations (SCBA, ladder, parquet, etc.) in two respects:
\begin{enumerate}
\item It contains explicit products of multiple full resolvents already in the bare expressions for the self-energy-like terms $\mathcal{G}_{\mu i}(z)$.
\item The recursive substitution of $\mathcal{R}$ into these products generates nested algebraic structures that are not bounded by any finite set of skeleton topologies.
\end{enumerate}
These features are essential for the framework's ability to capture tail distributions, branch splitting, and higher-order fluctuations without invoking a proliferation of diagrammatic structures. Consequently, the methodology presented in this work represents a genuine departure from conventional diagrammatic approaches.

\subsection{Spectral consequences of multi-resolvent structure}\label{SCMS}
To make the physical distinction more concrete, we examine how the multi-resolvent structure affects spectral properties, in particular the behavior of spectral tails.

\paragraph{Spectral representation.}
According to \cref{PEEX,SPIR}, for a diagonal resolvent \(\mathcal{R}_{\mu i}(z)\), we have
\begin{align}
 p_{\mu i}(\omega) &= \frac{1}{\pi e^{S(\omega)}}\,\Im\,\mathcal{R}_{\mu i}(\omega -  \mathrm{i}0^+),\\
 \mathcal{R}_{\mu i}(z) &= \int d\omega\; e^{S(\omega)}\frac{ p_{\mu i}(\omega)}{z - \omega}.
\end{align}

\paragraph{Single-resolvent structure (SCBA-type).}
In self-consistent Born-type approximations, the self-energy depends linearly on a single resolvent:
\begin{equation}
\mathcal{G}_{\mu i}(z) = \sum_{\nu j\neq \mu i} |V_{\mu i,\nu j}|^2\,\mathcal{R}_{\nu j}(z).
\end{equation}
Taking the imaginary part gives
\begin{equation}
\Im\,\mathcal{G}_{\mu i}(\omega) = \sum_{\nu j\neq \mu i} |V_{\mu i,\nu j}|^2\,\pi e^{S(\omega)} p_{\nu j}(\omega).
\end{equation}
Thus the self-energy at frequency \(\omega\) is determined by the spectral densities of all states \(\nu j\) at the \emph{same} frequency \(\omega\), weighted by the squared matrix elements. In typical situations where \(|V_{\mu i,\nu j}|^2\) decays with energy difference, the dominant contribution comes from states with \(\epsilon_\nu \approx \epsilon_\mu\). Under the ETH assumption of self-averaging, the \( p_{\nu j}\) for such states are approximately equal, leading to an effectively local relation in frequency. Nevertheless, the structure remains that of a single resolvent factor.

\paragraph{Multi-resolvent structure.}
In contrast, the present hierarchy contains higher-order terms that are products of resolvents, e.g. from \cref{eq:G3}, we have
\begin{equation}
\mathcal{G}_{\mu i}^{(3),\text{D}}(z) = \sum_{\xi k\neq\nu j\neq\mu i} V^{(3)}_{\mu i,\xi k,\nu j}\;
\mathcal{R}_{\xi k}(z)\, \mathcal{R}_{\nu j}(z).
\end{equation}
To analyze the imaginary part on the real axis, we follow \cref{TTG3}, which gives
\begin{align}
	\Im\big[\mathcal{R}_{\xi k} \mathcal{R}_{\nu j}\big](\omega) = \pi^2\Big[ e^{S(\omega)} p_{\nu j}(\omega)\,\mathcal{H}[e^{S} p_{\xi k}](\omega)\notag\\
	 + e^{S(\omega)}p_{\xi k}(\omega)\,\mathcal{H}[e^{S} p_{\nu j}](\omega)\Big],
\end{align}
where \(\mathcal{H}[f](\omega)=\frac{1}{\pi}\dashint d\omega'\,f(\omega')/(\omega-\omega')\) is the Hilbert transform. 

\paragraph{Nonlocal frequency structure.}
The key feature of the above expression is the \emph{difference} between two terms, each containing a Hilbert transform. This structure is fundamentally different from the simple additive form that would arise from a naive product of boundary values. The minus sign originates from the discrete spectral sum over distinct poles and reflects interference between different intermediate states. When this expression is inserted into \(\mathcal{G}_{\mu i}^{(3)}\), the resulting contribution to the self-energy becomes nonlocal in frequency: it involves Hilbert transforms that couple different spectral regions. Such nonlocality is absent in single-resolvent (SCBA-type) approximations.

\paragraph{Implications for spectral tails.}
This convolution-type structure implies that the self-energy at frequency \(\omega\) receives contributions from a broad range of frequencies \(\omega'\) via the Hilbert transform. As a result, the spectral function is no longer governed by a purely local self-consistent equation; its asymptotic behavior is controlled by integral transforms of \(e^{S(\omega)} p(\omega)\) rather than its local value. Such nonlocal structures are known to produce non-Lorentzian line shapes in many contexts, depending on the detailed form of the interaction and spectral density.

\paragraph{Summary.}
The essential distinction is therefore not a specific power-law exponent, but the mechanism:
single-resolvent structures lead to frequency-local self-energies and Lorentzian-type broadening, whereas multi-resolvent structures generate nonlocal (convolution-type) contributions with a characteristic \emph{difference} structure, which can naturally give rise to non-Lorentzian spectral tails. The explicit construction of a minimal self-consistent model incorporating this nonlocality is presented in \cref{app:const_selfenergy}.

\section{Hilbert Transform of Enhanced Ansatz}
\label{HTEA}

The principal-value integral appearing in Eq.~\eqref{REGui_sym} can be evaluated by exploiting the analytic (causal) structure of the resolvent, as expressed in Eq.~\eqref{RkkR}. In particular, the Hilbert transform can be obtained from the boundary value of an analytic function in the lower half-plane.

To evaluate the principal-value integral in Eq.~\eqref{REGui_sym}, we write
\[
I(\lambda') = \operatorname{PV} \int_{-\infty}^{\infty} d\lambda \,
\frac{G(\lambda-\mu_1;\sigma)\,L(\lambda-\mu_2;\chi)}{  \lambda'-\lambda}.
\]
We exploit the resolvent representation of the Lorentzian,
\[
L(\lambda-\mu_2;\chi)
=
\frac{1}{\pi}\,\Im\frac{1}{\lambda-\mu_2 - \mathrm{i}\chi}.
\]
This allows us to rewrite the integral as
\[
I(\lambda')
=
\frac{1}{\pi}\,
\Im
\dashint_{-\infty}^{\infty}
d\lambda\;
\frac{G(\lambda-\mu_1;\sigma)}
{(\lambda-\mu_2 - \mathrm{i}\chi)(\lambda'-\lambda)}.
\]
The integrand can now be simplified using partial fraction decomposition:
\[
\frac{1}{(\lambda-a)(\lambda-b)}
=
\frac{1}{a-b}
\left(
\frac{1}{\lambda-a}
-
\frac{1}{\lambda-b}
\right),
\]
with \(a=\mu_2+\mathrm{i}\chi\) and \(b=\lambda'\). This yields
\[
I(\lambda')
=
\frac{1}{\pi}\,
\Im
\left[
\frac{J(\lambda') - J(\mu_2+\mathrm{i}\chi)}
{\lambda' - \mu_2 - \mathrm{i}\chi}
\right],
\]
where we have introduced
\[
J(z)
=
\dashint_{-\infty}^{\infty}
\frac{G(\lambda-\mu_1;\sigma)}{z-\lambda }\,d\lambda.
\]
The function \(J(z)\) is the Hilbert transform of a Gaussian and admits a closed form in terms of the Faddeeva function:
\[
J(z)
=-
\mathrm{i}\sqrt{\frac{\pi}{2}}\frac{1}{\sigma}\,
w\!\left(\frac{z-\mu_1}{\sqrt{2}\sigma}\right).
\]
Substituting this result, we obtain the compact expression
\[
I(\lambda')
=-
\Im\left[\mathrm{i}
\frac{
w\!\left(\dfrac{\lambda'-\mu_1}{\sqrt{2}\sigma}\right)
-
w\!\left(\dfrac{\mu_2+\mathrm{i}\chi-\mu_1}{\sqrt{2}\sigma}\right)
}
{(\lambda' - \mu_2 - \mathrm{i}\chi)\sqrt{2\pi}\sigma}
\right].
\]
This form makes explicit that the principal-value integral is governed by a difference of analytic continuations of the Gaussian resolvent. The second term, evaluated at the complex pole \(\mu_2+i\chi\), ensures the correct analytic structure and guarantees convergence.

For practical use, one may further expand the imaginary part. Writing
\[
\frac{\mathrm{i}}{\lambda' - \mu_2 - \mathrm{i}\chi}
= \frac{\mathrm{i}\Delta -\chi}{\Delta^2 + \chi^2},
\qquad
\Delta = \lambda' - \mu_2,
\]
the result can be expressed as
\begin{equation}\label{TTI}
	I(\lambda')=\frac{-\Delta\, \Re\!\bigl[w(z') - w(z_0)\bigr]+\chi\, \Im\!\bigl[w(z') - w(z_0)\bigr]}{(\Delta^2+\chi^2)\sqrt{2\pi}\sigma},
\end{equation}
where
\[
z' = \frac{\lambda'-\mu_1}{\sqrt{2}\sigma},
\qquad
z_0 = \frac{\mu_2-\mu_1+i\chi}{\sqrt{2}\sigma}.
\]
This representation is fully consistent with the causal structure of the resolvent and automatically satisfies the Kramers--Kronig relations. It provides a compact and well-defined expression for the Hilbert transform of the enhanced (Voigt-type) ansatz without invoking uncontrolled factorization approximations.

It is useful to make explicit the equivalence between the Voigt–dispersion form derived above and the compact representation in terms of the Faddeeva function. Let
\[
w(z)=V(z)+\mathrm{i}D(z),
\]
where $V$ and $D$ denote, respectively, the Voigt profile and its associated dispersion function (Hilbert transform). For the arguments introduced above, define $x_0:=\mu_2-\mu_1$ and $x':=\lambda'-\mu_1$,
one has on the real axis $w(z')=\mathrm{i}D(x';\sigma,0)$, while at the complex pole
\[
w(z_0)/(\sqrt{2\pi}\sigma)=V(x_0;\sigma,\chi)+\mathrm{i}D(x_0;\sigma,\chi).
\]
Therefore,
\[
\frac{w(z')-w(z_0)}{\sqrt{2\pi}\sigma}
=
-\,V(x_0;\sigma,\chi)
+\mathrm{i}\bigl[D(x';\sigma,0)-D(x_0;\sigma,\chi)\bigr].
\]
Substituting this into \cref{TTI}, one directly recovers
\[
I(\lambda')
=
\frac{\Delta\,V(x_0;\sigma,\chi)
+\chi\,[D(x';\sigma,0)-D(x_0;\sigma,\chi)]}
{\Delta^2+\chi^2}.
\]
This establishes the complete equivalence between the compact Faddeeva form and the explicit Voigt-dispersion representation used in the main text.

\section{Analytic Foundations of the Effective Self-Energy Representation}
\label{app:analytic_foundations}

In this appendix we establish the rigorous basis for the effective self-energy
representation introduced in Sec.~\ref{sec:ansatz_closure}. The analysis proceeds
in two steps. First, we determine the class of admissible self-energy functions
from the requirement of causality and analyticity. Second, we show how the
resulting spectral function relates to the normalized Voigt-type profile used in
the hierarchical ansatz strategy, and provide a concrete procedure for fixing the
effective parameters.

\subsection{Admissible forms of the self-energy ansatz}

We examine the basic requirements for constructing an ansatz for the self-energy
$\mathcal{G}_{\mu i}(z)$.

\medskip
\noindent\textbf{Assumption.}
We assume that the self-energy function $\mathcal{G}_{\mu i}(z)$ is analytic in
the lower half complex plane and admits the boundary representation
\begin{equation}\label{assumpG_app}
    \mathcal{G}_{\mu i}(\lambda-\mathrm{i}0^+)
    = C_{\mu i}
    + \mathrm{i}\chi^{\mathrm{eff}}_{\mu i}\, g^{\mu i}(\lambda),
\end{equation}
where $C_{\mu i}$ is a real, $\lambda$-independent constant and
$g^{\mu i}(\lambda)$ is the boundary value of an analytic function satisfying
\begin{equation}
    \Re g^{\mu i}(\lambda)\ge 0,
    \qquad
    \int d\lambda\, \Re g^{\mu i}(\lambda)<\infty .
\end{equation}
These conditions ensure causality and the existence of a well-defined spectral
density. The constant $C_{\mu i}$ represents the subtraction freedom associated
with the high-energy behavior of $\mathcal{G}_{\mu i}$ and can be absorbed into a
redefinition of the bare energy.

\medskip
\noindent\textbf{Proposition.}
Under the assumption \eqref{assumpG_app}, the real and imaginary parts of
$\mathcal{G}_{\mu i}$ are related by the Kramers--Kronig relations in the form
\begin{equation}\label{propKK_app}
    -\Im g^{\mu i}(\lambda)
    =H\!\left(\Re g^{\mu i}\right)(\lambda).
\end{equation}
Consequently, the admissible function $g^{\mu i}$ must belong to a class closed
under the Hilbert transform.

\medskip
\noindent\textbf{Proof.}
Since $\mathcal{G}_{\mu i}(z)$ is analytic in the lower half-plane, its boundary
values on the real axis obey the (once-subtracted) Kramers--Kronig relations
\begin{equation}\label{KKsub_app}
    \Re\,\mathcal{G}_{\mu i}(\lambda)
    = C_{\mu i}
    + \dashint dx\,\frac{1}{\pi}
    \frac{\Im\,\mathcal{G}_{\mu i}(x-\mathrm{i}0^+)}{\lambda-x}.
\end{equation}
Note that Eq.~\eqref{REGAP} is equivalent to a once-subtracted Kramers-Kronig
relation. The constant $C_{\mu i}$ accounts for the high-energy subtraction
constant, ensuring the convergence of the principal value integral even when
$\Re\,\mathcal{G}$ does not vanish at infinity.
Substituting the ansatz \eqref{assumpG_app} into \eqref{KKsub_app} and using
\[
    \Im\,\mathcal{G}_{\mu i}(x)
    = \chi^{\mathrm{eff}}_{\mu i}\Re g^{\mu i}(x),
\]
we obtain
\[
    \Re\,\mathcal{G}_{\mu i}(\lambda)
    = C_{\mu i}
    + \chi^{\mathrm{eff}}_{\mu i}
    H\!\left(\Re g^{\mu i}\right)(\lambda).
\]
On the other hand, taking the real part of \eqref{assumpG_app} directly yields
\[
    \Re\,\mathcal{G}_{\mu i}(\lambda)
    = C_{\mu i}
    -\chi^{\mathrm{eff}}_{\mu i}\Im g^{\mu i}(\lambda).
\]
Equating the two expressions leads to \eqref{propKK_app}, completing the proof.

The simplest form satisfying \eqref{propKK_app} is $g^{\mu i}(\lambda)=1$, which
recovers the standard Lorentzian spectral function. If one wishes to introduce
Gaussian decay in the high-energy tail without violating the causal structure,
the natural choice is the Faddeeva function:
\begin{equation}
    g^{\mu i}(\lambda)= w\!\Big(\frac{-\delta \lambda'_{\mu i}}{\sqrt{2}\sigma_{\mu i}}\Big),
\end{equation}
where $w(z)=e^{-z^2}\mathrm{erfc}(-\mathrm{i}z)$ is analytic throughout the
complex plane. This choice leads directly to the effective self-energy
representation in the main text.

\subsection{Effective self-energy representation of Voigt-normalized spectral profiles}

We now show how the resulting spectral function relates to the normalized
Voigt-type profile, and provide a procedure for determining the effective
parameters $\epsilon_{\mathrm{eff}}$ and $\chi_{\mathrm{eff}}$.

We start from the normalized Voigt-type profile
\begin{equation}
    f_V(\lambda)
    = \frac{L(\lambda-\epsilon_L;\chi)\,G(\lambda-\epsilon_G;\sigma)}
           {V(\epsilon_L-\epsilon_G;\sigma,\chi)}.
\end{equation}
We approximate $f_V(\lambda)$ by a spectral function with an effective,
energy-dependent imaginary self-energy, as derived from the admissible form
above:
\begin{equation}\label{eq:voigt_selfenergy_equiv_app}
    f_V(\lambda)
    \simeq \frac{1}{\pi}\Im
    \frac{1}{\lambda-\epsilon_{\mathrm{eff}}
             -\mathrm{i}\chi_{\mathrm{eff}}\,
             w\!\left(\frac{-\lambda+\epsilon_G}{\sqrt{2}\sigma}\right)},
\end{equation}
In this parametrization, $\Re\Sigma(\lambda)=0$ at $\lambda=\epsilon_G$, and the
effective parameters $\epsilon_{\mathrm{eff}}$ and $\chi_{\mathrm{eff}}$ are
determined self-consistently from the properties of the spectral peak.

\medskip
\noindent\textbf{Peak position.}
The peak position $\lambda_p$ of the Voigt-type profile is obtained from
$\partial_\lambda f_V(\lambda)=0$. For
$|\epsilon_L-\epsilon_G|\lesssim \chi$ one finds, to leading order,
\begin{equation}
    \lambda_p
    \approx \frac{2\sigma^2\,\epsilon_L+\chi^2\,\epsilon_G}{2\sigma^2+\chi^2},
    \label{eq:voigt_peak_app}
\end{equation}
which corresponds to a weighted average of the Lorentzian and Gaussian centers.

On the other hand, the peak position of the spectral representation
\eqref{eq:voigt_selfenergy_equiv_app} follows from extremizing the denominator.
Expanding the Dawson function entering $w(z)$ linearly around the peak, one
obtains
\begin{equation}
    \lambda_p
    \approx \frac{\epsilon_{\mathrm{eff}}
                 -\epsilon_G\dfrac{2\chi_{\mathrm{eff}}}{\sqrt{2\pi}\sigma}}
                 {1-\dfrac{2\chi_{\mathrm{eff}}}{\sqrt{2\pi}\sigma}}.
    \label{eq:selfenergy_peak_app}
\end{equation}
Equating Eqs.~\eqref{eq:voigt_peak_app} and \eqref{eq:selfenergy_peak_app} yields
the first relation between $\epsilon_{\mathrm{eff}}$ and $\chi_{\mathrm{eff}}$.

\medskip
\noindent\textbf{Peak height.}
A second independent relation follows from matching the peak heights. Evaluating
Eq.~\eqref{eq:voigt_selfenergy_equiv_app} at $\lambda=\lambda_p$ and using the
small-argument expansion of the Faddeeva function, one finds
\begin{align}
    f_V(\lambda_p)
    = \frac{1}{\pi}\Im
      \frac{1}{\lambda_p-\epsilon_{\mathrm{eff}}
               -\mathrm{i}\chi_{\mathrm{eff}}\,
               w\!\left(\frac{-\lambda_p+\epsilon_G}{\sqrt{2}\sigma}\right)}\notag\\
    \approx e^{(-\lambda_p+\epsilon_G)^2/2\sigma^2}/(\pi\,\chi_{\mathrm{eff}}).
    \label{eq:peak_height_app}
\end{align}
Requiring equality with the exact peak height of $f_V(\lambda)$ provides a second
equation relating $\epsilon_{\mathrm{eff}}$ and $\chi_{\mathrm{eff}}$.

\medskip
\noindent\textbf{Determination of effective parameters.}
The effective parameters $\epsilon_{\mathrm{eff}}$ and $\chi_{\mathrm{eff}}$ are
thus uniquely fixed by solving Eqs.~\eqref{eq:voigt_peak_app},
\eqref{eq:selfenergy_peak_app}, and \eqref{eq:peak_height_app}. This procedure
ensures that both the peak position and peak height of the normalized Voigt
profile are reproduced by the effective spectral representation. Numerical
comparisons further confirm that the agreement extends beyond the immediate peak
region to the logarithmic decay regime.

Physically, the apparent peak displacement in the Voigt-normalized profile does
not originate from a static real-part energy shift, but rather from the energy
dependence of the imaginary self-energy, which redistributes spectral weight away
from the bare Lorentzian center.

\medskip
\noindent\textbf{Summary of the Appendix.}
Together, the two parts of this appendix establish a complete theoretical
foundation for the effective self-energy representation. The first part
identifies the Faddeeva function as the minimal causal extension of the constant
self-energy that introduces a non-Lorentzian scale. The second part provides the
quantitative link to the Voigt-type profile, fixing the effective parameters in
terms of physically measurable peak properties. This justifies the use of
\cref{afg_final} as a computationally efficient yet physically faithful
parametrization in the main text.

\section{Limitations of the constant self-energy approximation in the presence of third-order terms}
\label{app:const_selfenergy}

In this appendix we first recall how the SCBA leads to a Lorentzian spectral function under the wide-band limit (constant self-energy approximation). We then show that the same approximation fails when the third-order multi-resolvent contribution \(\mathcal{G}^{(3)}\) is included, because the resulting self-consistency equation for the imaginary part admits no positive solution. This forces us to abandon the constant self-energy assumption and to allow a frequency-dependent self-energy, which naturally leads to the effective representation introduced in Sec.~\ref{ESER}.
\subsection{SCBA and the Lorentzian solution}

In the SCBA (which corresponds to the mean-field approximation \(\mathcal{G}_{\mu i}\approx\mathcal{G}_{\mu i}^{\mathrm{OD}}\) of the main text), the self-energy for a single degree of freedom coupled to a continuum of states can be written as
\begin{equation}
    \Sigma(z) = \int_{-\infty}^{\infty} d\omega\, f(\omega)\,\frac{|V(\omega)|^2}{z-\omega},
    \qquad
    \mathcal{R}(z)=\frac{1}{z-\epsilon_0-\Sigma(z)},
\end{equation}
where \(f(\omega)\) is the density of states of the continuum, \(|V(\omega)|^2\) is the squared coupling matrix element (assumed to depend only on energy), and we have absorbed all indices into a single effective level \(\epsilon_0\). The wide-band limit consists of the following two assumptions:
\begin{itemize}
    \item The density of states is constant over the relevant energy range: \(f(\omega)=f_0\).
    \item The coupling matrix element is energy-independent: \(|V(\omega)|^2 = |V|^2\).
\end{itemize}
These approximations are justified when the bandwidth \(W\) of the continuum is much larger than any other energy scale (e.g., the level broadening), and when the spectral features of interest lie deep inside the band.

Under these assumptions, the self-energy becomes
\begin{equation}
    \Sigma(z) = f_0|V|^2 \int_{-W/2}^{W/2} \frac{d\omega}{z-\omega},
\end{equation}
where we have introduced a finite bandwidth \(W\) which will eventually be taken to infinity. For \(z=\omega'+\mathrm{i}0^+\) (with \(\omega'\) real), we evaluate the integral using the Sokhotski–Plemelj identity:
\begin{equation}
    \frac{1}{\omega'-\omega+\mathrm{i}0^+} = \mathcal{P}\frac{1}{\omega'-\omega} - \mathrm{i}\pi\delta(\omega'-\omega).
\end{equation}
The imaginary part is immediate:
\begin{equation}
    \Im\Sigma(\omega'+\mathrm{i}0^+) = -\pi f_0|V|^2,
\end{equation}
which is a constant (independent of \(\omega'\)). The real part is given by the principal-value integral:
\begin{equation}
    \Re\Sigma(\omega'+\mathrm{i}0^+) = f_0|V|^2\;\mathcal{P}\int_{-W/2}^{W/2} \frac{d\omega}{\omega'-\omega}.
\end{equation}
This integral can be evaluated exactly:
\begin{equation}
    \mathcal{P}\int_{-W/2}^{W/2} \frac{d\omega}{\omega'-\omega} = \ln\left|\frac{W/2-\omega'}{W/2+\omega'}\right|.
\end{equation}
In the wide-band limit \(W\to\infty\), the right-hand side tends to zero for any fixed \(\omega'\) (since the argument of the logarithm approaches 1). More precisely, expanding for large \(W\) gives
\begin{equation}
    \ln\left|\frac{W/2-\omega'}{W/2+\omega'}\right| = -\frac{2\omega'}{W} + O(W^{-2}) \to 0.
\end{equation}
Thus, in the limit of infinite bandwidth, the real part of the self-energy also becomes a constant (zero). In a more realistic treatment, any finite constant contribution from the principal-value integral can be absorbed into a redefinition of the bare level \(\epsilon_0\) (the so-called Lamb shift). Hence, we obtain
\begin{equation}
    \Sigma(z) = \Delta - \mathrm{i}\chi,\qquad \chi = \pi f_0|V|^2>0,
\end{equation}
where \(\Delta\) is a real constant (which may be set to zero by shifting \(\epsilon_0\)).

Substituting this constant self-energy into the Dyson equation yields
\begin{equation}
    \mathcal{R}(\omega-\mathrm{i}0^+) = \frac{1}{\omega-\epsilon_0-\Delta+\mathrm{i}\chi},
\end{equation}
and the spectral function is Lorentzian:
\begin{equation}
    f(\omega) = \frac{1}{\pi}\Im\mathcal{R}(\omega-\mathrm{i}0^+) = \frac{1}{\pi}\frac{\chi}{(\omega-\epsilon_0-\Delta)^2+\chi^2}.
\end{equation}
The self-consistency of the approximation is automatically satisfied: the constant self-energy we started with is exactly the one obtained from the integral over the continuum, because the integral's imaginary part is constant and its real part vanishes (or is absorbed). No additional “peak condition” is needed. This is the standard derivation of the Lorentzian line shape in the wide-band limit.

\subsection{Absence of spontaneous skewness in the mean-field (SCBA) self-consistency}
\label{app:no_skewness_mf}

We show that, under standard symmetry assumptions, the mean-field (SCBA) 
self-consistent equation cannot spontaneously generate a skewed spectral function. 
The argument relies on the parity-preserving structure of the mean-field functional 
and the stability of the iterative solution.

From Eqs.~\eqref{eq:CSEQ} and \eqref{eq:GRE}, the mean-field resolvent for state $\phi_{\mu i}$ satisfies
\begin{align}
  \mathcal{R}_{\mu i}(z) = \frac{1}{z - a_{\mu i} - V_{\mu i} - \mathcal{G}_{\mu i}^{\mathrm{OD}}(z)},\notag\\
  \mathcal{G}_{\mu i}^{\mathrm{OD}}(z) = \sum_{\nu j\neq \mu i} |V_{\mu i,\nu j}|^{2}\,\mathcal{R}_{\nu j}(z).
\end{align}
The mean-field self-energy consequently takes the form
\begin{equation}\label{eq:mf_self_parity}
  \mathcal{G}_{\mu i}^{\mathrm{OD}}(\omega-\mathrm{i}0^{+})
  =\sum_{\nu j\neq \mu i} |V_{\mu i,\nu j}|^{2}\,
    \bigl[ H[f_{\nu j}](\omega) + \mathrm{i} f_{\nu j}(\omega) \bigr].
\end{equation}

Assume that there exists a center energy $\omega_{0}$ such that the combined kernel
\begin{equation}
  K(\omega,\epsilon) \equiv \rho(\epsilon)\,|V_{\mu i,\nu j}|^{2}
\end{equation}
is invariant under reflection $\epsilon \to 2\omega_{0}-\epsilon$, and that the 
relevant spectral functions decay sufficiently fast at large $|\omega|$ so that 
the Hilbert transform preserves parity.

Define the reflection operator $\mathcal{P}$ acting on a function $g(\omega)$ as
\begin{equation}
  (\mathcal{P}g)(\omega) \equiv g(2\omega_{0}-\omega).
\end{equation}
\textbf{Claim:} If a set of spectral functions $\{f_{\mu i}\}$ is symmetric, i.e.\ 
$\mathcal{P}f_{\mu i} = f_{\mu i}$, then one iteration of the mean-field equations 
produces new spectral functions that are also symmetric.
\textbf{Proof of the claim.} For symmetric $f_{\mu i}$, we have:
\begin{itemize}
  \item $f_{\mu i}(\omega_{0}+\delta) = f_{\mu i}(\omega_{0}-\delta)$ (even),
  \item $H[f_{\mu i}](\omega_{0}+\delta) = -H[f_{\mu i}](\omega_{0}-\delta)$ (odd).
\end{itemize}
Thus the mean-field self-energy satisfies
\begin{align}
  \Re\mathcal{G}_{\mu i}^{\mathrm{OD}}(\omega_{0}+\delta) 
  = -\Re\mathcal{G}_{\mu i}^{\mathrm{OD}}(\omega_{0}-\delta), \notag\\
  \Im\mathcal{G}_{\mu i}^{\mathrm{OD}}(\omega_{0}+\delta) 
  = +\Im\mathcal{G}_{\mu i}^{\mathrm{OD}}(\omega_{0}-\delta).
\end{align}
Define $\tilde{\mathcal{R}}_{\mu i}(\omega) \equiv 
\mathcal{R}_{\mu i}^{*}(2\omega_{0}-\omega)$. Using the symmetry of the kernel, 
one verifies that $\tilde{\mathcal{R}}_{\mu i}$ satisfies the same Dyson equation 
as $\mathcal{R}_{\mu i}$. Therefore, for solutions obtained from symmetric initial 
conditions within the same basin of attraction, we have 
$\tilde{\mathcal{R}}_{\mu i} = \mathcal{R}_{\mu i}$, which implies
\begin{equation}
  f_{\mu i}^{\mathrm{new}}(\omega_{0}+\delta) 
  = f_{\mu i}^{\mathrm{new}}(\omega_{0}-\delta).
\end{equation}
Thus the symmetry is preserved under iteration.

Starting from symmetric initial spectral functions, each iteration preserves 
reflection symmetry. If the iteration converges to a stable fixed point within 
this symmetric sector, the resulting spectral function must also be symmetric. 
We do not exclude the formal existence of symmetry-broken solutions; however, 
such solutions are not generated within the SCBA iteration starting from 
physically relevant symmetric initial conditions.
Consequently, the spectral function admits an expansion around $\omega_{0}$:
\begin{equation}
  f_{\mu i}(\omega) = f^{(0)}_{\mu i} + f^{(2)}_{\mu i}(\omega-\omega_{0})^{2} 
  + \mathcal{O}((\omega-\omega_{0})^{4}),
\end{equation}
and the imaginary part of the self-energy contains only even powers:
\begin{equation}
  \Im\mathcal{G}_{\mu i}^{\mathrm{OD}}(\omega) = \Gamma_{0} + \Gamma_{2} x^{2} 
  + \mathcal{O}(x^{4}), \qquad x=\omega-\omega_{0}.
\end{equation}

Under the above symmetry conditions, the SCBA self-consistent equation is 
parity-preserving and cannot spontaneously generate spectral asymmetry. 
Skewness requires multi-resolvent correlations, which introduce intrinsically 
nonlocal frequency couplings absent in any single-resolvent mean-field theory.

\subsection{Beyond the constant approximation: necessity of frequency dependence}

In the wide-band SCBA limit the self-energy becomes constant,
$\Sigma(\omega)=\Delta-\mathrm{i}\Gamma$, leading to a Lorentzian spectral
function (\cref{app:no_skewness_mf}).  When the leading
multi-resolvent term $\mathcal{G}^{(3)}$ is included, the same constant
ansatz yields $\Gamma^2=-g$, which has no physical solution for $g>0$.
Hence a frequency-independent self-energy is mathematically inconsistent
with the multi-resolvent hierarchy.

The effective self-energy representation of Sec.~\ref{ESER} resolves this
inconsistency with the minimal causal extension
\begin{equation}
  \mathcal{G}_{\mu i}^{\mathrm{eff}}(\lambda)
  = \Delta_{\mu i}^{\mathrm{eff}} - V_{\mu i}
    + \mathrm{i}\chi_{\mu i}^{\mathrm{eff}}\,
      w\!\Bigl(\frac{-\delta\lambda_{\mu i}'}{\sqrt{2}\,\sigma_{\mu i}}\Bigr).
\end{equation}
This form is analytic in the lower half-plane, reduces to the constant
self-energy when $\sigma_{\mu i}\to\infty$, and produces a finite Gaussian
scale as soon as $\sigma_{\mu i}$ is finite.  The closure condition
Eq.~(\ref{gacon_final}) then provides a self-consistent scheme that
automatically incorporates the required frequency dependence and guarantees
positivity.  The remainder of this appendix therefore serves only to
highlight that the constant approximation must be abandoned once
multi-resolvent correlations are taken into account, a conclusion already
built into the effective self-energy framework.

\section{Skewness from Odd Components of the Self-Energy: A Minimal Bulk Ansatz}
\label{app:skewed_cauchy}

Using Eq.~\eqref{RkkR}, the coarse-grained probability distribution \eqref{PAG} can be written as
\[
p_{\mu i}(\lambda) = \frac{1}{\pi e^{S(\lambda)}} \Im
\frac{1}{\lambda - a_{\mu i} - V_{\mu i} - \mathcal{G}_{\mu i}(\lambda - i0^+)}.
\]

We decompose the self-energy as 
\[
\mathcal{G}_{\mu i} = \mathcal{G}_{\mu i}^{\mathrm{OD}} + \mathcal{G}_{\mu i}^{(3)} + \mathcal{G}_{\mu i}^{(5)} + \cdots,
\]
and expand around the mean-field peak $\lambda_0$, defining $x = \lambda - \lambda_0$.

At leading order, the even component $\mathcal{G}^{\mathrm{OD}}$ produces a symmetric Lorentzian profile,
\[
p^{(0)}(\lambda) \propto \frac{1}{(x + \delta_0)^2 + \Gamma_0^2},
\]
where $\delta_0$ and $\Gamma_0$ arise from the real and imaginary parts of the self-energy.

More generally, the self-energy can be decomposed into even and odd parts with respect to $x$,
\[
\Im \mathcal{G}(\lambda) = \Gamma_0 + \Gamma_2 x^2 + \cdots \;+\; \gamma_1 x + \gamma_3 x^3 + \cdots,
\]
where the coefficients $\gamma_{2n+1}$ originate from odd-order contributions in the hierarchy (e.g., $\mathcal{G}^{(3)}$, $\mathcal{G}^{(5)}$, etc.).

Substituting into the resolvent and expanding to leading order in $x$, one finds that even terms renormalize the symmetric denominator, while odd terms generate an asymmetric correction in the numerator. As a result, the probability density takes the universal form
\[
p_{\mu i}(\lambda) \propto 
\frac{1 + \alpha x}{(x + \delta)^2 + \Gamma^2},
\]
where $\delta$ and $\Gamma$ include renormalizations from even contributions, and the skewness parameter satisfies
\[
\alpha \sim \frac{\gamma_1}{\Gamma_0},
\qquad \gamma_1 \sim \mathcal{O}(V^3).
\]

This demonstrates that skewness in the coarse-grained spectral distribution is a direct consequence of the odd component of the self-energy. In particular, the third-order term $\mathcal{G}^{(3)}$ provides the leading contribution to $\gamma_1$, yielding a skewed Cauchy profile as the minimal bulk ansatz beyond mean field.

Thus, within the resolvent-based formulation, the emergence of asymmetric line shapes is not an ad hoc assumption but follows generically from the parity structure of the self-energy and the analytic properties encoded in Eq.~\eqref{RkkR}.

\medskip
\noindent
\textit{Remark.} This structure is analogous to a Gram--Charlier-type expansion, but is here derived directly from the resolvent and self-energy hierarchy rather than imposed phenomenologically.

\section{Fourth-order contribution and the recursive structure}
\label{app:fourth_order}

In this appendix we extend the recursive construction based on  Eq.~\eqref{CCRes} to the next level of the hierarchy, thereby generating the fourth-order contribution to the cross-correlated self-energy. The derivation follows the same algebraic steps as for the third-order term, but applies the projection identity twice.

Starting from the remainder term in Eq.~\eqref{CCRes}, we use \cref{offRexp} and  obtain
\begin{widetext}
\begin{equation}
\mathcal{G}^{\mathrm{(res)}}_{\mu i}(z)
= \sum_{\nu j\neq \mu i}\sum_{\xi k\neq \nu j \neq \mu i}
\bra{\phi_{\mu i}} V \Phi_{\mu i} \Phi_{\nu j} \Phi_{\xi k} \frac{1}{z -  \Phi_{\mu i} \Phi_{\nu j}H \Phi_{\mu i} \Phi_{\nu j}}\ket{\phi_{\xi k}}
\, V_{\xi k,\nu j} V_{\nu j,\mu i}\,\mathcal{R}^{(\mu i)}_{\nu j}(z) = \mathcal{G}^{(4)}_{\mu i}(z) + \mathcal{G}^{\mathrm{(4res)}}_{\mu i}(z),
\end{equation}
where we have separated the term in which all projection operators are distinct.  The explicit fourth-order contribution reads
\begin{equation}
\mathcal{G}^{(4)}_{\mu i}(z)
=\sum_{\eta l\neq \xi k \neq \nu j\neq \mu i}
V_{\mu i,\eta l} V_{\eta l,\xi k} V_{\xi k,\nu j} V_{\nu j,\mu i}
\;\mathcal{R}^{(\mu i, \nu j, \xi k)}_{\eta l}(z)\,\mathcal{R}^{(\mu i, \nu j)}_{\xi k}(z)\,\mathcal{R}^{(\mu i)}_{\nu j}(z) .
\end{equation}
\end{widetext}

\noindent\textbf{Remark on the notion of order.}
The expressions derived above, such as $\mathcal{G}^{(4)}_{\mu i}(z)$, contain products of \emph{exact} diagonal resolvents $\mathcal{R}(z)$.  Each resolvent already sums contributions from all orders in the interaction.  Therefore the superscript $(4)$ indicates the number of \emph{explicit} interaction vertices (i.e., the power of $V$) appearing in the product, not the perturbative order in the usual Dyson expansion.  This is a self-consistent reorganization of the correlation hierarchy, analogous to the treatment of the third-order term in the main text.  The recursive application of \cref{PRID} generates a systematic ladder of contributions of the form
\begin{equation}\label{TTGl}
	\mathcal{G}_{\mu i}^{(\ell)}(z) \sim 	\mathcal{G}_{\mu i}^{(\ell),\text{D}}(z) =\sum \bigl(\text{product of $\ell$ $V$'s}\bigr)\;
\prod_{a=1}^{\ell-1} \mathcal{R}_{\alpha_a}(z),
\end{equation}
which remains closed within the space of diagonal resolvents under the diagonal closure approximation. Hence the projected hierarchy is exact; the replacement of projected resolvents by full diagonal resolvents provides a systematically improvable route to incorporate higher-order correlations beyond mean-field theory, though the closure itself is not rigorously controlled in the absence of a small parameter.

Referring to \cref{TTG3}, \cref{TTGl} can be further expressed in terms of the distribution as
\begin{equation}
	\mathcal{G}_{\mu i}^{(\ell)}(z)/\pi^{\ell-1} \sim \sum \bigl(\text{product of $\ell$ $V$'s}\bigr)\;
\prod_{a=1}^{\ell-1}[H(f^{\alpha_a})+\mathrm{i} f^{\alpha_a}] .
\end{equation}

\section{Non-perturbative reformulation and continued fraction structure of the resolvent hierarchy}
\label{app:cf_closure}

In the main text, the self-energy is expanded as
\begin{align}
    \mathcal{G}_{\mu i}(z)=\mathcal{G}_{\mu i}^{\mathrm{OD}}(z)+\Delta\mathcal{G}_{\mu i}(z) +\mathcal{G}_{\mu i}^{\mathrm{CC}}(z), \notag\\
    \mathcal{G}_{\mu i}^{\mathrm{CC}}(z)=\sum_{\ell\ge3}\mathcal{G}_{\mu i}^{(\ell)}(z),
\end{align}
where \(\mathcal{G}^{(\ell)}\propto V^{\ell}\). This is an explicit expansion in powers of the interaction.
While formally exact, it becomes problematic in dense-spectrum correlated systems: the effective expansion parameter is not simply the bare coupling but the product \(|V|^2 e^{S(\lambda)}\) (density of states times squared matrix element). Because the Hilbert space dimension grows exponentially with system size, this product can be of order unity even for moderate \(V\). Consequently, higher-order terms \(\mathcal{G}^{(3)},\mathcal{G}^{(4)},\dots\) are not guaranteed to be smaller than \(\mathcal{G}^{(2)}\); truncating the series at any finite order may lead to uncontrolled errors, violation of causality, or unphysical spectral features. 

To overcome this limitation, we present an exact reformulation of the resolvent hierarchy that eliminates the explicit expansion in powers of the interaction and leads to a fully non-perturbative reformulation that enables closure under additional assumptions. We show that the hierarchy can be equivalently expressed either as a matrix continued fraction or as a closed nonlinear equation for the diagonal resolvent. This continued-fraction representation encodes the same multi-resolvent correlations as the hierarchy in \cref{sec:cross_expansion}, but reorganized into a one-dimensional recursive structure.

\subsection{Exact Feshbach structure}

We start from the exact Feshbach projection:
\begin{align}
\mathcal{R}_{\mu i}(z)
&=
\frac{1}{z - a_{\mu i} - V_{\mu i} - \mathcal{G}_{\mu i}(z)},\\
\mathcal{G}_{\mu i}(z)
&=
\bra{\phi_{\mu i}} V \Phi_{\mu i} \frac{1}{z-\Phi_{\mu i}H\Phi_{\mu i}} \Phi_{\mu i} V \ket{\phi_{\mu i}}.
\end{align}
Using $\Phi_{\mu i} H_0 \ket{\phi_{\mu i}}=0$, one can rewrite the self-energy purely in terms of the full Hamiltonian:
\begin{equation}
\mathcal{G}_{\mu i}(z)
=
\bra{\phi_{\mu i}} H \Phi_{\mu i} \frac{1}{z-\Phi_{\mu i}H\Phi_{\mu i}} \Phi_{\mu i} H \ket{\phi_{\mu i}}.
\end{equation}

\subsection{Lanczos representation and continued fraction}

The above structure becomes particularly transparent when the Hamiltonian is represented in a Krylov (Lanczos) basis. This representation provides a non-perturbative reformulation that is exact and avoids any expansion in the interaction strength.

Starting from an initial state $\ket{\phi_0}$ (which in our context could be any $\ket{\phi_{\mu i}}$), the Lanczos algorithm generates an orthonormal basis $\{\ket{\phi_n}\}$ such that the Hamiltonian becomes tridiagonal:
\begin{equation}
H \ket{\phi_n} = b_n \ket{\phi_{n-1}} + a_n \ket{\phi_n} + b_{n+1} \ket{\phi_{n+1}},
\end{equation}
with $b_0 = 0$ and $a_n = \bra{\phi_n} H \ket{\phi_n}$, $b_{n+1} = \bra{\phi_{n+1}} H \ket{\phi_n}$. In this basis the matrix of $z-H$ reads
\begin{equation}
z-H = 
\begin{pmatrix}
z-a_0 & -b_1 & 0 & \cdots \\
-b_1 & z-a_1 & -b_2 & \cdots \\
0 & -b_2 & z-a_2 & \cdots \\
\vdots & \vdots & \vdots & \ddots
\end{pmatrix}.
\end{equation}
Define the diagonal resolvent
\begin{equation}\label{RNdef}
R_n(z) = \bra{\phi_n} (z-H^{[\ge n]})^{-1} \ket{\phi_n}.
\end{equation}
Our goal is to derive an exact recurrence relation for $R_n(z)$.

Partition the Hilbert space into two subspaces: $P = \mathrm{span}\{\ket{\phi_n}\}$ (one-dimensional) and $Q = \mathrm{span}\{\ket{\phi_{n+1}}, \ket{\phi_{n+2}}, \dots\}$. In this block decomposition,
\begin{equation}
z-H = \begin{pmatrix}
A & B^\dagger \\
B & D
\end{pmatrix},
\end{equation}
where $A = z-a_n$ (a scalar), $B = (-b_{n+1}, 0, 0, \dots)$ (a row vector), and $D = z - H_{\text{tail}}$ with $H_{\text{tail}}=Q H^{[\ge n]} Q=H^{[\ge n+1]}$. The $(n,n)$ element of the inverse is given by the Schur complement:
\begin{equation}
\bra{\phi_n} (z-H^{[\ge n]})^{-1} \ket{\phi_n} = \bigl(A - B^\dagger D^{-1} B\bigr)^{-1}.
\end{equation}
Thus
\begin{equation}
R_n(z) = \frac{1}{z - a_n - B^\dagger D^{-1} B}.
\end{equation}
Now compute the coupling term:
\begin{equation}
B^\dagger D^{-1} B = b_{n+1}^2 \; \bra{\phi_{n+1}} (z - H^{[\ge n+1]})^{-1} \ket{\phi_{n+1}}.
\end{equation}

Because the Lanczos basis is tridiagonal, the state $\ket{\phi_{n+1}}$ has no overlap with any basis vector $\ket{\phi_m}$ for $m < n+1$ except through the term $b_{n+1}\ket{\phi_n}$ which is already excluded from $Q$. Consequently, the resolvent matrix element in the tail subspace coincides with the projected resolvent element:
\begin{equation}
R_{n+1}(z)=\bra{\phi_{n+1}} (z -  H^{[\ge n+1]})^{-1} \ket{\phi_{n+1}}.
\end{equation}
Substituting this equivalence yields the exact continued-fraction recurrence:
\begin{equation}
R_n(z) = \frac{1}{z - a_n - b_{n+1}^2 R_{n+1}(z)} .
\label{eq:cf_recurrence}
\end{equation}
This relation is not an approximation; it follows rigorously from the Lanczos tridiagonalization and the Schur complement formula. Iterating it gives a continued fraction representation of the diagonal resolvent:
\begin{equation}
R_0(z) = \cfrac{1}{z - a_0 - \cfrac{b_1^2}{z - a_1 - \cfrac{b_2^2}{z - a_2 - \ddots}}}.
\end{equation}
Although the recurrence appears local, the nonlocal frequency dependence identified in \cref{sec:cross_expansion} is implicitly contained in the energy dependence of $R_n(z)$, which reproduces the Hilbert-transform structure upon expansion.

Equation \eqref{eq:cf_recurrence} can be viewed as a one-dimensional Dyson equation:
\begin{equation}
R_n(z) = \frac{1}{z - a_n - \Sigma_n(z)},\qquad \Sigma_n(z) = b_{n+1}^2 R_{n+1}(z),
\end{equation}
where $\Sigma_n(z)$ plays the role of a self-energy that describes propagation from level $n$ to level $n+1$ and back. This is the exact non-perturbative analog of the mean-field expression $\mathcal{G}_{\mu i}^{\mathrm{OD}}(z) = \sum_{\nu j} |V_{\mu i,\nu j}|^2 \mathcal{R}_{\nu j}(z)$: the sum over intermediate states is replaced by a single “hopping” to the next Lanczos state, and the infinite sum is replaced by the continued fraction.

Crucially, the continued fraction representation does not rely on any expansion in the interaction. The Lanczos coefficients $a_n$ and $b_n$ are determined by the full Hamiltonian and contain all orders of $V$ exactly. Therefore, this reformulation provides a non-perturbative closure of the resolvent hierarchy: instead of truncating an expansion in $V$, one truncates the continued fraction at a finite depth (which is equivalent to approximating the Hamiltonian by a finite band matrix) and then solves exactly. This is particularly advantageous in dense-spectrum correlated systems where the naive $V$-expansion fails.

\subsection{Connection to the ETH and self-consistency}

Under the eigenstate thermalization hypothesis, the Lanczos coefficients for a typical initial state exhibit a universal behavior: $a_n$ approaches the mean energy of the spectrum, and $b_n^2$ becomes proportional to the energy variance. In the thermodynamic limit, one may postulate that $R_{n+1}(z) \approx R_n(z)$ for large $n$ (self-similarity), which leads to a closed quadratic equation:
\begin{equation}
R_n(z) = \frac{1}{z - a_n - b_{n+1}^2 R_n(z)}.
\end{equation}
Solving this gives $R_n(z) = \frac{z - a_n}{2b_{n+1}^2} \left(1 - \sqrt{1 - \frac{4b_{n+1}^2}{(z-a_n)^2}}\right)$, which yields the famous semi-circle law when $b_n$ is constant and $a_n=0$, and more generally produces non-Lorentzian spectral shapes. This self-consistent continued fraction thus provides a powerful non-perturbative framework that directly links the Lanczos coefficients to the spectral function, and it naturally generates the Gaussian tails and asymmetric line shapes discussed in the main text. The hierarchical ansatz strategy (Lorentzian, Gaussian, Voigt) can be seen as approximate solutions of this exact continued-fraction equation under specific asymptotic regimes of the Lanczos coefficients.

\subsubsection{Connection to SCBA and the next-order correction from the Feshbach projection}
We continue the Lanczos construction initiated in the main text for the initial state $\ket{\phi_0}\equiv\ket{\phi_{\mu i}}$.  
Recall the definitions:
\begin{align}
a_0 = \bra{\phi_0}H\ket{\phi_0}=a_{\mu i}+V_{\mu i},\notag\\ 
\ket{\tilde\phi_1}= \Phi_{\mu i}H\ket{\phi_0},\qquad 
b_1^2 = \bra{\phi_0}H\Phi_{\mu i}H\ket{\phi_0},	
\end{align}
and the normalized state $\ket{\phi_1}= \ket{\tilde\phi_1}/b_1$.  
The exact continued-fraction relation reads
\begin{equation}
R_0(z)=\frac{1}{z-a_0-b_1^2 R_1(z)}
\end{equation}
which is equivalent to \cref{eq:CSEQ}.

Using $\ket{\phi_1}=b_1^{-1}\Phi_{\mu i}H\ket{\phi_0}$,
\[
a_1=\bra{\phi_1}H\ket{\phi_1}= \frac{1}{b_1^2}\bra{\phi_0}H\Phi_{\mu i}H\Phi_{\mu i}H\ket{\phi_0}.
\]
Inserting the resolution of the identity $\sum_{\nu j}\ket{\phi_{\nu j}}\bra{\phi_{\nu j}}$ (with $\ket{\phi_{\nu j}}$ the unperturbed eigenbasis) and using $\Phi_{\mu i}=I-\ket{\phi_0}\bra{\phi_0}$, we obtain
\begin{align}
a_1&=\frac{1}{b_1^2}\sum_{\nu j\neq\mu i}\sum_{\xi k\neq\mu i}
V_{\mu i,\nu j}\; \bra{\phi_{\nu j}}H\ket{\phi_{\xi k}}\; V_{\xi k,\mu i}.
\end{align}
The operator $\Phi_{\mu i}$ removes the component along $\ket{\phi_0}$, which does not contribute because $\ket{\phi_{\nu j}}$ and $\ket{\phi_{\xi k}}$ are orthogonal to $\ket{\phi_0}$.  
To second order in the interaction $V$, we may approximate $H\approx H_0$ inside the matrix element (since $H_0$ gives the leading contribution and $V$ would produce terms of order $V^3$ or higher). Then
\[
\bra{\phi_{\nu j}}H_0\ket{\phi_{\xi k}}
= a_{\nu j}\,\delta_{\nu j,\xi k},
\]
yielding
\begin{equation}
a_1 \approx \frac{1}{b_1^2}\sum_{\nu j\neq\mu i} |V_{\mu i,\nu j}|^2\, a_{\nu j}.
\end{equation}
In many physically relevant situations the unperturbed energies $a_{\nu j}$ are close to a mean value $\bar a$; one then has $a_1\approx\bar a$ up to corrections of order $V^2$.
 
The next Lanczos vector is constructed as
\[
\ket{\tilde\phi_2}= H\ket{\phi_1}-a_1\ket{\phi_1}-b_1\ket{\phi_0},
\qquad 
\ket{\phi_2}= \ket{\tilde\phi_2}/b_2,
\]
with $b_2^2=\langle\tilde\phi_2|\tilde\phi_2\rangle$.  
Equivalently,
\[
b_2^2 = \bra{\phi_1}H\Phi_1 H\ket{\phi_1},
\]
where $\Phi_1 = I-\ket{\phi_0}\bra{\phi_0}-\ket{\phi_1}\bra{\phi_1}$ projects onto the orthogonal complement of the first two Krylov vectors.  
Substituting the expression for $\ket{\phi_1}$ gives
\begin{equation}
b_2^2 = \frac{1}{b_1^2}\bra{\phi_0}H\Phi_{\mu i}H\Phi_1 H\Phi_{\mu i}H\ket{\phi_0}.
\end{equation}
This is manifestly of order $V^4$ because each $H\Phi_{\mu i}H$ contains at least two powers of $V$ when acting on $\ket{\phi_0}$ (the $H_0$ part is eliminated by $\Phi_{\mu i}$). Hence $b_2^2$ contributes only at fourth order and beyond in the self-energy.

If we truncate the continued fraction after the first level, i.e., set $R_1(z)\approx 1/(z-a_1)$, we obtain
\begin{equation}
R_0(z) \approx \frac{1}{\displaystyle z-a_0-\frac{b_1^2}{z-a_1}}.
\end{equation}
Inserting $b_1^2=\sum_{\nu j\neq\mu i}|V_{\mu i,\nu j}|^2$ and treating $a_1$ as an average unperturbed energy $\bar a$ (or, more accurately, as the energy of the intermediate state in a single-site approximation), this becomes
\begin{equation}
R_0(z) \approx \frac{1}{z-a_{\mu i}-V_{\mu i}-\sum_{\nu j\neq\mu i}\dfrac{|V_{\mu i,\nu j}|^2}{z-\bar a}}.
\end{equation}
This is the non-self-consistent Born approximation (or the leading-order self-energy in the SCBA without self-consistency). The full SCBA is recovered by promoting $\bar a$ to a $z$-dependent quantity that includes the self-energy, which corresponds to replacing $1/(z-\bar a)$ with $R_1(z)$ and solving the coupled equations self-consistently. Thus, SCBA can be interpreted as a lowest-level truncation under additional averaging assumptions. Higher-order coefficients $b_2,b_3,\dots$ systematically incorporate multi-step scattering processes beyond SCBA.

This continued-fraction representation provides a non-perturbative realization of the resolvent hierarchy introduced in \cref{sec:cross_expansion}.

\section{Hierarchy-constrained non-perturbative closure of the resolvent}
\label{app:hierarchy_closure}

We now construct a non-perturbative closure scheme that combines the continued-fraction structure of the resolvent (\cref{app:cf_closure}) with the hierarchy expansion developed in \cref{sec:cross_expansion} and \cref{app:fourth_order}. The goal is not to truncate the continued fraction, but to express the deeper-level contributions as a functional of $R_{\mu i}$, thereby obtaining a self-consistent equation.

From the explicit form of the first Krylov vector (\cref{app:cf_closure}),
\begin{equation}
|\phi_{\mu i,1}\rangle
\propto
\sum_{\nu j \neq \mu i}
V_{\nu j,\mu i} |\phi_{\nu j}\rangle,
\end{equation}
one finds that $R_1(z)$ contains both diagonal and off-diagonal contributions in the original basis. The latter generate the multi-resolvent terms similar to $\mathcal{G}^{(\ell\ge 3)}$.

To make this explicit, we decompose the higher-level diagonal resolvent $R_{\mu i,n}(z)$ by inserting a complete set of projectors $\Pi_{\nu j}=|\phi_{\nu j}\rangle\langle\phi_{\nu j}|$ and $\Phi_{\nu j}=I-\Pi_{\nu j}$ into the definition (\ref{RNdef}):
\begin{align}
	R_{\mu i,n}(z)=\sum_{\nu j}\Big[ \bra{\phi_{\mu i,n}} \Phi_{\nu j}(z-H^{(\mu i)[\ge n]})^{-1} \Pi_{\nu j} \ket{\phi_{\mu i,n}}\notag\\
	+\abs{\braket{\phi_{\nu j}|\phi_{\mu i,n}}}^2 R^{(\mu i)[\ge n]}_{\nu j}(z)\Big],
\end{align}
where $H^{(\mu i)[\ge n]}$ denotes the semi-infinite tail subspace generated with $\phi_{\mu i}$ as the initial state.
The first term, containing $\Phi_{\nu j}$, can be systematically expanded using the same recursive projection technique as in \cref{sec:cross_expansion}, ultimately generating the multi-resolvent hierarchy. The second term, playing a role analogous to $\mathcal{G}^{\mathrm{OD}}_{\mu i}(z)+\Delta\mathcal{G}_{\mu i}(z)$ in the original basis, can denoted by $\mathcal{G}^{\mathrm{OD}}_{\mu i,n}(z)+\Delta\mathcal{G}_{\mu i,n}(z)$, where $\mathcal{G}^{\mathrm{OD}}_{\mu i,n}(z):=\sum_{\nu j}\abs{\braket{\phi_{\nu j}|\phi_{\mu i,n}}}^2 R_{\nu j}(z)$. The result mirrors the structure of (\ref{HHOT}):
\begin{equation}\label{eq:kernel_form}
	R_{\mu i,n}(z)=\mathcal{G}^{\mathrm{OD}}_{\mu i,n}(z)+\Delta\mathcal{G}_{\mu i,n}(z)+ \sum_{\ell \ge 3} \mathcal{G}^{(\ell)}_{\mu i,n}(z).
\end{equation}
Schematically substituting Eq.~(\ref{eq:kernel_form}) into the continued-fraction structure leads to a representation of the form
\begin{equation}
\mathcal{R}_{\mu i}(z) = \cfrac{1}{z - a_{\mu i,0} - \ddots \cfrac{\ddots}{z - a_{\mu i,n-1} - b_{\mu i,n}^2  R_{\mu i,n}(z)}},
\end{equation}
which provides a formally self-consistent closure for $R_{\mu i}$ once the deeper-level functions are specified.

Here we consider the continued-fraction expansion to second order, which corresponds to
\begin{equation}
\mathcal{R}_{\mu i}(z)
=
\frac{1}{
z - a_{\mu i,0}
-
\frac{b_{\mu i,1}^2}{
z - a_{\mu i,1}
- b_{\mu i,2}^2 [ \mathcal{G}^{\mathrm{OD}}_{\mu i,2}(z)+ \Delta\mathcal{G}_{\mu i,2}(z)+\dots]}
},
\label{eq:2oclosure}
\end{equation}
where $b_{\mu i,1}^2=\sum_{\nu j \neq \mu i} |V_{\mu i,\nu j}|^2$ and 
\begin{equation}
	a_{\mu i,1} = \frac{1}{b_{\mu i,1}^2}[\sum_{\nu j \neq \mu i} |V_{\mu i,\nu j}|^2 H_{\nu j}
+ \sum_{\nu j \neq \xi k \neq\mu i} V_{\mu i,\nu j}  V_{\nu j,\xi k}V_{\xi k,\mu i} ].
\end{equation}
To the lowest order,
\begin{align}
	b_{\mu i,2}^2\mathcal{G}^{\mathrm{OD}}_{\mu i,2}(z)= \frac{1}{b_{\mu i,1}^2}\sum_{\nu j\neq \mu i } [\mathcal{R}_{\nu j}(z)\notag \\
	\times\abs{V_{\nu j,\mu i}(H_{\nu j}-a_{\mu i,1})+\sum_{\xi k\neq \mu i,\nu j}V_{\nu j,\xi k}V_{\xi k, \mu i}}^2].
\end{align}

To close the equation, we invoke ETH-type self-averaging at sufficiently deep levels of the hierarchy. The central assumption is that coarse-grained spectral functions become statistically equivalent, differing only by a shift in their energy arguments:
\begin{equation}
\mathcal{R}_{\nu j}(z)
\;\rightarrow\;
\mathcal{R}_{\mu i}(z - \delta_{\nu j}),
\qquad
\delta_{\nu j} = a_{\nu j} - a_{\mu i}.
\end{equation}
This is not an exact identity but a controlled approximation within the ETH regime, where statistical fluctuations between different basis states are suppressed. Analogous to \cref{APPR2DR}, the statistical decorrelation renders the difference between the projected resolvent and the full resolvent negligible. Under this projection, the multi-resolvent contributions can be reorganized as a functional expansion in powers of the shifted resolvent:
\begin{equation}
	 b_{\mu i,n}^2R_{\mu i,n}(z)
\approx
\sum_{\ell \ge 1}
\int d\omega\,
K^{(\ell)}_{\mu i,n}(\omega)\,
\bigl[ \mathcal{R}_{\mu i}(z - \omega) \bigr]^\ell,
\end{equation}
where the kernels $K^{(\ell)}_{\mu i,n}(\omega)$, corresponding to the resummation of all $\mathcal{G}^{(\ell)}$ terms at order $\ell$, encode the statistical structure of interaction matrix elements. Note that the $\ell=1$ (single-resolvent) contribution has been separated out as $\mathcal{G}^{\mathrm{OD}}_{\mu i,n}(z)$, while $\ell\ge2$ captures genuine multi-resolvent correlations.

The closure is thus achieved not by identifying higher-level resolvents with $R_{\mu i}$ itself, but by expressing them as functionals of $R_{\mu i}$ under the ETH assumption. This yields the final, closed nonlinear equation:
\begin{equation}
\mathcal{R}_{\mu i}(z)
=
\frac{1}{
z - a_{\mu i,0}
-
\frac{b_{\mu i,1}^2}{
z - a_{\mu i,1}
-
\sum_{\ell \ge 1}
\int d\omega\,
K^{(\ell)}_{\mu i,1}(\omega)\,
[ \mathcal{R}_{\mu i}(z - \omega) ]^\ell
}
}.
\label{eq:final_closure}
\end{equation}
On the real axis $z=\omega - i0^+$, according to \cref{RkkR}
\begin{equation}
\mathcal{R}_{\mu i}(\omega - i0^+)/\pi
=
H[f_{\mu i}](\omega) + i f_{\mu i}(\omega),
\end{equation}
the functional expansion reduces to combinations of $f_{\mu i}$ and its Hilbert transform, consistent with the structure derived in \cref{app:fourth_order}.

\textbf{Remarks.}
\begin{itemize}
    \item[(i)] The continued fraction provides a non-perturbative backbone that preserves the analytic structure of the resolvent.
    \item[(ii)] The hierarchy expansion determines the kernels $K^{(\ell)}$, encoding nonlocal correlations beyond SCBA.
    \item[(iii)] The closure is achieved not by identifying $R_n$ with $\mathcal{R}_{\mu i}$, but by expressing them as functionals of $\mathcal{R}_{\mu i}$ under ETH-type assumptions.
    \item[(iv)] The hierarchical expansion in Eq.~\eqref{eq:kernel_form} is exact as long as no truncation in $\ell$ is imposed. The closure to a single resolvent via the convolution form in Eq.~\eqref{eq:final_closure} is not exact, but provides a controlled approximation under ETH-type self-averaging, where statistical fluctuations between different basis states are suppressed and spectral functions become translationally equivalent in energy.
    \item[(v)]  The continued-fraction representation is an exact algebraic consequence of the Lanczos procedure. It does not rely on diagrammatic expansions or perturbative assumptions, and it provides a non-perturbative backbone that is distinct from self-consistent diagrammatic resummations.
\end{itemize}

\section{Finite-$\eta$ regularization and discrete-spectrum effects}
\label{sec:finite_eta_limitations}

The self-consistent hierarchy developed in this work is formulated in terms of the analytic resolvent
\begin{equation}
\mathcal{R}_{\mu i}(z)
=
\bra{\phi_{\mu i}} (z-H)^{-1} \ket{\phi_{\mu i}} .
\end{equation}
For finite systems or sparse spectra, it is natural to introduce a finite broadening parameter $\eta>0$ and consider the regularized resolvent
\begin{equation}
\mathcal{R}_{\mu i}(\omega-\mathrm{i}\eta)
=
\sum_n
\frac{
p_n^{\mu i}
}{
\omega-\lambda_n-\mathrm{i}\eta
},
\label{eq:finite_eta_resolvent}
\end{equation}
where
\begin{equation}
p_n^{\mu i}
=
\abs{\braket{\psi_n|\phi_{\mu i}}}^2 .
\end{equation}
Equation~\eqref{eq:finite_eta_resolvent} remains exact for arbitrary finite systems and preserves the analytic structure underlying the continued-fraction hierarchy.

In this finite-$\eta$ formulation, the hierarchy-constrained closures and the continued-fraction recursion can formally be generalized through the analytic continuation
\begin{equation}
z \rightarrow z-\mathrm{i}\eta .
\end{equation}
In particular, the recursive resolvent relation
\begin{equation}
R_n(z)
=
\frac{
1
}{
z-a_n-b_{n+1}^2 R_{n+1}(z)
}
\label{eq:cf_eta}
\end{equation}
remains structurally unchanged. Likewise, the hierarchy-constrained off-diagonal corrections (e.g., $\mathcal{G}^{(3)}_{\mu i}(z)$ in \cref{eq:G3}) may still be written in terms of products of broadened resolvents.

However, although the formal analytic structure survives, finite-$\eta$ regularization introduces an important practical distinction between two different procedures.

\vspace{0.3em}

\noindent
{\bf (i) Direct finite-$\eta$ approach.}
One may directly use \cref{eq:finite_eta_resolvent} in the self-consistent equations, without introducing any continuum approximation or spectral fitting.

\vspace{0.3em}

\noindent
{\bf (ii) Indirect coarse-grained approach.}
Alternatively, one may first fit the discrete spectrum using a smooth spectral ansatz, and then reconstruct a coarse-grained resolvent from the fitted distribution. In this case, the resulting smooth function effectively provides an approximation to the $\eta\to0^+$ continuum limit, even for finite systems.

The second approach is often advantageous because the nonlinear hierarchy structures developed in this work are intrinsically coarse-grained objects. Their improvement over simpler approximations relies on correlated multi-resolvent structures surviving the spectral averaging process. Consequently, the choice of fitting function becomes important. In particular, the fitting ansatz should remain compatible with the nonperturbative structure being analyzed. For example, if the hierarchy closure contains skewness-generating terms (such as $\mathcal{G}^{(3)}_{\mu i}$, which explicitly mixes parity sectors; see \cref{TTG3} and Appendix~\ref{app:no_skewness_mf}), fitting the spectrum using a purely symmetric line shape may artificially suppress the very structure that the hierarchy attempts to capture.

\subsubsection{Large-$\eta$ limitation of nonlinear hierarchy closures}

Although the hierarchy-constrained structures and deeper continued fractions can formally be extended to finite $\eta$, they possess an intrinsic limitation in the large-$\eta$ regime.

To see this, consider the exact resolvent expansion
\begin{equation}
\mathcal{R}(z)
=
\frac{1}{z-H},
\qquad
z=\omega-\mathrm{i}\eta .
\end{equation}
For
\begin{equation}
\eta \gg \abs{\omega},\,\norm{H},
\end{equation}
one obtains the asymptotic expansion
\begin{equation}
\mathcal{R}(z)
=
\frac{\mathrm{i}}{\eta}
+
\frac{H-\omega}{\eta^2}
-
\mathrm{i}\frac{(H-\omega)^2}{\eta^3}
+
O(\eta^{-4}).
\label{eq:R_large_eta}
\end{equation}

Therefore, in the large-$\eta$ regime the resolvent itself becomes dominated by the universal leading contribution
\begin{equation}
\mathcal{R}(z)\sim \frac{\mathrm{i}}{\eta},
\end{equation}
while the sensitivity to higher spectral moments is progressively suppressed.

For the mean-field (SCBA-type) closure,
\begin{equation}
\mathcal{G}^{\mathrm{OD}}_{\mu i}(z)
=
\sum_{\nu j \neq \mu i} \abs{V_{\mu i,\nu j}}^2 \mathcal{R}_{\nu j}(z),
\end{equation}
\cref{eq:R_large_eta} yields
\begin{equation}
\mathcal{G}^{\mathrm{OD}}_{\mu i}(z)
\sim
\mathrm{i}\frac{\Gamma_2}{\eta}
+
O(\eta^{-2}),
\qquad
\Gamma_2=\sum_{\nu j\neq\mu i} \abs{V_{\mu i,\nu j}}^2 .
\end{equation}
Thus the mean-field self-energy depends primarily on the lowest local variance scale and naturally reduces to a stable coarse-grained damping structure.

By contrast, hierarchy-constrained corrections involve products of multiple resolvents and consequently higher-order spectral moments. The leading such term,
\begin{equation}
\mathcal{G}^{(3),\text{D}}_{\mu i}(z)
=
\sum_{\xi k\neq\nu j\neq\mu i}
V_{\mu i,\xi k} V_{\xi k,\nu j} V_{\nu j,\mu i}\,
\mathcal{R}_{\xi k}(z)\,\mathcal{R}_{\nu j}(z),
\end{equation}
contains an explicit product of two resolvents. In the large-$\eta$ expansion, each factor behaves as $\mathcal{R}\sim\mathrm{i}/\eta+O(\eta^{-2})$, giving
\begin{equation}
\mathcal{G}^{(3)}_{\mu i}(z)
\sim
-\frac{\Gamma_3}{\eta^2}
+
O(\eta^{-3}),
\end{equation}
where $\Gamma_3$ collects the triple vertex sums. Higher-order terms $\mathcal{G}^{(\ell)}_{\mu i}$ (see \cref{HHOT}) scale as $\eta^{-\ell+1}$ or higher. Likewise, deeper continued-fraction structures become sensitive to higher Krylov moments through the recursion coefficients $b_n$, which encode energy-mismatch information that requires spectral resolution finer than $\eta$ to be physically meaningful.

\Cref{eq:R_large_eta} shows that large $\eta$ suppresses the spectral resolution required to reliably resolve these higher-order structures. The resolvent becomes nearly energy-independent on scales smaller than $\eta$, while the hierarchy corrections still attempt to distinguish increasingly fine spectral information. Consequently, the nonlinear corrections become systematically biased in the large-$\eta$ regime.

This effect becomes particularly pronounced in self-consistent iterations. The mean-field self-energy depends linearly on the resolvent,
\begin{equation}
\mathcal{G}^{\mathrm{OD}}_{\mu i}
\propto \mathcal{R},
\end{equation}
whereas nonlinear hierarchy corrections contain products of broadened propagators. As a result, the hierarchy corrections become increasingly sensitive to regularization-induced distortions, resonance broadening, and finite-size fluctuations. Concretely, in a self-consistent cycle the SCBA feeds a smeared $\mathcal{R}$ into a linear functional, which preserves the coarse-grained structure; the hierarchy, in contrast, multiplies smeared resolvents together, thereby squaring the regularization error and propagating it nonlinearly through the iterative loop.

From the continued-fraction viewpoint, finite $\eta$ also suppresses deep Krylov-space propagation. Indeed, the recursion
\begin{equation}
R_n(z)
=
\frac{
1
}{
\omega-a_n-\mathrm{i}\eta-b_{n+1}^2 R_{n+1}(z)
}
\end{equation}
shows that each Krylov layer acquires an effective damping scale $\eta$. Consequently, deep continued-fraction structures become progressively less physically resolvable as $\eta$ increases, even though the formal recursion remains exact.

Therefore, in the large-$\eta$ regime, the nonlinear hierarchy corrections may over-renormalize spectral structures that have already been smeared out by the broadening itself. In this situation, the effective advantage of hierarchy corrections may diminish, and the coarse-grained SCBA structure can become comparatively more robust.

\subsubsection{Practical implication}

The above analysis suggests that the direct finite-$\eta$ approach is most suitable for sufficiently large systems, where the typical level spacing $\delta_{\rm LS}$ remains much smaller than the physically relevant broadening scale (cf.\ the scale-separation condition \eqref{eq:eta_window}). In this regime, one may use relatively small $\eta$, allowing the hierarchy structures to survive coarse graining without introducing large regularization bias.

For smaller systems, however, direct finite-$\eta$ self-consistency may become unreliable because the nonlinear hierarchy structures become overly sensitive to pole broadening and finite-size fluctuations. In such situations, an indirect coarse-grained approach---namely fitting the spectrum using a physically motivated smooth ansatz before performing the self-consistent analysis---can be substantially more stable.

Importantly, the fitting ansatz should remain compatible with the nonperturbative structure being studied. Otherwise, the fitting procedure itself may artificially suppress the hierarchy structure that the self-consistent theory attempts to capture.

\section{Diagonal Closure Approximation}
\label{app:DCA}

In the main text, the recursive expansion of the cross-correlated self-energy
\cref{HHOT} yields terms of the generic form
\begin{equation}\label{DCAdef}
    \mathcal{G}^{(\ell)}_{\mu i}(z)
    = \sum_{\alpha_1\neq\cdots\neq\alpha_{\ell-1}\neq\mu i}
    V^{(\ell)}_{\mu i,\alpha_{\ell-1},\dots,\alpha_1}
    \prod_{a=1}^{\ell-1}
    \mathcal{R}_{\alpha_a}^{(\mu i,\alpha_1,\dots,\alpha_{a-1})}(z)
\end{equation}
for $ \ell\ge 2$. When $\ell=2$, its corresponds to $\mathcal{G}^{\mathrm{OD}}_{\mu i}(z) + \Delta\mathcal{G}_{\mu i}(z)$ and $\ell\ge3$
to the multi-resolvent cross-correlated terms.
Each projected diagonal resolvent is defined by
\begin{equation}
    \mathcal{R}_{\nu j}^{(\mu i,\dots)}(z)
    = \bra{\phi_{\nu j}}
    \frac{1}{z-\Phi_{\mu i}\cdots\Phi H\,\Phi\cdots\Phi_{\mu i}}
    \ket{\phi_{\nu j}},
\end{equation}
with $\Phi_{\mu i}=I-|\phi_{\mu i}\rangle\langle\phi_{\mu i}|$ the projector
that removes the basis state $|\phi_{\mu i}\rangle$ from the Hilbert space.
Equation~\eqref{DCAdef} is exact; however, the projected resolvents couple to the
cavity Hamiltonians $H^{(\mu i,\dots)}:=\Phi_{\mu i}\cdots H\cdots\Phi_{\mu i}$
and are not directly expressible in terms of quantities computable from the
mean-field theory alone.
The diagonal closure approximation (DCA) closes this hierarchy by replacing
every projected diagonal resolvent with the corresponding full diagonal resolvent:
\begin{equation}\label{DCAapprox}
    \mathcal{R}_{\alpha}^{(\mu i,\alpha_1,\dots,\alpha_{m})}(z)
    \;\approx\;
    \mathcal{R}_{\alpha}(z).
\end{equation}
Under this approximation, each term of the hierarchy reduces to
\begin{align}
    \mathcal{G}^{(\ell)}_{\mu i}(z)
    &\approx
    \sum_{\alpha_1\neq\cdots\neq\alpha_{\ell-1}\neq\mu i}
    V^{(\ell)}_{\mu i,\dots,\alpha_1}
    \prod_{a=1}^{\ell-1}
    \mathcal{R}_{\alpha_a}(z)
    \equiv \mathcal{G}^{(\ell),\text{D}}_{\mu i}(z),
    \label{DCAred}
\end{align}
which is expressed entirely in terms of ordinary (full) diagonal resolvents.

For the Krylov-continued-fraction hierarchy at level $k\geq 2$ (corresponding
to the tail subspace $H^{[\ge n]}$ with $n=k-1$ in the Lanczos recurrence),
the $\ell^{\text{th}}$-order term retains the same structural form of $\ell-1$
projected-resolvent factors.  Its effective coupling, however, is no longer
a pure product of interaction matrix elements but acquires contributions from
the Lanczos coefficients $a_{\mu i,m},b_{\mu i,m}^2$ and diagonal shifts
$(H_{\nu j}-a_{\mu i,m})$.  We denote the generic term by 
\begin{align}\label{eq:full_hierarchy_k}
    \mathcal{G}^{(\ell)}_{\mu i,k}(z)
    =\sum_{\alpha_1\neq\cdots\neq\alpha_{\ell-1}}
    \mathfrak{C}^{(\ell)}_{\mu i,k;\,\alpha_1,\dots,\alpha_{\ell-1}}
    \prod_{a=1}^{\ell-1}
    \mathcal{R}_{\alpha_a}^{(\mu i,\dots,\alpha_{a-1})[\ge n]}(z),
\end{align}
where the effective coupling coefficients $\mathfrak{C}^{(\ell)}_{\mu i,k}$
have the following properties:
    For $k=1$ (the base level, $n=0$), 
    $\mathfrak{C}^{(\ell)}_{\mu i,1}=V^{(\ell)}_{\mu i,\alpha_{\ell-1},\dots,\alpha_1}
    =\prod_{r=1}^{\ell}V_{\alpha_{r-1},\alpha_r}$ (with $\alpha_0=\alpha_\ell=\mu i$),
    i.e., the pure $V$-product structure given explicitly in
    \cref{app:fourth_order}.
    For $k\ge2$, $\mathfrak{C}^{(\ell)}_{\mu i,k}$ contains, besides
    products of $V$ matrix elements, inverse powers of the Lanczos coefficients
    $b_{\mu i,m}$ and diagonal-energy shifts of the type
    $(H_{\nu j}-a_{\mu i,m})$.
    The precise form follows from the expansion coefficients of the Krylov
    vectors in the original basis together with the Lanczos recurrence
    (see \cref{app:hierarchy_closure} for the explicit $k=2,\ell=1$
    example).

Under the DCA, the projected resolvents are replaced by full resolvents,
yielding
\begin{align}
    \mathcal{G}^{(\ell)}_{\mu i,k}(z)
    \approx \sum_{\alpha_1\neq\cdots\neq\alpha_{\ell-1}}
    \mathfrak{C}^{(\ell)}_{\mu i,k;\dots}
    \prod_{a=1}^{\ell-1}
    \mathcal{R}_{\alpha_a}(z)
    \;:=\; \mathcal{G}^{(\ell),\text{D}}_{\mu i,k}(z).
\end{align}
The mixed $V$-$H$ structure of $\mathfrak{C}^{(\ell)}_{\mu i,k}$ is
\emph{unchanged} by the DCA---the closure approximation acts only on the
resolvent arguments, not on the coupling coefficients.
The DCA thus converts the formally exact but practically intractable projected
hierarchy into a closed system of equations for diagonal resolvents.

\subsection{Physical origin: single-state backreaction in the ETH regime}

The DCA relies on delocalized, self-averaging propagation in exponentially 
large Hilbert spaces---a defining characteristic of nonintegrable many-body 
systems in the ETH regime. In such systems, the off-diagonal propagator
\begin{equation}
    G_{\nu\mu}(z) := \bra{\phi_\nu} (z-H)^{-1} \ket{\phi_\mu},
    \qquad \nu\neq\mu,
\end{equation}
is expected to be strongly suppressed relative to the diagonal resolvent,
$|G_{\nu\mu}|\ll|G_{\nu\nu}|$, because propagation rapidly spreads over 
exponentially many states and the return amplitude to any single removed 
state is diluted by the entropy of the bath.

The physical content of the DCA is most transparent when expressed through
the exact cavity correction identity (derived rigorously in 
 \cref{sec:DCA_resolvent_id}):
\begin{align}\label{CavityCorr}
    \mathcal{R}_{\nu j}(z) - \mathcal{R}^{(\mu i)}_{\nu j}(z)
    = \mathcal{R}_{\nu j,\mu i}(z)\,
      \bra{\phi_{\mu i}} H\,G^{(\mu i)}(z)\ket{\phi_{\nu j}},
\end{align}
where terms involving \(\bra{\phi_{\nu j}} G^{(\mu i)} \ket{\phi_{\mu i}}\) 
vanish because \(\Phi_{\mu i}\ket{\phi_{\mu i}}=0\). 
Expanding \(H=H_0+V\) and using 
\(\bra{\phi_{\mu i}}G^{(\mu i)}\ket{\phi_{\nu j}}=0\) (for \(\nu j\neq\mu i\)),
the remaining factor becomes
\(\bra{\phi_{\mu i}} H G^{(\mu i)}\ket{\phi_{\nu j}}
 = \sum_{\gamma k} V_{\mu i,\gamma k}\,\mathcal{R}_{\gamma k,\nu j}^{(\mu i)}(z)\).
Equation~\eqref{CavityCorr} shows that the cavity correction factorizes into 
an off-diagonal full resolvent \(\mathcal{R}_{\nu j,\mu i}\) and a 
matrix element \(\bra{\phi_{\mu i}} H G^{(\mu i)}\ket{\phi_{\nu j}}\).
The latter expands as \(\sum_{\gamma k}V_{\mu i,\gamma k}\mathcal{R}_{\gamma k,\nu j}^{(\mu i)}\);
its dominant contribution comes from the diagonal channel \(\gamma k=\nu j\), 
giving \(V_{\mu i,\nu j}\mathcal{R}_{\nu j}^{(\mu i)}\sim e^{-S/2}\).
Combined with \(\mathcal{R}_{\nu j,\mu i}\sim e^{-S/2}\) (ETH), 
the cavity correction scales as
\begin{equation}\label{CCS}
    \mathcal{R}^{(\mu i)}_{\nu j}(z) - \mathcal{R}_{\nu j}(z) 
    \sim e^{-S},
\end{equation}
with off-diagonal (\(\gamma k\neq\nu j\)) channels further suppressed 
under incoherent-summation (random-walk) arguments. \Cref{CCS} suggests the removal of a single basis state from the Hilbert space is expected
to produce a correction that vanishes exponentially with the entropy $S$.
This entropy-dilution picture provides the physical motivation for the DCA:
single-state cavity backreaction is suppressed by the exponentially large
phase space of the bath, rather than being assumed small on dimensional
grounds alone.

However, a crucial subtlety must be addressed. Although the dominant 
(diagonal-\(\gamma k\)) return channel contributes \(\sim e^{-S}\) 
(from \(\mathcal{R}_{\nu j,\mu i}V_{\mu i,\nu j}\mathcal{R}_{\nu j}^{(\mu i)}\)), 
the number of off-diagonal (\(\gamma k\neq\nu j\)) channels grows as \(e^{S}\)---the same exponential factor that suppresses individual
amplitudes also proliferates the number of contributing paths.  If all
return amplitudes were to add \emph{coherently} (with the same phase),
the total cavity correction would be $O(1)$ and the DCA would fail
catastrophically.  The resolution lies in the statistical structure of
the off-diagonal resolvent elements under the ETH: the phases (or signs)
of $G_{\nu\mu}$ for different intermediate states $\nu$ fluctuate
pseudo-randomly and are mutually incoherent~\cite{AKP16,FK19}.
Consequently, the sum over $e^{S}$ return channels behaves as a random
walk rather than a coherent superposition: individual terms add in
\emph{quadrature}, suggesting a typical fluctuation scale of order
\(\sim \sqrt{e^{S}}\cdot e^{-S} = e^{-S/2}\) for the off-diagonal
propagator, and the cavity correction 
\(\mathcal{R}_{\nu j,\mu i}\bra{\phi_{\mu i}}HG^{(\mu i)}\ket{\phi_{\nu j}}\)
is correspondingly suppressed to \(e^{-S}\) by the product of two 
entropy-diluted factors.  The cavity correction is therefore
\emph{self-averaging} rather than coherently amplified---a direct
consequence of the random-matrix-like structure of off-diagonal
matrix elements in nonintegrable systems.

The $e^{-S}$ scaling can be motivated from the ETH structure of the 
interaction matrix elements~\cite{AKP16}:
\begin{equation}
    |V_{\mu i,\nu j}|^2 \sim e^{-S(\epsilon^+_{\mu\nu})}
    f^2(\epsilon^+_{\mu\nu},\delta),\qquad
    \mu\neq\nu,
\end{equation}
together with the fact that the number of states connected to a given
intermediate state $\nu j$ grows as $e^{S}$, so that
the mean-field self-energy
\begin{equation}
    \mathcal{G}^{\mathrm{OD}}_{\mu i}(z)
    = \sum_{\nu j\neq\mu i} |V_{\mu i,\nu j}|^2 \mathcal{R}_{\nu j}(z)
    \sim O(1)
\end{equation}
while the contribution of a single intermediate state to this sum is
\begin{equation}
    |V_{\mu i,\nu j}|^2 \mathcal{R}_{\nu j}(z) \sim e^{-S}.
\end{equation}
For the Krylov hierarchy ($k\ge2$), the removed subspace
$\mathcal{K}_{\mu i}^{(n)}$ has dimension $n$, and the same scaling
argument applies with the substitution
$e^{-S}\to n\,e^{-S}$---still exponentially suppressed since
$n$ is finite while $S\to\infty$; a similar incoherent-summation picture is expected to apply to the $n$ Krylov directions.

In the thermodynamic limit $S\to\infty$, the feedback of any single
removed state onto the remaining bath is expected to become
negligible---a manifestation of self-averaging in nonintegrable
quantum systems.
This is analogous to the physical mechanism that underlies
the cavity construction in dynamical mean-field theory (DMFT)
and the Bethe-lattice Green function: removing one node from an
infinitely connected graph does not alter the global environment.
We emphasize, however, that the above scaling argument is an
ETH-inspired expectation rather than a rigorous theorem; the precise
prefactor and the validity of the incoherent-summation picture depend
on the system being in a delocalized, chaotic regime with 
exponentially large connectivity (see~\cref{sec:DCA_validity}
for detailed validity conditions and failure mechanisms).

\subsection{Rigorous structure from the resolvent identity}\label{sec:DCA_resolvent_id}

The DCA can be given a precise algebraic foundation using the
resolvent (second-resolvent) identity.
Define the cavity Hamiltonian
\begin{equation}
    H^{(\mu i)} := \Phi_{\mu i} H \Phi_{\mu i},
\end{equation}
and its resolvent
\begin{equation}
    G^{(\mu i)}(z) := \frac{1}{z - H^{(\mu i)}}.
\end{equation}
The full Hamiltonian is related to the cavity Hamiltonian by
\begin{equation}
    H = H^{(\mu i)} + \delta H_{\mu i},
\end{equation}
where the perturbation
\begin{equation}
    \delta H_{\mu i}
    = |\phi_{\mu i}\rangle\langle\phi_{\mu i}| H
    + H|\phi_{\mu i}\rangle\langle\phi_{\mu i}|
    - |\phi_{\mu i}\rangle\langle\phi_{\mu i}|H|\phi_{\mu i}\rangle
      \langle\phi_{\mu i}|
\end{equation}
is a rank-2 operator on the Hilbert space.
The resolvent identity $G = G^{(\mu i)} - G^{(\mu i)}\,\delta H_{\mu i}\,G$
(applied to both $G$ and $G^{(\mu i)}$) yields the exact relation
\begin{equation}\label{exactdiff}
    \mathcal{R}_{\nu j}(z) - \mathcal{R}^{(\mu i)}_{\nu j}(z)
    =
    \bra{\phi_{\nu j}} G(z)\,\delta H_{\mu i}\,G^{(\mu i)}(z)\ket{\phi_{\nu j}}.
\end{equation}
Because $\delta H_{\mu i}$ contains projectors onto $|\phi_{\mu i}\rangle$,
the right-hand side is controlled by the off-diagonal resolvent elements
\begin{equation}
    \mathcal{R}_{\nu j,\mu i}(z)
    := \bra{\phi_{\nu j}} G(z) \ket{\phi_{\mu i}}.
\end{equation}
Explicit evaluation of~\eqref{exactdiff} gives
\begin{align}\label{bounds}
    \mathcal{R}_{\nu j}(z) - \mathcal{R}^{(\mu i)}_{\nu j}(z)
    = \mathcal{R}_{\nu j,\mu i}(z)\,
      \bra{\phi_{\mu i}} H\,G^{(\mu i)}(z)\ket{\phi_{\nu j}},
\end{align}
where terms involving \(\bra{\phi_{\nu j}} G^{(\mu i)} \ket{\phi_{\mu i}}\) 
vanish because \(\Phi_{\mu i}\ket{\phi_{\mu i}}=0\), and for \(\nu j\neq\mu i\)
the identity \(\bra{\phi_{\mu i}}G^{(\mu i)}\ket{\phi_{\nu j}}=0\) reduces the
\(H\)-matrix element to \(\sum_{\gamma k}V_{\mu i,\gamma k}\mathcal{R}_{\gamma k,\nu j}^{(\mu i)}\).
Under the ETH, \(\mathcal{R}_{\nu j,\mu i}\sim e^{-S/2}\) and 
\(V_{\mu i,\nu j}\sim e^{-S/2}\); the dominant (\(\gamma k=\nu j\)) 
contribution to the sum yields an overall scaling \(e^{-S}\) 
for the cavity correction. Off-diagonal (\(\gamma k\neq\nu j\)) channels 
are further suppressed under the incoherent-summation picture.
Hence the product in~\eqref{bounds} is of order $e^{-S}$, and the DCA
$\mathcal{R}^{(\mu i)}_{\nu j}\approx\mathcal{R}_{\nu j}$ becomes
increasingly accurate in the thermodynamic limit, with the error
controlled by the exponentially small off-diagonal return amplitude.
We caution that this scaling is an ETH-based estimate; the DCA is not
claimed as a rigorous theorem but as a physically motivated closure
whose validity conditions are systematically examined in
\cref{sec:DCA_validity}.

For the Krylov hierarchy the cavity Hamiltonian is
$H_{\mu i}^{[\ge n](\mu i)}:=\Pi_{\mu i}^{[\ge n]}\Phi_{\mu i}
H\,\Phi_{\mu i}\Pi_{\mu i}^{[\ge n]}$,
where $\Pi_{\mu i}^{[\ge n]}$ projects onto $H_{\mu i}^{[\ge n]}$.
The resolvent identity generalizes with \(\delta H\) now of rank
\(2n\) rather than \(2\). The extra terms involve 
\(\mathcal{R}_{\nu j,\mu i}\) multiplied by matrix elements of 
\(H G^{(\mu i)[\ge n]}\); the dominant diagonal contributions again 
scale as \(e^{-S}\), while off-diagonal channels proliferate as \(e^{S}\) 
and require the same incoherent-summation caveat.

\subsection{Krylov recursion perspective}

The continued-fraction representation~\eqref{eq:cf_recurrence}
builds a one-dimensional chain of Krylov vectors from the seed
$\ket{\phi_{\mu i}}$ via the Lanczos recurrence. The tail resolvent
$\mathcal{R}_{\mu i,n}(z)$ describes propagation in the subspace
$H_{\mu i}^{[\ge n]}$, which is orthogonal to the first $n$
Krylov directions $\ket{\phi_{\mu i}},\ket{\phi_{\mu i,1}},\dots,\ket{\phi_{\mu i,n-1}}$.

In the projected hierarchy~\eqref{eq:full_hierarchy_k} at level $k\ge 2$
($n=k-1\ge 1$), the cavity resolvent that appears---and which the DCA
replaces by the full resolvent---is precisely
$\mathcal{R}_{\nu j}^{(\mu i)[\ge n]}(z)$:
the resolvent of $\ket{\phi_{\nu j}}$ in the subspace from which
\emph{both} the seed state $\ket{\phi_{\mu i}}$
and its first $n$ Krylov descendants have been projected out.
Since $\ket{K_{\mu i,0}}=\ket{\phi_{\mu i}}$, this amounts to
removing a total of $n$ linearly independent Krylov directions.
The DCA error is therefore measured by
\begin{equation}\label{DCA_Krylov_diff}
    \mathcal{R}_{\nu j}(z)
    - \mathcal{R}_{\nu j}^{(\mu i)[\ge n]}(z),
\end{equation}
which captures the contribution of all propagation paths of
$\ket{\phi_{\nu j}}$ that visit the $n$-dimensional Krylov subspace
$\mathcal{K}_{\mu i}^{(n)}:=\operatorname{span}\{\ket{\phi_{\mu i}},\dots,\ket{\phi_{\mu i,n-1}}\}$:

\[
    \ket{\phi_{\nu j}}
    \;\longrightarrow\;
    \mathcal{K}_{\mu i}^{(n)}
    \;\longrightarrow\;
    \ket{\phi_{\nu j}}.
\]

The physical suppression of this difference follows from two
concurrent mechanisms.
First, \emph{entropic dilution}: the $n$-dimensional Krylov subspace
occupies a fractional volume $n\,e^{-S}$ of the total Hilbert space,
so the measure of return paths through it vanishes in the thermodynamic
limit for any finite $n$.
Second, \emph{Krylov delocalization}: in a nonintegrable (ergodic)
system, each Lanczos step mixes the Krylov vector with an exponentially
growing number of original basis states~\cite{AKP16};
the vectors $\ket{\phi_{\mu i,m}}$ for $m\ge 1$ are superpositions of
$\sim e^{S}$ basis states with pseudo-random coefficients.
Return amplitudes from $\ket{\phi_{\nu j}}$ to these specific
superpositions are therefore suppressed by an additional factor
$e^{-S/2}$ from each off-diagonal transition.

Consequently,
\begin{equation}
    \mathcal{R}_{\nu j}^{(\mu i)[\ge n]}(z)
    \;\approx\;
    \mathcal{R}_{\nu j}(z),
\end{equation}
and the approximation \emph{improves with increasing $n$}:
deeper Krylov levels project out more directions, further diluting
the cavity subspace and shrinking the measure of the excised
return paths.
In the base level $k=1$ ($n=0$), only the single seed direction
$\ket{\phi_{\mu i}}$ is removed, and the
difference~\eqref{DCA_Krylov_diff} reduces to the elementary
DCA~\eqref{DCAapprox} discussed in the preceding subsections.

\subsection{Behavior under the hierarchy}

The accuracy of the DCA is \emph{not uniform} across the hierarchy.
Consider a projected resolvent at depth $m$:
\begin{equation}
    \mathcal{R}_{\alpha}^{(\mu i,\alpha_1,\dots,\alpha_{m})}(z).
\end{equation}
The corresponding cavity Hamiltonian has $m$ basis states removed from
the Hilbert space.
Since the total Hilbert space dimension is $e^{S}$ and $m$ is a finite
integer, the fraction of removed states is $m e^{-S}$, which remains
exponentially small in the thermodynamic limit.
More importantly, at higher levels of the hierarchy the propagator
has already explored a large portion of the Hilbert space; the
additional removal of a finite number of states further dilutes the
cavity information.
Thus, counter-intuitively, the DCA becomes \emph{more} accurate at
higher orders of the hierarchy, not less. The same conclusion holds, and is in fact reinforced, for the
Krylov-continued-fraction hierarchy ($k\ge 2$).  There the cavity
subspace is reduced not only by a finite set of basis states but
additionally by $n=k-1$ Krylov directions, further shrinking the
fractional volume of the excised subspace to
$(m+n)\,e^{-S}$.  The entropic suppression of residual backreaction
is thus stronger than in the base-level ($k=1$) case for the same
number of hierarchy steps.
This property is shared with Bethe-lattice cavity methods:
locally removing finitely many nodes does not alter an infinite graph.

However, this reasoning assumes that propagation is delocalized.
If the system is in a localized phase or if certain rare paths
dominate the transport, the DCA may fail at any level, as will be discussed later.

\subsection{Inter-resolvent correlations and improved DCA closure}
\label{sec:DCA_multiR_correction}
The continued-fraction (Lanczos) closure discussed in the preceding
sections already embodies a nonperturbative resummation of propagation
paths {\em within} the single-resolvent sector: each diagonal resolvent
$\mathcal{R}_\alpha(z)$ encodes infinitely many recursive returns,
memory effects, and intrinsic non-Lorentzian spectral scales.  The
multi-resolvent hierarchy introduces a qualitatively distinct
ingredient---{\em inter-resolvent interference} between different
resolvent factors $\mathcal{R}_\nu\mathcal{R}_\xi$---which is absent
from any single-resolvent closure and is the microscopic origin of
parity mixing and spectral skewness.  The DCA~\eqref{DCAapprox}, which
replaces projected resolvents by full resolvents, affects both sectors
but with structurally different consequences: in the single-resolvent
sector the error is entropically diluted, whereas in multi-resolvent
products it disrupts the nested mutual-exclusion constraints among the
resolvent factors.  This section analyzes the latter regime and
develops systematically improvable corrections to the pure DCA.

\subsubsection{Topological origin of the DCA error in multi-resolvent products}

The DCA~\eqref{DCAapprox} replaces each projected diagonal resolvent
by its full counterpart.  To understand why this replacement is
structurally more consequential for multi-resolvent products than for
single-resolvent quantities, one must examine {\em what} the projection
actually removes.

A projected resolvent $\mathcal{R}_{\alpha}^{(\mathcal{S})}(z)$, with
$\mathcal{S}=\{\mu i,\alpha_{1},\dots,\alpha_{m}\}$ the set of removed
basis states, is the diagonal element of
$(z-\Phi_{\mathcal{S}}H\Phi_{\mathcal{S}})^{-1}$, where
$\Phi_{\mathcal{S}}$ projects onto the orthogonal complement of
$\operatorname{span}\{|\phi_{\beta}\rangle:\beta\in\mathcal{S}\}$.
The crucial point is that the projection does not merely remove
external summation indices from the hierarchy expansion; it modifies
the internal Dyson resummation topology of each propagator itself.
Through its own Dyson (or continued-fraction) structure, this propagator
internally resums an infinite class of propagation paths---{\em but
exclusively within the cavity Hilbert space}.  All paths that visit any
state $\beta\in\mathcal{S}$ and return to $\alpha$ are topologically
forbidden: they are not suppressed by a small parameter but identically
absent from the internal resummation.

The full resolvent $\mathcal{R}_{\alpha}(z)$, by contrast, is the diagonal
element of $(z-H)^{-1}$ and internally resums propagation in the
{\em unrestricted} Hilbert space.  Its Dyson resummation inherently
contains the backtracking channels
$|\phi_{\alpha}\rangle\to|\phi_{\beta}\rangle\to|\phi_{\alpha}\rangle$
for {\em every} $\beta$, including those that should have been excluded.

Consequently, replacing $\mathcal{R}_{\alpha}^{(\mathcal{S})}$ by
$\mathcal{R}_{\alpha}$ is {\em not} merely a local magnitude
approximation of order $e^{-S}$; it is a {\bf topological mutation}
of the propagator: the class of allowed paths changes from
``propagation avoiding $\mathcal{S}$'' to ``propagation unrestricted.''
We emphasize that this topological distinction concerns the
{\em inter-resolvent} path constraints: the backbone of recursive
returns internal to each resolvent factor is already captured by the
continued-fraction structure of the full resolvent.  What the
projection removes and the DCA erroneously restores are the
cross-resolvent backtracking channels that couple distinct factors in
the hierarchy product.
The error resides in the internal Dyson resummation of each resolvent
factor, hidden beneath its compact analytic form.  In this sense, the
projected hierarchy can be understood as a {\bf nonbacktracking
resolvent hierarchy} in a generalized sense (with progressively
expanding forbidden sets), and the DCA as the ordinary (unconstrained)
resolvent approximation to it.

In single-resolvent quantities such as the mean-field self-energy
$\mathcal{G}_{\mu i}^{\mathrm{OD}}$, this topological error affects
one propagator and is diluted by the entropy of the bath, justifying
the DCA. Note that the continued-fraction structure of each resolvent already
encodes nontrivial single-channel path correlations: recursive
returns, non-Lorentzian broadening, and memory effects, even at the
single-resolvent level; the topological error under discussion here
concerns the {\em additional} inter-resolvent interference that
distinguishes multi-resolvent products.

In the $\ell$-th order hierarchy term
$\mathcal{G}_{\mu i}^{(\ell)}$ of \cref{DCAdef}, however, the product
$\prod_{a=1}^{\ell-1}\mathcal{R}_{\alpha_{a}}^{(\mathcal{S}_{a})}$
encodes a {\em nested} set of topological constraints:
\begin{equation}\label{eq:nested_constraints}
    \mathcal{S}_{1}\subset\mathcal{S}_{2}\subset\cdots\subset\mathcal{S}_{\ell-1},
    \qquad
    \mathcal{S}_{a}=\{\mu i,\alpha_{1},\dots,\alpha_{a-1}\}.
\end{equation}
Each successive propagator is restricted to a strictly smaller Hilbert
space, reflecting the physical requirement that the propagation path
$0\to 1\to 2\to\cdots\to (\ell-1)\to 0$ must never revisit an earlier
intermediate state.
The DCA simultaneously lifts {\em all} $\ell-1$ of these topological
constraints, converting a chain of mutually exclusive propagators
into a product of fully connected ones.  The resulting error is
therefore not additive in $\ell-1$ independent $e^{-S}$ factors but
involves the reopening of cross-talk channels among the resolvent
factors, whose number grows with the hierarchy depth.

\subsubsection{Cavity-inspired relation for the projected resolvent}

To estimate the effect of restoring the correct topological constraints,
we compare the inverse of the projected resolvent to that of the full
resolvent.
From the Feshbach identity~\eqref{eq:CSEQ} applied to the cavity
Hamiltonian, one has
\begin{equation}\label{eq:cavity_inverse}
    \bigl[\mathcal{R}_{\alpha}^{(\mathcal{S})}(z)\bigr]^{-1}
    = z - a_{\alpha} - V_{\alpha}
    - \mathcal{G}_{\alpha}^{(\mathcal{S})}(z),
\end{equation}
where $\mathcal{G}_{\alpha}^{(\mathcal{S})}(z)$ is the self-energy
evaluated with the states in $\mathcal{S}$ removed.
The difference from the full inverse is governed by the return channels
to $\mathcal{S}$ that the projection forbids.  This difference admits
the physically motivated decomposition
\begin{align}\label{eq:cavity_diff_approx}
    \bigl[\mathcal{R}_{\alpha}^{(\mathcal{S})}(z)\bigr]^{-1}
    - \bigl[\mathcal{R}_{\alpha}(z)\bigr]^{-1}
    &\approx
    \sum_{\beta\in\mathcal{S}} |V_{\alpha,\beta}|^{2}\,
       \mathcal{R}_{\beta}(z)
       \;+\;
       \Delta\mathcal{G}_{\alpha}^{\mathcal{S},\,\mathrm{CC}}(z),
\end{align}
where the first term isolates the direct single-scattering return
channels, and $\Delta\mathcal{G}_{\alpha}^{\mathcal{S},\,\mathrm{CC}}(z)$
collects cross-correlated return processes involving multiple
intermediate states and interference channels.  Its quantitative
importance depends on the degree of coherent multi-resolvent
interference and therefore cannot generally be reduced to a simple
local self-energy correction; in strongly resonant sectors it may
be comparable to the direct-return contribution.
As a physically motivated starting point, we retain the direct-channel
contribution and obtain the approximate cavity-inspired inverse relation
\begin{equation}\label{eq:cavity_diff_leading}
    \bigl[\mathcal{R}_{\alpha}^{(\mathcal{S})}(z)\bigr]^{-1}
    \approx
    \bigl[\mathcal{R}_{\alpha}(z)\bigr]^{-1}
    + \sum_{\beta\in\mathcal{S}} |V_{\alpha,\beta}|^{2}\,
      \mathcal{R}_{\beta}(z).
\end{equation}
Equation~\eqref{eq:cavity_diff_leading} expresses algebraically the
topological statement: the full propagator's inverse contains additional
contributions from return paths to $\mathcal{S}$ that the projected
propagator lacks.

\subsubsection{Leading local-return correction}

Expanding \eqref{eq:cavity_diff_leading} to first order in the return
couplings motivates the following leading local-return subtraction:
\begin{equation}\label{eq:linear_cavity}
    \mathcal{R}_{\alpha}^{(\mathcal{S})}(z)
    \approx
    \mathcal{R}_{\alpha}(z)
    - \mathcal{R}_{\alpha}(z)\,
      \Bigl(\sum_{\beta\in\mathcal{S}} |V_{\alpha,\beta}|^{2}\,
            \mathcal{R}_{\beta}(z)\Bigr)\,
      \mathcal{R}_{\alpha}(z).
\end{equation}
Diagrammatically, the subtraction term removes the leading class of
forbidden paths: those in which the propagator $\alpha$ makes a single
excursion to a state $\beta\in\mathcal{S}$ and returns
($\alpha\to\beta\to\alpha$).
Inserting \eqref{eq:linear_cavity} into the hierarchy product
\cref{DCAdef} and retaining only the first-order subtraction yields
\begin{align}\label{eq:DCA_additive_correction}
    \mathcal{G}_{\mu i}^{(\ell)}(z)
    \approx
    \mathcal{G}_{\mu i}^{(\ell),\,\mathrm{D}}(z)
    \;-\;
    \Delta_{\mathrm{ret}}^{(\ell)}(z),
    \\
    \Delta_{\mathrm{ret}}^{(\ell)}(z)
    :=
    \sum_{\alpha_{1}\neq\cdots\neq\alpha_{\ell-1}\neq\mu i}
    V^{(\ell)}_{\mu i,\alpha_{\ell-1},\dots,\alpha_{1}}\notag\\
  \times   \sum_{a=1}^{\ell-1}\;
    \sum_{\beta\in\mathcal{S}_{a}}
    |V_{\alpha_{a},\beta}|^{2}\,
    \mathcal{R}_{\alpha_{a}}(z)\,
    \mathcal{R}_{\beta}(z)\,
    \mathcal{R}_{\alpha_{a}}(z)
    \prod_{\substack{b=1\\ b\neq a}}^{\ell-1}
    \mathcal{R}_{\alpha_{b}}(z),
    \notag
\end{align}
with $\mathcal{S}_{a}=\{\mu i,\alpha_{1},\dots,\alpha_{a-1}\}$.

The term $\Delta_{\mathrm{ret}}^{(\ell)}$ enumerates all configurations
in which exactly one intermediate propagator $\mathcal{R}_{\alpha_{a}}$
violates the topological exclusion constraint by visiting a historically
forbidden state $\beta\in\mathcal{S}_{a}$, while all other propagators
retain their (unconstrained) full-resolvent form at this level of
approximation.
The number of such single-return channels grows with the hierarchy
depth, increasing the sensitivity of the pure DCA to the erroneous
reopening of inter-resolvent backtracking channels that the projected
hierarchy was designed to exclude. Whether this growth can
compensate the ETH entropy suppression depends on the statistical
structure of the corresponding amplitudes; in delocalized regimes
with strong self-averaging, the individual channels remain mutually
incoherent and their net contribution continues to be controlled by
the random-walk cancellation discussed in \cref{sec:DCA_resolvent_id}.

\subsubsection{Renormalized cavity propagator: Z-factor representation}

The subtraction~\eqref{eq:linear_cavity} removes the leading forbidden
return topology but truncates the internal resummation of repeated
returns ($\alpha\to\beta\to\alpha\to\beta\to\alpha\to\cdots$) at first
order.  In spectral regions where individual return amplitudes are not
negligibly small---near resonances, in the far tails, or in systems
with reduced connectivity---this truncation may be insufficient.

A more complete topological restoration at the local-return level is
obtained by inverting \eqref{eq:cavity_diff_leading} without expanding,
yielding a {\em renormalized cavity propagator}:
\begin{align}\label{eq:Z_factor}
    \mathcal{R}_{\alpha}^{(\mathcal{S})}(z)
    \;\approx\;
    \mathcal{R}_{\alpha}(z)\;\mathcal{Z}_{\alpha}^{(\mathcal{S})}(z),\\
    \mathcal{Z}_{\alpha}^{(\mathcal{S})}(z)
    :=
    \Bigl[\,1
          + \mathcal{R}_{\alpha}(z)
            \sum_{\beta\in\mathcal{S}} |V_{\alpha,\beta}|^{2}\,
            \mathcal{R}_{\beta}(z)
        \Bigr]^{-1}.\notag
\end{align}
The Z-factor $\mathcal{Z}_{\alpha}^{(\mathcal{S})}$ captures an infinite
subclass of local return processes: it keeps the geometric series of
all ladder returns $\alpha\to\beta\to\alpha$ for every $\beta\in\mathcal{S}$
subtracted from the propagator at the level of the denominator.
Expanding the denominator reproduces \eqref{eq:linear_cavity} at leading
order.  For sufficiently weak return couplings and away from pathological
resonances, the denominator is expected to remain analytic in the same
half-plane as the original resolvent, thereby preserving the causal
structure inherited from the spectral representation.

Under this renormalized closure, the hierarchy term acquires a
locally renormalized approximation to the topologically constrained
hierarchy:
\begin{align}\label{eq:DCA_multiplicative}
    \mathcal{G}_{\mu i}^{(\ell)}(z)
    &\approx
    \sum_{\alpha_{1}\neq\cdots\neq\alpha_{\ell-1}\neq\mu i}
    V^{(\ell)}_{\mu i,\alpha_{\ell-1},\dots,\alpha_{1}}
    \prod_{a=1}^{\ell-1}
    \Bigl[
        \mathcal{R}_{\alpha_{a}}(z)\;
        \mathcal{Z}_{\alpha_{a}}^{(\mathcal{S}_{a})}(z)
    \Bigr],
\end{align}
with $\mathcal{S}_{a}$ given by \eqref{eq:nested_constraints}.
Each factor $\mathcal{R}_{\alpha_{a}}\mathcal{Z}_{\alpha_{a}}^{(\mathcal{S}_{a})}$
is the full propagator stripped of its local return topology to the
forbidden set $\mathcal{S}_{a}$; the product over $a$ approximately
restores the nested mutual-exclusion constraints order by order in
the hierarchy.  We emphasize that this construction restores the
{\em local} return topology; the full nonbacktracking hierarchy would
additionally require correlated multi-loop exclusion and nonlocal
interference contributions beyond the scope of the present closure.

The Z-factor representation has two structural advantages.  First,
it is algebraically homologous to the continued-fraction recursion
\eqref{eq:cf_recurrence}, where each level contributes an inverse
denominator $(z-a_{n}-b_{n+1}^{2}R_{n+1})^{-1}$; this homology
suggests that the renormalized cavity factors can in principle be
absorbed into effective Lanczos coefficients, providing a bridge
between the multi-resolvent topological constraints and the
single-resolvent Krylov backbone.  Second, unlike truncated additive
subtractions, the denominator form avoids introducing spurious
singularities that could violate the analytic structure of the
self-energy, provided the return couplings remain moderate.

\subsection{Validity conditions and failure mechanisms of the DCA}
\label{sec:DCA_validity}

The DCA~\eqref{DCAapprox} rests on several interconnected conditions, 
each of which can fail in physically distinct regimes.  When these 
conditions are violated, the multi-resolvent topological corrections 
discussed in~\cref{sec:DCA_multiR_correction} become quantitatively 
non-negligible and must be incorporated through the additive 
closure~\eqref{eq:DCA_additive_correction} or the Z-factor 
renormalization~\eqref{eq:DCA_multiplicative}.

\subsubsection*{Condition 1: Delocalized propagation with exponentially large connectivity}

The DCA requires that each intermediate state $\nu j$ be connected to 
the removed state $\mu i$ through exponentially many independent 
propagation paths.  Two distinct mechanisms can violate this requirement.

In the ergodic (ETH) phase, delocalization is guaranteed by the 
exponential proliferation of Fock-space paths, and the number of 
states connected by $V$ scales as $N_{\rm conn}\sim e^{S}$, ensuring 
self-averaging of the self-energy through entropic dilution 
(\cref{sec:DCA_resolvent_id}).

In a localized (many-body localized, MBL) phase, propagation is
confined to a finite number of paths and the off-diagonal resolvent
decays exponentially with distance:
\begin{equation}
    |\mathcal{R}_{\nu j,\mu i}| \sim e^{-|\nu-\mu|/\xi},
\end{equation}
where $\xi$ is the localization length.
Removing a state then produces an $O(1)$ change in the cavity resolvent,
and the DCA fails.

Even in the absence of localization, if the connectivity is sparse 
(e.g., in integrable systems with selection rules or low-connectivity 
Fock graphs), then $\delta\Sigma\sim |V|^2 R$ is no longer suppressed 
by entropy, and removing a single state can significantly alter the 
self-energy.  In such low-connectivity or quasi-integrable regimes, 
return-path contributions may become leading-order effects even at 
moderate hierarchy depth.

\subsubsection*{Condition 2: Spectral continuity and self-averaging 
in the relevant spectral region}

The DCA itself requires only $e^{-S}\ll 1$ (Condition~1);
it does not intrinsically need a scale-separation window for
the broadening $\eta$.
The $\eta$-window~\eqref{eq:eta_window},
$\delta_{\rm LS}\ll\eta\ll\Gamma_{\mu i}$,
concerns the continuum (coarse-graining) approximation---a
distinct step that regularizes the discrete resolvent for
finite systems.
In the strict thermodynamic limit, $\eta=0^+$ suffices and the
window is moot.
When $\Gamma_{\mu i}\lesssim\delta_{\rm LS}$ (e.g.\ near spectral
edges), the continuum description breaks down and the self-consistent
equations must be applied to the discrete spectrum directly;
the DCA may nevertheless remain valid if $S\gg 1$, as the two
conditions are independent.

The DCA is most reliable in the spectral bulk, where the self-energy
receives contributions from exponentially many channels and the
central limit theorem guarantees self-averaging.
In the far tails of the distribution, the spectral weight is
dominated by a small number of rare propagation paths.
In this regime, $\mathcal{R}_{\alpha}\sim 1/(\omega-\omega_{*})$ with
$|\omega-\omega_{*}|\lesssim\Gamma$, single-resolvent amplitudes
are parametrically enhanced, individual return paths can
become non-negligible, and the DCA may become quantitatively inaccurate.
This is consistent with the Gaussian-to-Lorentzian crossover 
discussed in~\cref{SLGA}, where the tails require a qualitatively 
different (Gaussian) ansatz compared to the bulk (Lorentzian).

\subsubsection*{Condition 3: Strong coherent overlap among resolvent factors}

As the hierarchy itself reveals, terms of the form
$\mathcal{R}_\nu\mathcal{R}_\xi$ in $\mathcal{G}^{(3)}$
involve products of resolvents from different channels.
When $\mathcal{R}_\nu$ and $\mathcal{R}_\xi$ exhibit
strong coherent overlap (near degeneracies, collective
modes, or low-rank resonant subspaces), the projected
resolvent $\mathcal{R}_\xi^{(\mu,\nu)}$ can differ
substantially from $\mathcal{R}_\xi$.
At higher orders of the hierarchy ($\ell\gtrsim 3$), the nested
product of resolvents can develop coherent overlap when
intermediate states are nearly degenerate or belong to a
low-rank resonant subspace.
In such regimes the leading
local-return correction~\eqref{eq:DCA_additive_correction} or
the Z-factor renormalization~\eqref{eq:DCA_multiplicative}
provide a systematically improvable route to enforce the
topological nonbacktracking constraints beyond the pure DCA.
This implies that the DCA is most accurate in the mean-field
(single-resolvent) sector, where the continued-fraction backbone
already captures the essential path correlations.  The DCA may
gradually deteriorate when inter-resolvent interference becomes strong,
precisely because the pure DCA fails to respect the nested
mutual-exclusion constraints among resolvent factors.  In such regimes,
the improved closures~\eqref{eq:DCA_additive_correction}
and~\eqref{eq:DCA_multiplicative} provide the necessary
topological restoration, while the physical inter-resolvent
interference effects---parity mixing, nonlocal frequency coupling, and
skewness---are supplied by the hierarchy itself rather than by the
closure approximation.

\subsubsection*{Additional failure mechanism: Resonant single-channel 
dominance}

When $\lambda\approx a_{\mu i}$ and a single channel dominates
the resolvent, $\mathcal{R}\sim 1/(\lambda-\lambda_*)$,
cavity removal can shift the pole by $O(\Gamma)$, and the
DCA error is amplified by resonance. This is relevant near
mobility edges, quantum scar states, and rare resonances.

\subsubsection*{When does the multi-R correction become quantitatively 
necessary?}

Summarizing the above conditions, the pure DCA~\eqref{DCAapprox} is 
expected to be accurate when (i)~propagation is delocalized 
(Condition~1), (ii)~one operates in the spectral bulk with 
$\Gamma_{\mu i}\gg\delta_{\rm LS}$ and away from the far tails 
(Condition~2), and (iii)~multi-resolvent interference is weak 
(Condition~3).  The improved closures~\eqref{eq:DCA_additive_correction} 
and~\eqref{eq:DCA_multiplicative} become relevant when any of these 
conditions is weakened.

We emphasize that both the additive and multiplicative closures
presented in~\cref{sec:DCA_multiR_correction} are {\em physically 
motivated approximation schemes}, not rigorously controlled expansions.  
Their quantitative accuracy depends on the validity of the 
direct-channel dominance assumption~\eqref{eq:cavity_diff_leading} 
and on the statistical self-averaging of the neglected cross-correlated 
return processes.  Nevertheless, they capture the essential structural 
distinction between projected and full resolvents: projected 
propagators resum only cavity-compatible paths within their 
respective forbidden sets, whereas full resolvents implicitly restore 
forbidden return topology through their internal Dyson structure.

\subsection{Cavity Z-factor expansion and subdominant $|V|^{4}$ structure}
\label{app:cavity_V4}

The exact $|V|^{2}$ self-energy involves the cavity (projected)
propagator,
\begin{equation}\label{eq:exact_g2}
  \mathcal{G}^{(2)}_{\mu i}(z)
  = \sum_{\nu j \neq \mu i} |V_{\mu i,\nu j}|^{2}\,
    \mathcal{R}_{\nu j}^{(\mu i)}(z),
\end{equation}
where $\mathcal{R}_{\nu j}^{(\mu i)}(z)$ is the propagator with the
state $\mu i$ projected out.  The standard SCBA replaces
$\mathcal{R}_{\nu j}^{(\mu i)} \approx \mathcal{R}_{\nu j}$ (the
Diagonal Closure Approximation), which Appendix~\ref{app:no_skewness_mf}
proves is strictly parity-preserving.  Retaining the cavity constraint
through the Z-factor representation,
\begin{equation}\label{eq:Zfactor_cavity}
  \mathcal{R}_{\nu j}^{(\mu i)}(z)
  \approx \frac{\mathcal{R}_{\nu j}(z)}
         {1 + \mathcal{R}_{\nu j}(z)\,|V_{\nu j,\mu i}|^{2}\,
              \mathcal{R}_{\mu i}(z)},
\end{equation}
and expanding the denominator yields
\begin{align}\label{eq:g2_multires}
  \mathcal{G}^{(2)}_{\mu i}(z)
  \approx \sum_{\nu j \neq \mu i} |V_{\mu i,\nu j}|^{2}\,\mathcal{R}_{\nu j}(z)\notag\\
        - \sum_{\nu j \neq \mu i}
          |V_{\mu i,\nu j}|^{2}\,|V_{\nu j,\mu i}|^{2}\,
          \mathcal{R}_{\nu j}^{2}(z)\,\mathcal{R}_{\mu i}(z)
        + \cdots .
\end{align}
The $|V|^{4}$ correction is structurally a multi-resolvent product
$\mathcal{R}_{\nu j}^{2}\mathcal{R}_{\mu i}$, whose imaginary part
(through the Kramers--Kronig decomposition) contains interference
terms that mix even and odd parity sectors under frequency
reflection---qualitatively analogous to the $\mathcal{G}^{(3)}$
mechanism discussed in Sec.~\ref{sec:cross_expansion}.  This
illustrates that multi-resolvent correlations are already seeded at
the level of the cavity-dressed $|V|^{2}$ sector, without requiring
Z-factor expansion.

Numerically, however, the explicit $|V|^{4}$ suppression renders this
contribution quantitatively subdominant for the skewness observed in
typical many-body spectra.  The leading microscopic mechanism for
spectral asymmetry therefore remains the explicit multi-resolvent
hierarchy $\mathcal{G}^{(\ell\ge 3)}$ developed in
Sec.~\ref{sec:cross_expansion}, whose leading term
$\mathcal{G}^{(3)}_{\mu i}$ enters at order $|V|^{3}$ and generates
odd-parity self-energy components through Hilbert-transform
convolutions of distinct resolvent channels.

\end{document}